\documentclass[prd,onecolumn,notitlepage,eqsecnum,nofootinbib,floatfix]{revtex4-1}
\usepackage{amssymb}
\usepackage{amsmath}
\usepackage{amsfonts} 
\usepackage{bm}
\usepackage{graphicx}
\usepackage{comment} 
\newcommand{\Lie}{{\pounds}}

\newcommand{\E}{{\cal E}} 
\newcommand{\EE}{({\cal EE})} 
\newcommand{\bEE}{(\bar{\cal E} \bar{\cal E})} 
\newcommand{\stf}[1]{{\langle #1 \rangle}} 
\newcommand{\BH}{\mbox{\sc bh}} 
\newcommand{\n}{\mbox{\sc n}} 
\newcommand{\pn}{\mbox{\sc pn}} 
\newcommand{\gothm}{{\mathfrak{M}}} 
\newcommand{\gothn}{{\mathfrak{N}}} 
\newcommand{\gotho}{{\mathfrak{O}}} 
\newcommand{\gotha}{{\mathfrak{A}}}
\newcommand{\gothb}{{\mathfrak{B}}} 
\newcommand{\gothc}{{\mathfrak{C}}} 
\newcommand{\gothd}{{\mathfrak{D}}}
\newcommand{\gothf}{{\mathfrak{F}}}
\newcommand{\gothg}{{\mathfrak{G}}}
\newcommand{\dilog}{{\mbox{dilog}}} 
\newcommand{\polylog}{{\mbox{polylog}}} 
\allowdisplaybreaks
\begin{document}
\title{Compact body in a tidal environment: New types of relativistic Love numbers, and a post-Newtonian operational definition for tidally induced multipole moments} 
\author{Eric Poisson}  
\affiliation{Department of Physics, University of Guelph, Guelph,
  Ontario, N1G 2W1, Canada} 
\date{February 7, 2021} 
\begin{abstract} 
We examine the tidal deformation of a nonrotating compact body (material body or black hole) in general relativity. The body's exterior metric is calculated in a simultaneous expansion in powers of the ratio between the distance to the body and three distinct length scales. The first is the radius of curvature of the external spacetime in which the body is inserted, the second is the scale of spatial inhomogeneity of the curvature, and the third is the scale of temporal variation. The metric is valid in the body's immediate neighborhood, which excludes the external matter responsible for the tidal environment. The body's tidal response is encapsulated in four types of relativistic Love numbers. The first is $k_\ell$, the familiar Love number that measures the linear response to a static tidal field. A second is $p_\ell$, which measures the quadratic response to the tidal field. A third is $\dot{k}_\ell$, which is associated with the first time derivative of the tidal field. And the fourth is $\ddot{k}_\ell$, associated with the second time derivative of the tidal field. All Love numbers are gauge-invariant constants that appear in the body's exterior metric. Their computation (not carried out here, except for black holes) requires extending the metric to the body's interior. The Love numbers acquire an operational meaning through the definition of tidally induced multipole moments. Previously proposed definitions for the moments suffer from ambiguities associated with the subtraction of a ``pure tidal field'' from the full metric; such subtraction prescriptions are artificial and subjective. A robust operational definition is proposed here. It relies on inserting the body's local metric within a global metric constructed in post-Newtonian theory; the global metric includes the external matter responsible for the tidal environment. When viewed in the post-Newtonian spacetime, the compact body appears as a skeletonized object with a specific multipole structure, moving on a given world line. The tidally induced multipole moments provide a description of this multipole structure. They manifest themselves, for example, in the body's tidal acceleration, which is nonlinear in the tidal field. At leading order in the tidal interaction, the acceleration is proportional to the $k_2$ Love number as calculated in full general relativity. The computation of the multipole moments is carried out to the first post-Newtonian order, but the general method can in principle be extended to higher orders. 
\end{abstract} 
\maketitle

\section{Introduction and summary} 
\label{sec:intro} 

\subsection{Tidal deformation of compact bodies and gravitational-wave astronomy} 

Our knowledge of the internal constitution of neutron stars is limited by a poor understanding of the physical properties of nuclear matter at extremely high densities \cite{ozel-freire:16, oertel-etal:17, baym:18}. To make progress on this front it is proving necessary to combine the best models from nuclear theory with reliable astronomical observations, in an effort to constrain the equation of state and mass-radius relation of neutron stars. Two recent developments are adding momentum to this quest. The first is the deployment of the NICER instrument (Neutron star Interior Composition Explorer) on board the International Space Station, which estimates the mass and radius of pulsars from a measurement of their X-ray curves \cite{miller-etal:19}. The second is the dawn of gravitational-wave astronomy.  

The first observation of the coalescence of a neutron-star binary occurred on 17 August 2017, with a near-simultaneous measurement in gravitational waves (GW170817 \cite{GW170817:17}) and gamma rays \cite{GW170817fermi:18, GW170817integral:17}. The resulting afterglow was then observed in a multitude of wavelengths of the electromagnetic spectrum (see Ref.~\cite{GW170817multi:17} for a summary), and evidence was presented that such events could hold the key to the production of heavy nuclei through the $r$-process \cite{siegel-metzger:17}. Another binary neutron-star merger may also be the source of the GW190425 event \cite{GW198425:20}, but no electromagnetic emissions were reported in this case. It is conceivable (though perhaps not likely) that GW190814 \cite{GW190814:20} was the result of a heavy neutron star merging with a black hole.    

The measurement of gravitational waves from merging neutron stars will become routine in the near future, and we may hope that these signals will soon deliver a reliable estimate of the tidal deformability of a neutron star. The method to achieve this was first identified by Flanagan and Hinderer \cite{flanagan-hinderer:08}: each neutron star deforms under the tidal forces exerted by its companion, the deformation affects the orbital motion and produces its own emission of gravitational waves, and the combination of these effects manifests itself in the phasing of the waves. Because the tidal deformation of a body depends sensitively on the mass distribution, a measurement in gravitational waves can disclose some precious clues regarding the star's interior. This method can thus be exploited to constrain the equation of state, and it complements in a nice way the constraints provided by X-ray observations \cite{landry-essick-reed-chatziioannou:20}. A measurement of the tidal deformability  was already attempted for GW170817, and it delivered an upper bound of astrophysical  significance \cite{GW170817:18}. 

The observational work rests on a large body of theoretical investigations. These range from providing a proper formulation of the relativistic theory of tidal deformability \cite{damour-lecian:09, damour-nagar:09, binnington-poisson:09, landry-poisson:14}, to determining how well it can be measured in gravitational waves, and how much light it can shed on the properties of matter at super-nuclear densities (see Ref.~\cite{chatziioannou:20} for a recent review). 

This work is concerned with the foundations of the theory of tidal deformability in general relativity. My aims are twofold. First, I wish to extend the theory by introducing new classes of relativistic Love numbers, which can be used to characterize the tidal deformation of a compact body (either a material body such as a neutron star, or a black hole). Second, I wish to supply these Love numbers with a robust operational definition, which should allay doubts regarding their true ability to give a meaningful measure of tidal deformation. The new Love numbers are beyond the reach of current measurement, but they nevertheless enrich the relativistic theory of tidal deformability. The operational definition provides what I hope is a helpful and constructive way to view these quantities.     

\subsection{New types of Love numbers} 
\label{subsec:newlove} 

I calculate the exterior metric of a tidally deformed, nonrotating compact body in Sec.~\ref{sec:GR}; the computation is carried out in general relativity. This calculation generalizes previous work \cite{poisson-vlasov:10, taylor-poisson:08, poisson-corrigan:18} that was concerned exclusively with black holes. Here the compact object can be a black hole or a material body such as a neutron star. It has a mass $M$ and an unperturbed radius $R$.  

The metric is presented as a simultaneous expansion in powers of the ratio of $r$, the distance to the body, to three distinct length scales. The first is ${\cal R}$, the radius of curvature of the external spacetime in which the body is inserted, the second is ${\cal L}$, the scale of spatial inhomogeneity of the curvature, and the third is ${\cal T}$, the scale of temporal variation. The metric is valid in the body's immediate neighborhood, which excludes the external matter responsible for the tidal environment. The tidal field is characterized by multipole moments $\E_{ab}(t)$, $\E_{abc}(t)$, and so on; these are symmetric and tracefree Cartesian tensors that depend on time only. The tidal multipole moments appear in the metric as freely-specifiable functions that are not determined by the Einstein field equations; they can be loosely interpreted as components of the ``external Weyl tensor'' --- the curvature tensor produced by the external matter --- and its derivatives. The expansion in powers of $r/{\cal R}$ is manifested in terms nonlinear in the tidal moments, the expansion in $r/{\cal L}$ coincides with the multipole series, and the expansion in $r/{\cal T}$ introduces time derivatives of the tidal moments. The metric is obtained in full general relativity; there is no expansion in powers of $M/r$.  

A representative sample of the metric of Sec.~\ref{sec:GR} is 
\begin{align} 
g_{tt} &= -1 + \frac{2M}{r} 
- \biggl[ r^2 (1 + \cdots) + 2 k_2 R^5 \frac{1}{r^3} (1 + \cdots) \biggr]\, \E_{ab} \Omega^a \Omega^b 
\nonumber \\ & \quad \mbox{}
- \biggl[ \frac{2}{7} r^4 (1 + \cdots) + 2 p_2 \frac{R^8}{M} \frac{1}{r^3} (1 + \cdots) 
+ \frac{8}{7} k_2 R^5 \frac{1}{r} (1 + \cdots)
+ \frac{8}{7} k_2^2 R^{10} \frac{1}{r^6} (1 + \cdots) 
\biggr]\, \E_{c\langle a} \E^c_{\ b\rangle} \Omega^a \Omega^b 
\nonumber \\ & \quad \mbox{}
- 2 \dot{k}_2 M R^5 \frac{1}{r^3} (1 + \cdots)\, \dot{\E}_{ab} \Omega^a \Omega^b 
- \biggl[ \frac{11}{42} r^4 (1 + \cdots) + 2 \ddot{k}_2 \frac{R^8}{M} \frac{1}{r^3} (1 + \cdots) 
+ k_2 R^5 \frac{1}{r} (1 + \cdots) \biggr]\, \ddot{\E}_{ab} \Omega^a \Omega^b. 
\label{metric_sample} 
\end{align} 
In the second group of terms, the angular brackets around tensor indices are an instruction to symmetrize the indices and remove all traces (with respect to the Euclidean metric); the operation returns another symmetric and tracefree tensor. In the third and fourth groups of terms, an overdot on $\E_{ab}$ indicates differentiation with respect to time. 
The sample includes quadrupole terms ($\ell = 2$) only; the metric of Sec.~\ref{sec:GR} incorporates multipoles up to $\ell = 5$, and the remaining components of the metric are also provided there. 

The sample of Eq.~(\ref{metric_sample}) implicates the tidal quadrupole moment $\E_{ab}(t)$, which is joined with the unit radial vector $\Omega^a := [\sin\theta\cos\phi, \sin\theta\sin\phi, \cos\theta]$ to form a basis of spherical harmonics. The growing terms, those with a positive power of $r$, represent the applied tidal field. The decaying terms, with a negative power of $r$, represent the body's response to the tidal field. The response comes with dimensionless constants, $k_2$, $p_2$, $\dot{k}_2$, and $\ddot{k}_2$; these are Love numbers associated with the compact body. The first one, $k_2$, is the familiar number that characterizes the body's response to leading order in the simultaneous curvature, spatial-derivative, and time-derivative expansions. The second one, $p_2$, is a number associated with terms quadratic in the curvature. The third and fourth numbers, $\dot{k}_2$ and $\ddot{k}_2$, are associated with time-derivative terms. In Eq.~(\ref{metric_sample}), the factors denoted $1 + \cdots$ stand for functions of $M/r$ that are fully identified in Sec.~\ref{sec:GR}; each one admits the asymptotic expansion $1 + O(M/r)$ when $r \gg M$. 

At the order of accuracy maintained in Sec.~\ref{sec:GR}, the body's response to an applied tidal field is measured by the Love numbers $k_\ell$, $p_\ell$, $\dot{k}_\ell$, and $\ddot{k}_\ell$. These are a property of the exterior metric, and in this context the tidal deformation is measured in terms of the metric perturbation, the deviation of the actual metric with respect to the Schwarzschild form. The Love numbers are gauge invariant in the sense of perturbation theory: if we take the metric perturbation to be infinitesimal, then the value of each Love number is invariant under an infinitesimal coordinate transformation. This can be seen in a number of ways. The most direct is to point out that the metric of Sec.~\ref{sec:GR} is presented in the Regge-Wheeler gauge, which is unique\footnote{The Regge-Wheeler gauge requires an extension to account for the monopole ($\ell = 0$) and dipole ($\ell = 1$) sectors of the perturbation. These extensions are not unique, and the metric of Sec.~\ref{sec:GR} actually comes with two additional constants that reflect the residual gauge freedom. This freedom, however, has no impact on the Love numbers, which are defined for $\ell \geq 2$ only.}; because the metric is completely gauge-fixed, quantities that appear in it are necessarily gauge invariant. Another way to demonstrate gauge invariance involves mapping the metric perturbation to a gauge-invariant quantity such as the Zerilli-Moncrief master function or a Weyl scalar constructed from a null tetrad; in this situation the Love numbers are read off from a gauge-invariant quantity and are therefore themselves gauge invariant.

As was stated, the Love numbers $k_\ell$, $p_\ell$, $\dot{k}_\ell$, and $\ddot{k}_\ell$ provide a measure of the tidal deformation of a compact body, as manifested in the metric perturbation. While $k_\ell$ is well known, the nonlinear number $p_\ell$ and the time-derivative numbers $\dot{k}_\ell$, $\ddot{k}_\ell$ are not, and the first goal of this work is to introduce them and supply them with a proper relativistic definition. The numbers are defined here, but they are not computed for any type of material body, for example, a neutron star with a realistic equation of state. Such a computation would require the construction of an interior metric to complement the exterior metric of Sec.~\ref{sec:GR}, and the value of each Love number would be determined by joining the interior and exterior metrics smoothly at the body's deformed surface. These computations will be left for future work. 

It is possible, however, to compute the Love numbers for a black hole, and I take this on in Sec.~\ref{sec:GR}. 
In this case the exterior metric extends to the interior, and it must be smooth across the deformed event horizon. This requirement dictates that 
\begin{equation} 
k_\ell[\BH] = p_\ell[\BH] = \ddot{k}_\ell[\BH] = 0 
\label{BH_love1} 
\end{equation} 
but that 
\begin{equation} 
\dot{k}_\ell[\BH] = -\frac{(\ell-2)!(\ell-1)!\ell!(\ell+2)!}{2 (2\ell-1)!(2\ell+1)!}. 
\label{BH_love2} 
\end{equation} 
That $k_\ell = 0$ for a black hole is a well-known result \cite{binnington-poisson:09}. The vanishing of $p_\ell$ is compatible with the results of Ref.~\cite{gurlebeck:15}. The results for $\dot{k}_\ell$ and $\ddot{k}_\ell$ are new.  

\subsection{Geroch-Hansen multipole moments} 
\label{subsec:GH}

Having introduced the Love numbers as gauge-invariant properties of the metric perturbation, I turn to the question of how to give these numbers a useful operational definition, so that they can correctly be involved in the calculation of observables. To illustrate the obstacles involved in this identification, I first review what I consider to be a failed attempt at an operational definition. 

We wish to turn the Love number $k_2$ into a quadrupole moment $Q_{ab}$ that provides a meaningful measure of the body's deformation. (I leave the additional Love numbers out of the discussion; they will be reinstated at a later stage.) An approach for doing this was proposed in the literature \cite{pani-etal:15a, letiec-casals:20, letiec-casals-franzin:20}: take the perturbed metric, subtract out the tidal field, retain only the body's response, and calculate the multipole moments with the Geroch-Hansen method \cite{geroch:70, hansen:74}. This approach is artificial: the actual metric is a superposition of tidal field and body response, and both pieces are essential to an accurate description of the gravitational field. The approach is also ambiguous: any attempt to subtract ``the'' tidal field must be based on a subjective and arbitrary definition for what this field should be. The truth of the matter is that the metric is necessarily limited to a neighborhood of the body, and it is not asymptotically flat; any attempt to make it so introduces arbitrary elements that bring obscurity to the enterprise. The mathematical rigor that Geroch and Hansen brought to the definition of multipole moments in general relativity is wasted here, because it is overshadowed by the ambiguity of the subtraction prescription. Moreover, their method requires the spacetime to be stationary, which is not the case when the tidal environment varies with time (which is the norm). 

To better understand the issues involved in this approach, let us examine the leading term in the metric perturbation, given by 
\begin{equation} 
\delta g_{tt} = -f^2 \bigl( r^2\, A_2 + 2 k_2 R^5 r^{-3}\, B_2 \bigr) \, \E_{ab} \Omega^a \Omega^b, 
\label{deltag} 
\end{equation} 
where $f^2 A_2$ and $f^2 B_2$, with $f := 1-2M/r$, are the precise identities of the functions denoted $(1 + \cdots)$ in Eq.~(\ref{metric_sample}); explicit expressions are given in Appendix~\ref{app:ABCD}. This perturbation satisfies the vacuum field equations linearized with respect to the Schwarzschild solution. As was stated, the perturbed metric is not asymptotically flat, because of the term involving $r^2 A_2$. To produce one that is asymptotically flat, we must subtract out what might be defined as a tidal field. Because $r^2 A_2$ and $r^{-3} B_2$ constitute a basis of solutions to the perturbation equations, the tidal field must be of the form 
\begin{equation} 
\delta g^{\rm tidal}_{tt} = -f^2 \bigl( r^2\, A_2 + 2 \lambda^5 r^{-3} B_2 \bigr)\, \E_{ab} \Omega^a \Omega^b, 
\label{deltag_tidal} 
\end{equation} 
where $\lambda$ is an arbitrary constant with the dimension of length. Subtraction produces 
\begin{equation} 
\delta g^{\rm af}_{tt} = -2 f^2(k_2 R^5 - \lambda^5) r^{-3} B_2 \, \E_{ab} \Omega^a \Omega^b, 
\end{equation} 
and the metric is now asymptotically flat. The Geroch-Hansen procedure then assigns the body a tidally induced quadrupole moment given by 
\begin{equation} 
Q_{ab} = -\frac{2}{3}  (k_2 R^5 - \lambda^5)\, \E_{ab}.  
\end{equation} 
The conclusion is unavoidable: the quadrupole moment is ambiguous, because it depends on the parameter $\lambda$ introduced in the subtraction prescription. A scaling $\lambda^5 \sim R^5$ affects the quadrupole moment at Newtonian order, a scaling $\lambda^5 \sim M R^4$ affects it at the first post-Newtonian order, and so on. A scaling $\lambda^5 \sim M^5$ would perhaps be more sensible, because the candidate tidal field should not depend on the body's radius, and this scaling would affect the quadrupole moment at the fifth post-Newtonian order. The point remains that the need for a subtraction prescription brings ambiguity to the quadrupole moment, and this ambiguity cannot be resolved as a matter of principle.\footnote{The subtraction prescription rests on a linearized description of the tidal perturbation. Beyond linear order, a superposition of solutions is no longer a solution to the field equations, and the method fails entirely.} The argument generalizes to any multipole order $\ell$. 

An alternative operational definition for Love numbers is provided by an effective action for point particles \cite{damour-espositofarese:96, goldberger-rothstein:06a, bini-damour-faye:12, kol-smolkin:12, chakrabarti-delsate-steinhoff:13a, porto:16, kalin-liu-porto:20}, in which they appear as coupling constants in front of scalars constructed from the tidal moments. This approach also relies on a subtraction prescription, because the tidal moments inserted in the world-line action are meant to represent the external curvature only, and to leave out the particle's own contribution to the curvature, which is formally infinite.  

\subsection{Definition of a Love number} 
\label{subsec:redef} 

Sam Gralla pointed out \cite{gralla:18} that Love numbers can be defined in a multitude of ways, and that this also may bring ambiguity to a proposed operational definition for these numbers. I subscribe fully to Gralla's freedom of definition, but will show below that it produces no ambiguity when the tidally induced multipole moments are viewed in a proper way. 

The considerations are similar to, but subtly distinct from, those of the preceding subsection. The main point is that the definition of $k_2$ rests on a choice of basis $r^2 A_2$, $r^{-3} B_2$ of solutions to the (linearized) perturbation equations; the Love number is attached to the second member of the basis. In the ``standard'' definition of the Love number, $r^2 A_2$ is distinguished by the fact that it is a growing solution to the perturbation equation and a terminating polynomial in $M/r$. On the other hand, $r^{-3} B_2$ is a decaying solution that possesses terms proportional to $\ln(1-2M/r)$; its expansion in powers of $M/r$ does not terminate. The decaying solution is not smooth across $r = 2M$, and for this reason it must be eliminated when the compact body is a black hole; this is done by setting $k_2 = 0$.   

The ``standard'' definition of the Love number assigns mathematical significance to the surface $r=2M$, because it is the presence of terms proportional to $\ln(1-2M/r)$ that permits a precise decomposition of the metric perturbation in terms of growing and decaying pieces; the growing piece is not allowed to contain logarithms. The surface, however, has no physical significance when the compact object is a material body, because it is then situated inside the body, where the exterior metric does not apply. In such situations, the logarithms do not matter. 

Following Gralla, we may then consider a change to a new basis $r^2 \bar{A}_2$, $r^{-3} \bar{B}_2$ given by 
\begin{equation} 
r^2 \bar{A}_2  = r^2 A_2 + 2 \lambda^5 r^{-3} B_2, \qquad 
r^{-3} \bar{B}_2 = r^{-3} B_2. 
\label{basis_change} 
\end{equation} 
The first member, $r^2 \bar{A}_2$, is still growing, but it now contains logarithms and is no longer smooth across $r = 2M$; the second member, $r^{-3} \bar{B}_2$, is unchanged. The metric perturbation now takes the form 
\begin{equation} 
\delta g_{tt} = -f^2 \Bigl[ r^2\, \bar{A}_2 + 2 (k_2 R^5 - \lambda^5) r^{-3}\, \bar{B}_2 \Bigr] \, \E_{ab} \Omega^a \Omega^b, 
\label{deltag_new} 
\end{equation} 
and it comes with the new Love number $\bar{k}_2 = k_2 - (\lambda/R)^5$. This calculation, which generalizes easily to any multipole order $\ell$, shows very clearly that the numerical value of a Love number depends on its definition; a different definition --- a different choice of basis --- will give rise to a different value. In view of this, Gralla concludes that while individual Love numbers may not be well defined, differences in Love numbers for distinct bodies of the same mass are meaningful.  

I shall go another way: individual Love numbers are well defined and meaningful. Throughout the paper I adhere to the ``standard'' definition of these numbers.  

\subsection{Multipole moments from post-Newtonian matching} 
\label{subsec:pNmatch} 

And now a proposal for an operational definition of tidally induced multipole moments, one which eschews the subtraction ambiguity reviewed in Sec.~\ref{subsec:GH} and is invariant against the redefinitions of Sec.~\ref{subsec:redef}, up to a given post-Newtonian order. The ideas outlined here can be traced to Damour \cite{damour:83}; the implementation carried out here adds much flesh to these ideas. 

The metric of Sec.~\ref{sec:GR}, sampled in Eq.~(\ref{metric_sample}), applies to a small neighborhood around the body, and the external matter responsible for the tidal environment lies outside this neighborhood. Because the domain of consideration is thus limited, the body metric is expressed in terms of tidal multipole moments that are not determined by the Einstein field equations. To be of use, the metric should be inserted in a larger spacetime that contains the external matter; the wider domain of consideration then permits the determination of the tidal moments. This could be done in a number of ways. For example, the body metric could be matched to a global metric computed during a numerical simulation of the field equations. 

I shall construct a global metric in Sec.~\ref{sec:PN}, taking the mutual gravity between compact body and external matter to be weak, and relying on post-Newtonian theory. When viewed from the vantage point of the post-Newtonian spacetime, the compact body appears to be a skeletonized object --- I prefer not to use the phrase ``point particle'' --- with a specific multipole structure, moving on some world line. The domain of consideration of the post-Newtonian metric, however, cannot include the compact body, because its internal gravity is too strong to be adequately captured by a post-Newtonian series. The domain excludes the body, and as a consequence, the post-Newtonian metric provides only a partial description of the spacetime. The partial metrics carry complementary information: the body metric depends on unknown tidal moments, and the post-Newtonian metric depends on an unknown multipole structure for the skeletonized body, which moves on an unknown world line. 

The operational definition of the tidally induced multipole moments is this: the moments are a property of the compact body when viewed as a skeletonized object moving in a post-Newtonian spacetime. Their definition does not require an artificial subtraction of a tidal field. The definition, however, is an approximate one based on the post-Newtonian approximation to general relativity. In practice, the multipole moments are defined up to a given post-Newtonian order. 

The post-Newtonian metric and the body metric give different descriptions of the same gravitational field, and these must agree in a region of common validity; this region must be sufficiently far from the body that its gravity becomes weak, and sufficiently close to accommodate the expansion in tidal multipole moments. A detailed comparison between the metrics in this region permits the determination of the tidal moments from post-Newtonian information, and the determination of the multipole structure from relativistic information. Moreover, the comparison is performed after the post-Newtonian metric is transformed to the rest frame of the compact body, and the body's equations of motion are revealed by this coordinate transformation. Matching the metrics, therefore, permits a complete determination of the spacetime. 

The comparison reveals that the mass quadrupole moment of the skeletonized object is related to the tidal quadrupole moment by 
\begin{equation} 
Q_{ab} = -\frac{2}{3} k_2 R^5\, \E_{ab} - \frac{2}{3} p_2 \frac{R^8}{M}\, \E_{c\langle a} \E^c_{\ b\rangle} 
- \frac{2}{3} \ddot{k}_2 \frac{R^8}{M}\, \ddot{\E}_{ab} + \mbox{post-Newtonian corrections}. 
\label{Q_vs_E} 
\end{equation} 
This relation supplies the Love numbers $k_2$, $p_2$, and $\ddot{k}_2$, defined and calculated in full general relativity, with a robust operational meaning in post-Newtonian theory. The multipole moments refer to a specific coordinate system: the harmonic coordinates employed in the description of the post-Newtonian metric. With suitable boundary conditions imposed on the metric --- regularity at the spatial origin of the coordinates, zero ADM momentum at spatial infinity, outgoing waves at future null infinity --- the coordinates are unique up to a global rotation. The gauge is completely fixed, and the multipole moments are therefore gauge invariant. 

The relation of Eq.~(\ref{Q_vs_E}) is derived in Sec.~\ref{sec:PN}, along with all corrections of the first post-Newtonian order --- see Eq.~(\ref{Qab_bary}). The comparison between local and global metrics also provides relations between the tidal multipole moments and the potentials $U$, $U^a$, $\Psi$ that appear in the post-Newtonian metric --- see Eqs.~(\ref{Eab_bary}) and (\ref{Eabc_bary}). It turns out that the tidal moments depend exclusively on the potentials $U_{\rm ext}$, $U_{\rm ext}^a$, $\Psi_{\rm ext}$ that are created by the external matter; those created by the skeletonized object do not appear in these relations. This separation of potentials is an outcome of the comparison; it requires no prescription, and it involves no arbitrary ingredients. 

The quadrupole moment of Eq.~(\ref{Q_vs_E}) is invariant under the redefinitions examined in Sec.~\ref{subsec:redef}. To see this, it suffices to note that the comparison between metrics associates the quadrupole moment with all terms in Eq.~(\ref{metric_sample}) that come with a factor of $r^{-3}$ when expanded in powers of $M/r$. Now, inspection of Eq.~(\ref{deltag}) reveals that the $r^{-3}$ term is attached solely to $r^{-3} B_2$; $r^2 A_2$ does not contain such a term. For this choice of basis, therefore, the quadrupole moment is proportional to $k_2$. Under the change of basis described by Eq.~(\ref{basis_change}), $r^2 \bar{A}_2$ acquires a term proportional to $r^{-3}$, and this new term must be accounted for in the calculation of the quadrupole moment. What happens is that the new term cancels out the change in the Love number, and the overall coefficient in front of $r^{-3}$ remains proportional $k_2$. The relation of Eq.~(\ref{Q_vs_E}) is therefore invariant under the change of basis.  

The Love numbers $k_2$, $p_2$, and $\ddot{k}_2$ are all zero for a black hole. Does this mean that $Q_{ab}$ necessarily vanishes for a (nonrotating) black hole? The answer is no. The quadrupole moment of Eq.~(\ref{Q_vs_E}) is calculated to the first post-Newtonian order only. All that can be said is that when viewed from the vantage point of the post-Newtonian spacetime, the black hole appears as a skeletonized object with a vanishing $Q_{ab}$ at Newtonian and first post-Newtonian orders. Its multipole structure beyond these orders is not available given the confines of this work, but in principle, an extension to higher post-Newtonian orders should be able to deliver it. It would not be surprising to find that for a black hole, $Q_{ab}$ is nonzero at the fifth post-Newtonian order.        

The operational definition of tidally induced multipole moments, as developed here, is tied to the specific framework of post-Newtonian theory, and less essentially, to the theory's reliance on harmonic coordinates. Can one do better, and formulate a theory of these moments in full general relativity, without coordinates, a timelike Killing vector, asymptotic flatness, and a subtraction prescription? Perhaps so. Future will tell.  

\subsection{Tidal acceleration} 
\label{subsec:acceleration} 

The relativistic Love numbers can be tied to the tidal acceleration of the compact body. In Newtonian gravity, it is well known that a body with a quadrupole moment is subjected to an acceleration that results from the coupling between the moment and three derivatives of the external gravitational potential (refer, for example, to Sec.~1.6 of Poisson and Will's {\it Gravity} \cite{poisson-will:14}). When the quadrupole moment is induced by a tidal field (two derivatives of the external potential), the acceleration is proportional to the Love number $k_2$. 

The same is true in general relativity. It was previously stated that the comparison between the local body metric and the global post-Newtonian metric involves a coordinate transformation to the body's rest frame, and that this transformation reveals the body's equations of motion (when viewed in the post-Newtonian spacetime). I show in Sec.~\ref{sec:PN} that these equations contain a tidal acceleration term given by 
\begin{equation} 
g_a^{\rm tidal} = \frac{1}{3} k_2 \frac{R^5}{M}\, \E_{abc} \E^{bc} 
+ \mbox{post-Newtonian corrections}. 
\label{tidal_acc} 
\end{equation} 
The full statement of the equations of motion appears in Eqs.~(\ref{acceleration}) and (\ref{gtidal}); they are given in terms of the potentials $U_{\rm ext}$, $U_{\rm ext}^a$, $\Psi_{\rm ext}$ created by the external matter. More explicit versions of these equations were presented previously \cite{damour-nagar:10, vines-flanagan:13} and generalized to higher post-Newtonian orders \cite{bini-damour-faye:12, henry-faye-blanchet:20a}. While Eq.~(\ref{tidal_acc}) takes the same form as the Newtonian expression, it is important to note that $k_2$ is the Love number that is defined and calculated in full general relativity. 

The fact that $k_2 = 0$ for a black hole implies that its tidal acceleration vanishes at Newtonian and post-Newtonian orders. What happens beyond these orders, especially at the fifth post-Newtonian order, remains to be seen. The black hole may well undergo a tidal acceleration, but it is not clear that this could be distinguished from monopole terms in the equations of motion. The methods exploited in this paper, suitably extended to the fifth post-Newtonian order, can in principle provide answers to these open questions. 

\subsection{Organization of this paper} 
\label{subsec:organization} 

The heavy lifting is carried out in Secs.~\ref{sec:GR} and \ref{sec:PN}. The local metric of a tidally deformed compact body in general relativity, sampled briefly in Eq.~(\ref{metric_sample}), is described in full in Sec.~\ref{sec:GR}. As was stated, this metric is constructed as a simultaneous expansion in powers of $r/{\cal R}$, $r/{\cal L}$, and $r/{\cal T}$, and is given to a degree of accuracy that is sufficient to extract all the information reviewed above. In particular, the metric includes terms that are quadratic in the tidal moments, and terms that involve time derivatives of these moments. The global post-Newtonian metric is constructed in Sec.~\ref{sec:PN}. It is first presented in the barycentric frame of the entire system (compact body and external matter), and then transformed to the body's rest frame. The local and global metrics are next compared term by term in a domain of common validity, and as was summarized above, the procedure reveals the relation between mass and tidal multipole moments, the relation between the tidal moments and the post-Newtonian potentials of the external matter, and the body's equations of motion.  

The heavy lifting comes after a few warmup exercises. I set the stage in Sec.~\ref{sec:moments} with an informal
definition of the tidal moments $\E_{ab}$, $\E_{abc}$, and so on, the formation of relevant bilinear combinations of these, and their merger with the unit radial vector $\Omega^a$ to form a basis of angular functions --- spherical harmonics in disguise. The tidal moments are introduced more formally in Sec.~\ref{sec:worldline}, where I calculate the metric of a generic vacuum spacetime in a neighborhood of a timelike geodesic. In this context the tidal moments can be given a precise definition in terms of the spacetime's Weyl tensor and its derivatives, all evaluated on the world line. The construction of this metric is a useful undertaking, because it allows me to explain, in the simplest manner, the calculational strategies to be deployed in Sec.~\ref{sec:GR}. The metric also provides a valuable check on the forthcoming results: the metric of Sec.~\ref{sec:GR} reduces to it in the limit $M \to 0$. 

I examine the tidal deformation of a fluid body in Newtonian theory in Sec.~\ref{sec:constant-density}; for maximum simplicity the body is taken to have a uniform density. This excursion also provides useful guidance in preparation of the relativistic implementation of Sec.~\ref{sec:GR}. In particular, the nonlinear dependence of the Newtonian potential on the tidal moments is revealed most directly, as is its dependence on the time derivatives. The Newtonian treatment also discloses the natural scaling of the mass multipole moments in relation to the tidal moments. In addition, it identifies the tidal acceleration of Eq.~(\ref{tidal_acc}): it appears in the inertial potential that accounts for the body's motion in the field of the external matter.  

After the heavy lifting, the cool down. Appendices contain technical material that would hinder the flow of presentation in the bulk of the paper. I review the definition of the Regge-Wheeler gauge for perturbations of the Schwarzschild spacetime in Appendix~\ref{app:RW}, and show that the gauge is unique. In Appendix~\ref{app:Legendre}, I relate two representations of the radial functions that appear in the linearized and static piece of the metric perturbation. The first representation is in terms of hypergeometric functions, the second involves Legendre functions. The relation between the two is put to use in the computation of Love numbers for black holes in Sec.~\ref{subsec:BHLove}. I provide an explicit listing of all radial functions that appear in the metric of Sec.~\ref{sec:GR} in Appendices~\ref{app:ABCD}, \ref{app:MNO}, \ref{app:edot_tr}, and \ref{app:ABCDFG}.     

\subsection{No gravitomagnetic tidal moments} 

Throughout this work I set to zero all gravitomagnetic tidal moments ${\cal B}_{ab}$, ${\cal B}_{abc}$, and so on. This is not meant to represent a physical condition on the body's tidal environment; the actual gravitomagnetic moments do not vanish. Instead, the assignment is made because I wish to focus my attention exclusively on the gravitoelectric sector of the tidal perturbation. As will be clear from the preceding presentation, the main intent behind (part of) this work was to provide an operational definition to $k_2$ and $Q_{ab}$. This does not require consideration of the gravitomagnetic sector. This omission, which does represent an incompleteness of the work reported here, shall have to be remedied in the future. The payoff will be a similar operational definition for gravitomagnetic Love numbers and tidally induced current multipole moments.  

\section{Tidal moments and angular functions} 
\label{sec:moments} 

\subsection{Linear tidal moments} 

A compact body is immersed in a tidal environment characterized by gravitoelectric tidal multipole moments $\E_{ab}(t)$, $\E_{abc}(t)$, $\E_{abcd}(t)$, $\E_{abcde}(t)$, and so on. These Cartesian tensors are symmetric and tracefree (STF) with respect to all indices; they are functions of time only. The tidal tensors will be given a precise definition in Sec.~\ref{sec:worldline}. For the time being it suffices to mention that in Newtonian gravity, they are defined by 
\begin{equation} 
\E_L = -\frac{1}{(\ell-2)!} \partial_L U^{\rm ext}, 
\label{EL_Newton} 
\end{equation} 
where $U^{\rm ext}$ is the gravitational potential created by the external matter responsible for the tidal field, evaluated at the body's center of mass after differentiation. The multi-index $L$ contains a number $\ell$ of individual indices, and the factor of $(\ell-2)!$ is inherited from Zhang's choice of normalization \cite{zhang:86} for tidal multipole moments. If $M'$ is a characteristic mass scale associated with the external matter, and if $a$ is a characteristic distance scale between this and the reference body, then the tidal tensors scale as $\E_L \sim G M'/a^{\ell+1}$.  

We obtain building blocks for the construction of tidal potentials by combining the tidal multipole moments with factors of the unit radial vector 
\begin{equation} 
\Omega^a := [\sin\theta\cos\phi, \sin\theta\sin\phi, \cos\theta], 
\label{Omega_def} 
\end{equation} 
where $\theta$ and $\phi$ are polar angles associated in the usual way with the Cartesian coordinates $x^a = (x, y, z)$. We thus form 
\begin{subequations} 
\label{potentials_linear} 
\begin{align} 
\E^{\sf q} &:= \E_{ab}\, \Omega^a \Omega^b, \\ 
\E^{\sf o} &:= \E_{abc}\, \Omega^a \Omega^b \Omega^c, \\ 
\E^{\sf h} &:= \E_{abcd}\, \Omega^a \Omega^b \Omega^c \Omega^d, \\ 
\E^{\sf t} &:= \E_{abcde}\, \Omega^a \Omega^b \Omega^c \Omega^d \Omega^e. 
\end{align} 
\end{subequations} 
The angular functions $\E_L \Omega^L$ are spherical harmonics of degree $\ell$; explicit expressions are given in Tables VII and VIII of Ref.~\cite{poisson-vlasov:10}. The labels stand for ``quadrupole'' ($\ell=2$), ``octupole'' ($\ell=3$), ``hexadecapole'' ($\ell=4$), and ``triakontadipole'' ($\ell=5$), respectively. 

\subsection{Bilinear tidal moments} 

We shall require bilinear combinations of $\E_{ab}$ and $\E_{abc}$, organized in irreducible STF tensors. We thus introduce 
\begin{subequations}
\label{quad-quad_tidal} 
\begin{align} 
\EE &:= \E_{ab} \E^{ab}, \\ 
\EE_{ab} &:= \E_{c\langle a} \E^c_{\ b\rangle}, \\ 
\EE_{abcd} &:= \E_{\langle ab} \E_{cd \rangle} 
\end{align} 
\end{subequations} 
and 
\begin{subequations}
\label{quad-oct_tidal} 
\begin{align} 
\EE_a &:= \E_{abc} \E^{bc}, \\ 
\EE_{abc} &:= \E_{d\langle ab} \E^d_{\ c\rangle}, \\ 
\EE_{abcde} &:= \E_{\langle abc} \E_{de\rangle}. 
\end{align} 
\end{subequations} 
The angular brackets are an instruction to symmetrize all indices and remove all traces. The first set of bilinear moments, those with an even number of indices, scale as $(GM')^2/a^6$. The second set of moments, with an odd number of indices, scale as $(GM')^2/a^7$.  

From the bilinear moments we form the angular functions 
\begin{subequations} 
\label{quad-quad_potentials} 
\begin{align} 
\EE^{\sf m} &:= \EE, \\ 
\EE^{\sf q} &:= \EE_{ab}\, \Omega^a \Omega^b, \\ 
\EE^{\sf h} &:= \EE_{abcd}\, \Omega^a \Omega^b \Omega^c \Omega^d 
\end{align} 
\end{subequations} 
and 
\begin{subequations} 
\label{quad-oct_potentials} 
\begin{align} 
\EE^{\sf d} &:= \EE_a\, \Omega^a, \\ 
\EE^{\sf o} &:= \EE_{abc}\, \Omega^a \Omega^b \Omega^c, \\ 
\EE^{\sf t} &:= \EE_{abcde}\, \Omega^a \Omega^b \Omega^c \Omega^d \Omega^e.  
\end{align} 
\end{subequations} 
The new labels ``${\sf m}$'' and  ``${\sf d}$'' stand for ``monopole'' ($\ell=0$) and ``dipole'' ($\ell=1$), respectively. These constructions are equivalent to a composition of spherical harmonics using Clebsch-Gordan coefficients. We note that the angular functions are denoted ${\cal P}^{\sf m}$ and so on in Ref.~\cite{poisson-vlasov:10}  (see their Table~III); explicit expressions are provided in their Table~X.  

It is useful to record the decompositions 
\begin{align} 
\E_{(ab} \E_{bc)} &= \EE_{abcd} + \frac{2}{21} \bigl[ \delta_{ab} \EE_{cd} + \delta_{ac} \EE_{bd} 
+ \delta_{ad} \EE_{bc} + \delta_{bc} \EE_{ad} + \delta_{bd} \EE_{ac} + \delta_{cd} \EE_{ab} \bigr] 
\nonumber \\ & \quad \mbox{} 
+ \frac{2}{45} \EE \bigl( \delta_{ab} \delta_{cd} + \delta_{ac} \delta_{bd} + \delta_{ad} \delta_{bc} \bigr) 
\end{align} 
and 
\begin{align} 
\E_{(abc} \E_{de)} &= \EE_{abcde} + \frac{1}{15} \bigl[ \delta_{ab} \EE_{cde} + \delta_{ac} \EE_{bde} 
+ \delta_{ad} \EE_{bce} + \delta_{ae} \EE_{bcd} + \delta_{bc} \EE_{ade} + \delta_{bd} \EE_{ace} 
\nonumber \\ & \quad \mbox{} 
+ \delta_{be} \EE_{acd} + \delta_{cd} \EE_{abe} + \delta_{ce} \EE_{abd} + \delta_{de} \EE_{abc} \bigr] 
+ \frac{2}{175} \Bigl[ 
\bigl( \delta_{ab} \delta_{cd} + \delta_{ac} \delta_{bd} + \delta_{ad}\delta_{bc} \bigr) \EE_e 
\nonumber \\ & \quad \mbox{} 
+ \bigl( \delta_{ab} \delta_{ce} + \delta_{ac} \delta_{be} + \delta_{ae}\delta_{bc} \bigr) \EE_d 
+ \bigl( \delta_{ab} \delta_{de} + \delta_{ad} \delta_{be} + \delta_{ae}\delta_{bd} \bigr) \EE_c
\nonumber \\ & \quad \mbox{} 
+ \bigl( \delta_{ac} \delta_{de} + \delta_{ad} \delta_{ce} + \delta_{ae}\delta_{cd} \bigr) \EE_b
+ \bigl( \delta_{bc} \delta_{de} + \delta_{bd} \delta_{ce} + \delta_{be}\delta_{cd} \bigr) \EE_a \Bigr].
\end{align} 
From these it follows that 
\begin{equation} 
\E^{\sf q} \E^{\sf q} = \EE^{\sf h} + \frac{4}{7} \EE^{\sf q} + \frac{2}{15} \EE
\label{EqEq} 
\end{equation} 
and 
\begin{equation} 
\E^{\sf q} \E^{\sf o} = \EE^{\sf t} + \frac{2}{3} \EE^{\sf o} + \frac{6}{35} \EE^{\sf d}. 
\label{EqEo} 
\end{equation} 

\subsection{Axisymmetric tidal environment} 
\label{subsec:axi} 

It is helpful to apply the formalism deployed here to a special case. We consider an axisymmetric tidal environment, for which each tidal tensor $\E_L$ possesses a single independent component. The tensors can then be related to STF combinations of the unit vector $z^a = [0,0,1]$, which points in the direction of the symmetry axis (here aligned with the $z$-direction). We require 
\begin{subequations} 
\begin{align} 
z_\stf{ab} &:= z_a z_b - \frac{1}{3} \delta_{ab}, \\ 
z_\stf{abc} &:= z_a z_b z_c - \frac{1}{5} \bigl( \delta_{ab} z_c + \delta_{ac} z_b + \delta_{bc} z_a \bigr), \\ 
z_\stf{abcd} &:= z_a z_b z_c z_d - \frac{1}{7} \bigl( \delta_{ab} z_c z_d + \delta_{ac} z_b z_d 
+ \delta_{ad} z_b z_c + \delta_{bc} z_a z_d + \delta_{bd} z_a z_c + \delta_{cd} z_a z_b \bigr) 
\nonumber \\ & \quad \mbox{} 
+ \frac{1}{35} \bigl( \delta_{ab} \delta_{cd} + \delta_{ac} \delta_{bd} + \delta_{ad} \delta_{bc} \bigr), \\ 
z_\stf{abcde} &:= z_a z_b z_c z_d z_e - \frac{1}{9} \bigl( \delta_{ab} z_c z_d z_e + \delta_{ac} z_b z_d z_e 
+ \delta_{ad} z_b z_c z_e + \delta_{ae} z_b z_c z_d + \delta_{bc} z_a z_d z_e + \delta_{bd} z_a z_c z_e 
+ \delta_{be} z_a z_c z_d 
\nonumber \\ & \quad \mbox{} 
+ \delta_{cd} z_a z_b z_e + \delta_{ce} z_a z_b z_d + \delta_{de} z_a z_b z_c \bigr) 
+ \frac{1}{63} \Bigl[ \bigl( \delta_{ab} \delta_{cd} + \delta_{ac} \delta_{bd} + \delta_{ad} \delta_{bc} \bigr) z_e 
+ \bigl( \delta_{ab} \delta_{ce} + \delta_{ac} \delta_{be} + \delta_{ae} \delta_{bc} \bigr) z_d 
\nonumber \\ & \quad \mbox{} 
+ \bigl( \delta_{ab} \delta_{de} + \delta_{ad} \delta_{be} + \delta_{ae} \delta_{bd} \bigr) z_c 
+ \bigl( \delta_{ac} \delta_{de} + \delta_{ad} \delta_{ce} + \delta_{ae} \delta_{cd} \bigr) z_b
+ \bigl( \delta_{bc} \delta_{de} + \delta_{bd} \delta_{ce} + \delta_{be} \delta_{cd} \bigr) z_a \Bigr]. 
\end{align} 
\end{subequations} 
The axial symmetry ensures that each tidal tensor $\E_L$ is proportional to $z_\stf{L}$, and the constant of proportionality defines the tensor's sole independent component. We write 
\begin{subequations} 
\label{axi_components} 
\begin{align} 
\E_{ab} &= -3\, \E^{\sf q}_0\, z_\stf{ab}, \\ 
\E_{abc} &= -5\, \E^{\sf o}_0\, z_\stf{abc}, \\ 
\E_{abcd} &= \frac{35}{2}\, \E^{\sf h}_0\, z_\stf{abcd}, \\ 
\E_{abcde} &= 63\, \E^{\sf t}_0\, z_\stf{abcde}; 
\end{align} 
\end{subequations} 
the strange numerical factors are inserted for later convenience. 

The axisymmetric expressions for the tidal tensors can be inserted within Eqs.~(\ref{potentials_linear}) to form axisymmetric angular functions --- spherical harmonics with $m=0$, which depend on $\theta$ only.  For this purpose we introduce the (rescaled) Legendre polynomials (see Table~VII of Ref.~\cite{poisson-vlasov:10})
\begin{subequations} 
\label{axi_harmonics}
\begin{align} 
Y_0 &:= 1, \\ 
Y_1 &:= \cos\theta, \\ 
Y_2 &:= 1 - 3\cos^2\theta, \\ 
Y_3 &:= \cos\theta(3 - 5\cos^2\theta), \\ 
Y_4 &:= \frac{1}{2} (3 - 30\cos^2\theta + 35\cos^4\theta), \\ 
Y_5 &:= \cos\theta(15 - 70\cos^2\theta + 63\cos^4\theta), 
\end{align} 
\end{subequations} 
and note that 
\begin{subequations} 
\label{axi_stf}
\begin{align} 
z_\stf{a} \Omega^a &= Y_1, \\ 
z_\stf{ab} \Omega^a \Omega^b &= -\frac{1}{3} Y_2, \\ 
z_\stf{abc} \Omega^a \Omega^b \Omega^c &= -\frac{1}{5} Y_3, \\
z_\stf{abcd} \Omega^a \Omega^b \Omega^c \Omega^d &= \frac{2}{35} Y_4, \\ 
z_\stf{abcde} \Omega^a \Omega^b \Omega^c \Omega^d \Omega^e &= \frac{1}{63} Y_5. 
\end{align} 
\end{subequations} 
When we substitute Eqs.~(\ref{axi_components}) and (\ref{axi_stf}) into Eqs.~(\ref{potentials_linear}), we obtain 
\begin{subequations} 
\label{axi_potentials} 
\begin{align} 
\E^{\sf q} &= \E^{\sf q}_0\, Y_2, \\ 
\E^{\sf o} &= \E^{\sf o}_0\, Y_3, \\ 
\E^{\sf h} &= \E^{\sf h}_0\, Y_4, \\ 
\E^{\sf t} &= \E^{\sf t}_0\, Y_5.
\end{align} 
\end{subequations} 
The numerical factors in Eqs.~(\ref{axi_components}) were inserted specifically to simplify the form of the
angular functions, given the adopted normalization for the Legendre polynomials. 

The tidal moments of Eqs.~(\ref{axi_components}) can be inserted within Eqs.~(\ref{quad-quad_tidal}) and (\ref{quad-oct_tidal}) to obtain the bilinear tensors. We get 
\begin{subequations} 
\label{axi_quad_quad_tensors} 
\begin{align} 
\EE &= 6 (\E^{\sf q}_0)^2, \\ 
\EE_{ab} &= 3 (\E^{\sf q}_0)^2\, z_\stf{ab}, \\ 
\EE_{abcd} &= 9 (\E^{\sf q}_0)^2\, z_\stf{abcd} 
\end{align} 
\end{subequations}
and 
\begin{subequations} 
\label{axi_quad_oct_tensors} 
\begin{align} 
\EE_a &= 6\, \E^{\sf q}_0 \E^{\sf o}_0\, z_a \\
\EE_{abc} &= 4\, \E^{\sf q}_0 \E^{\sf o}_0\, z_\stf{abc} \\
\EE_{abcde} &= 15\, \E^{\sf q}_0 \E^{\sf o}_0\, z_\stf{abcde}. 
\end{align} 
\end{subequations} 
The bilinear angular functions are then 
\begin{subequations} 
\label{axi_quad_quad_potentials} 
\begin{align} 
\EE^{\sf m} &= 6 (\E^{\sf q}_0)^2\, Y_0, \\ 
\EE^{\sf q} &= -(\E^{\sf q}_0)^2\, Y_2, \\ 
\EE^{\sf h} &= \frac{18}{35} (\E^{\sf q}_0)^2\, Y_4
\end{align} 
\end{subequations} 
and 
\begin{subequations} 
\label{axi_quad_oct_potentials} 
\begin{align} 
\EE^{\sf d} &= 6\, \E^{\sf q}_0 \E^{\sf o}_0\, Y_1, \\ 
\EE^{\sf o} &= -\frac{4}{5}\, \E^{\sf q}_0 \E^{\sf o}_0\, Y_3, \\ 
\EE^{\sf t} &= \frac{5}{21}\, \E^{\sf q}_0 \E^{\sf o}_0\, Y_5. 
\end{align} 
\end{subequations} 
 
\section{World line in a tidal environment} 
\label{sec:worldline}  

The metric of a tidally deformed body will be constructed in Sec.~\ref{sec:GR}. To set the stage and outline the general calculational strategy, we first go through a simple warmup exercise. In this section we obtain the metric of a generic vacuum spacetime in a neighborhood of a timelike geodesic. It is useful to think of this as the $M \to 0$ limit of the metric of Sec.~\ref{sec:GR}, where $M$ is the body's mass; the geodesic is what remains when the body vanishes in the limit. The methods introduced in this section will be recycled (and generalized) in Sec.~\ref{sec:GR}.  

\subsection{Definition of tidal moments} 

To give a precise definition to the tidal multipole moments introduced in Sec.~\ref{sec:moments}, we consider a vacuum region of a generic spacetime in a neighborhood of a timelike geodesic $\gamma$. The world line is described by the parametric equations $x^\alpha = z^\alpha(t)$, in which $t$ is proper time; its tangent vector is $u^\alpha = dz^\alpha/dt$, and it is assumed to satisfy the geodesic equation. We complete the vectorial basis with a triad of unit vectors $e^\alpha_a(t)$, with $a = 1, 2, 3$, which we take to be mutually orthogonal and also orthogonal to $u^\alpha$; as $u^\alpha$, the triad is parallel transported on the world line. 

We use the tetrad $(e^\alpha_0 := u^\alpha, e^\alpha_a)$ to decompose tensors evaluated on the world line. For example, 
\begin{equation} 
C_{0a0b}(t) := C_{\mu\alpha\nu\beta}\Bigr|_{\gamma}\, u^\mu e^\alpha_a u^\nu e^\beta_b 
\end{equation} 
are frame components of the Weyl tensor evaluated at $\gamma$; because the Ricci tensor vanishes on the world line, the Weyl tensor is equal to the Riemann tensor. The notation extends to derivatives of tensors; for example,  
\begin{equation} 
C_{0a0b|c}(t) := \nabla_\gamma C_{\mu\alpha\nu\beta} \Bigr|_{\gamma}\, 
u^\mu e^\alpha_a u^\nu e^\beta_b e^\gamma_c  
\end{equation} 
are frame components of the first derivative of the Weyl tensor. 

The tidal moments are defined by \cite{zhang:86} 
\begin{equation} 
\E_{a_1\cdots a_\ell} := \frac{1}{(\ell-2)!} \bigl( C_{0 a_1 0 a_2 | a_3 \cdots a_\ell} \bigr)^{\rm STF}. 
\label{EL_def} 
\end{equation} 
In words, a tidal moment of order $\ell$ is obtained by differentiating the Weyl tensor $\ell-2$ times, evaluating the result on the world line, taking projections with members of the tetrad, symmetrizing with respect to all indices, and removing all traces with the Euclidean metric $\delta_{ab}$. The low-order moments are  
\begin{equation} 
\E_{ab} := \bigl( C_{0a0b} \bigr)^{\rm STF}, \qquad 
\E_{abc} := \bigl( C_{0a0b|c} \bigr)^{\rm STF}, \qquad 
\E_{abcd} = \frac{1}{2} \bigl( C_{0a0b|cd} \bigr)^{\rm STF}, \qquad 
\E_{abcde} = \frac{1}{6} \bigl( C_{0a0b|cde} \bigr)^{\rm STF}. 
\end{equation} 
In the Newtonian limit, the Weyl tensor is related to the gravitational potential $U$ by $C_{0a0b} = -\partial_{ab} U$, and the definition of the tidal moments reduces to Eq.~(\ref{EL_Newton}). 

When we move to the context of a tidally deformed body in Sec.~\ref{sec:GR}, the tidal multipole moments will no longer be defined by Eq.~(\ref{EL_def}). There are two reasons for this. First, the geodesic $\gamma$ is no longer defined. While one might choose any representative world line inside a material body as a candidate for $\gamma$, in general this world line will not be a geodesic, and it will not move in a vacuum spacetime; moreover, the world line clearly does not exist for a black hole. The second reason is that the Weyl tensor of Eq.~(\ref{EL_def}) is meant to represent a tidal field produced by external sources of gravitation, while the complete Weyl tensor of a tidally deformed body also includes the body's own contribution to the spacetime curvature; there is no meaningful and unambiguous way to decompose the complete tensor into tidal and body pieces. 

Nevertheless, the metric of a tidally deformed body will still be expressed in terms of tensors $\E_L(t)$, with the understanding that these provide a characterization of the tidal environment. In this new context, the tidal moments appear as freely-specifiable functions of time that are not determined by the Einstein field equations when the domain of consideration is limited to a neighborhood around the body. The determination of the tidal moments must instead be carried out by inserting the local metric within a global metric that includes the reference body in addition to all external objects that create the tidal environment. Such an exercise will be carried out in Sec.~\ref{sec:PN}, with a global metric presented as a post-Newtonian expansion. In the $M \to 0$ limit, the tidal moments of Sec.~\ref{sec:GR} reduce to the ones introduced here.   

\subsection{Metric near the world line: Goals and strategy} 
\label{subsec:M0_goals} 

We wish to construct the metric of a generic spacetime in a neighborhood of a timelike geodesic $\gamma$; the spacetime is assumed to be empty of matter in this neighborhood. This can be done in a number of ways, in any number of coordinate systems. A well-known construction, for example, utilizes Fermi normal coordinates \cite{manasse-misner:63}, and produces a metric expanded in powers of the spatial distance from $\gamma$. The approach pursued here is very much in the same spirit, but it utilizes tailor-made coordinates that can readily be generalized to the case of a tidally deformed body. 

Experience with Fermi coordinates teaches us that the metric can be expressed as a simultaneous expansion in powers of the ratio of $r$, the distance to the world line, to three distinct length scales. The first is ${\cal R}$, the spacetime's radius of curvature, as measured by a typical frame component of the Weyl tensor. The second is ${\cal L}$, the scale of curvature inhomogeneities, as measured by spatial derivatives of the Weyl tensor; these are derivatives in the direction of $e^\alpha_a$. And the third is ${\cal T}$, the scale of temporal variation in the curvature, as measured by time derivatives of the Weyl tensor --- derivatives in the direction of $u^\alpha$.  

The curvature expansion is of the schematic form 
\begin{equation} 
\mbox{metric} = 1 + (r/{\cal R})^2+ (r/{\cal R})^4 + (r/{\cal R})^6 + \cdots. 
\label{curvature-expansion} 
\end{equation} 
For the leading curvature term, the spatial-derivative expansion is of the schematic form 
\begin{equation} 
(r/{\cal R})^2 \bigl[ 1 + (r/{\cal L}) + (r/{\cal L})^2 + \cdots \bigr], 
\label{spatial-expansion} 
\end{equation} 
and the time-derivative expansion looks like
\begin{equation} 
(r/{\cal R})^2 \bigl[ 1 + (r/{\cal T}) + (r/{\cal T})^2 + \cdots \bigr].  
\label{time-expansion} 
\end{equation} 
Higher-order curvature terms also acquire these types of corrections. 

In terms of the scaling quantities introduced previously, the external mass $M'$ and the distance $a$, we have that 
\begin{equation} 
{\cal R} \sim \sqrt{a^3/M'}, \qquad 
{\cal L} \sim a. 
\label{RL_scales} 
\end{equation} 
For a gravitationally bound system of external bodies, the scale of temporal variation is 
\begin{equation} 
{\cal T} \sim \sqrt{a^3/M'}, 
\label{T_scale} 
\end{equation} 
and it is of the same order of magnitude as ${\cal R}$. In such a situation, the time-derivative expansion has to keep step with the curvature expansion in order to produce a consistent approximation for the metric. 

Throughout the paper we shall adopt the scalings of Eqs.~(\ref{RL_scales}) and (\ref{T_scale}), and we shall assume that $r \ll a$, so that $r/{\cal L} \ll 1$. We shall also assume that $a \gg M'$, so that ${\cal R}, {\cal T} \gg {\cal L}$, which ensures that $r/{\cal R}, r/{\cal T} \ll r/{\cal L}$. Under these conditions, corrections from the spatial-derivative expansion dominate over those from the curvature and time-derivative expansions. 

We aim to obtain the metric near the world line $\gamma$ in simultaneous curvature, spatial-derivative, and time-derivative expansions. Our first goal is to account for the spatial-derivative expansion of the leading curvature term, as expressed in Eq.~(\ref{spatial-expansion}). In this expansion, the leading term corresponds to a tidal multipole of order $\ell = 2$, and each successive power of $r/{\cal L}$ raises $\ell$ by one unit; we shall truncate beyond $\ell = 5$, and therefore expand through $O(r^5)$. Our second goal is to reveal curvature-squared terms in the metric, specifically those that result from bilinear couplings between ${\cal E}_{ab}$ and ${\cal E}_{abc}$. These terms are of order $(r/{\cal R})^4$ in the curvature expansion, and they include a spatial-derivative correction of relative order $r/{\cal L}$; here also the expansion proceeds through $O(r^5)$. Our third goal is to account for the time-derivative expansion of the leading curvature term, as expressed in Eq.~(\ref{time-expansion}), but generalized to incorporate also spatial derivatives. For a multipole of order $\ell$ we shall have terms of the form 
\begin{equation} 
(r/{\cal R})^2 (r/{\cal L})^{\ell - 2}\bigl[ 1 + (r/{\cal T}) + (r/{\cal T})^2 \bigr].
\end{equation} 
As it turns out, the contributions of order $r/{\cal T}$ all vanish. We truncate the expansion beyond the second time derivative, and for each $\ell$ we have an expansion through $O(r^{\ell + 2})$.    

\subsection{Transformation from Cartesian to spherical coordinates} 
\label{subsec:tranf_cart_sphe} 

In the following we shall make use of spherical polar coordinates $(r,\theta^A)$, with $\theta^A = (\theta,\phi)$, related in the usual way to a system of Cartesian coordinates $x^a$; we have that 
\begin{equation} 
x^a = r\, \Omega^a(\theta^A),
\label{cart_vs_spher} 
\end{equation} 
where $\Omega^a$ is the radial unit vector introduced in Eq.~(\ref{Omega_def}). We let $\Omega_{AB} := \mbox{diag}[1,\sin^2\theta]$ be the metric on the unit two-sphere, and $\Omega^{AB}$ is its matrix inverse. We use the notation 
\begin{equation} 
\Omega^a_A := \frac{\partial \Omega^a}{\partial \theta^A}, \qquad 
\Omega_a^A := \delta_{ab} \Omega^{AB} \Omega^b_B. 
\end{equation} 
The partial derivatives involved in a transformation between Cartesian and spherical coordinates are 
\begin{equation} 
\frac{\partial x^a}{\partial r} = \Omega^a, \qquad 
\frac{\partial x^a}{\partial \theta^A} = r \Omega^a_A, \qquad 
\frac{\partial r}{\partial x^a} = \Omega_a, \qquad 
\frac{\partial \theta^A}{\partial x^a} = \frac{1}{r} \Omega^A_a, 
\label{partials_cart_sphe} 
\end{equation} 
where $\Omega_a := \delta_{ab} \Omega^b$. The identities 
\begin{equation} 
\Omega_a \Omega^a_A = 0, \qquad 
\Omega_{AB} = (\delta_{ab} - \Omega_a \Omega_b) \Omega^a_A \Omega^b_B, \qquad 
\delta_{ab} - \Omega_a \Omega_b = \Omega_{AB} \Omega^A_a \Omega^B_b 
\label{identities_cart_sphe}
\end{equation} 
are sometimes useful in computations. 

\subsection{Metric construction: Linear and static tides}
\label{subsec:M0_linear-static} 

To construct the metric we make use of the angular functions introduced in Sec.~\ref{sec:moments}, which we combine with appropriate factors of $r$ to make each term dimensionless. The numerical factors are then determined by making sure that the metric is a solution to the Einstein field equations in vacuum, and that the tidal moments are related to the Weyl tensor by Eq.~(\ref{EL_def}). 

To begin we keep the metric to first order in the curvature expansion; second-order terms will be incorporated in the following subsections. And for the time being we take the tidal moments to be independent of time; time-derivative terms will be incorporated at a later stage. To implement the procedure described in the preceding paragraph, we must first make a choice of coordinate conditions. We choose to express the metric in the diagonal form 
\begin{subequations}
\label{M0_linear_metric} 
\begin{align}  
g_{tt} &= -1 + p^{\rm ls}_{tt}, \\ 
g_{rr} &= 1 + p^{\rm ls}_{rr}, \\ 
g_{AB} &= r^2 \Omega_{AB} (1 + p^{\rm ls}), 
\end{align} 
\end{subequations} 
where $p^{\rm ls}_{tt}$, $p^{\rm ls}_{rr}$, and $p^{\rm ls}$ (with the label standing for ``linear and static'') are the components of the metric perturbation. The coordinate conditions consist of setting $g_{rA} = 0$ and making $g_{AB}$ diagonal; the remaining conditions, $g_{tr} = 0 = g_{tA}$, are a consequence of the (momentary) assumption that the tidal environment is time-independent. In the broader context of the formalism of metric perturbations of the Schwarzschild spacetime (see, for example, Ref.~\cite{martel-poisson:05}), here applied to the special case $M=0$, the coordinate conditions amount to adopting the Regge-Wheeler gauge \cite{regge-wheeler:57} --- see Appendix~\ref{app:RW} for a refresher. It is known that this gauge is unique; there is no residual gauge freedom.  

We find that up to $\ell = 5$, the perturbations are given by 
\begin{equation}
p^{\rm ls}_{tt} = p^{\rm ls}_{rr} = p^{\rm ls} 
= -r^2\, \E^{\sf q} - \frac{1}{3} r^3\, \E^{\sf o} - \frac{1}{6} r^4\, \E^{\sf h} 
- \frac{1}{10} r^5\, \E^{\sf t}, 
\label{M0_linear_perturbation} 
\end{equation}   
The numerical coefficient in front of each term is determined by the definition of the tidal moments: a calculation of the Weyl tensor from the metric, linearized with respect to the perturbation, can indeed be shown to reproduce Eq.~(\ref{EL_def}). That the perturbations are all equal to each other is a consequence of the vacuum field equations. 

When we carry out these manipulations, we take advantage of the fact that the perturbation is linear in each component of the tidal moments $\E_L$, and that a complete decomposition in spherical harmonics is accomplished with the angular functions of Eq.~(\ref{potentials_linear}). It is therefore advantageous --- and sufficient --- to imagine (momentarily, for the purpose of this calculation) that the tidal environment is axisymmetric, and to adopt the simplified expressions listed in Sec.~\ref{subsec:axi}. There is no loss of generality in employing this stratagem.  

\subsection{Metric construction: Quadrupole-quadruple coupling} 
\label{subsec:M0_quad-quad} 

Next we proceed to second order in the curvature expansion, and focus for the time being on terms that arise from the coupling between $\E_{ab}$ and itself. The required angular functions were introduced in Eqs.~(\ref{quad-quad_potentials}), and these must all be multiplied by $r^4$ to produce valid candidate contributions to the metric perturbation. We see that the bilinear terms occur at monopole ($\ell = 0$), quadrupole ($\ell = 2$), and hexadecapole $(\ell = 4)$ orders. 

We may continue to exploit the same coordinate conditions as previously for the $\ell=2$ and $\ell=4$ sectors of the perturbation. For $\ell = 0$, however, we have the freedom to impose one additional condition, and we choose to eliminate the perturbation in the angular components of the metric. The metric of Eq.~(\ref{M0_linear_metric}) is updated to 
\begin{subequations}
\label{M0_quad-quad_metric} 
\begin{align}  
g_{tt} &= -1 + p^{\rm ls}_{tt}+p^{\rm qq}_{tt}, \\ 
g_{rr} &= 1 + p^{\rm ls}_{rr} +p^{\rm qq}_{rr}, \\ 
g_{AB} &= r^2 \Omega_{AB} (1 + p^{\rm ls}  + p^{\rm qq}), 
\end{align} 
\end{subequations} 
where the ``qq'' label on the new terms stands for ``quadrupole-quadrupole''. It is understood that $p^{\rm qq}_{tt}$, $p^{\rm qq}_{rr}$, and $p^{\rm qq}$ are superpositions of terms $r^4 \EE^{\sf m}$, $r^4 \EE^{\sf q}$, and $r^4 \EE^{\sf h}$, with unknown numerical coefficients.

We substitute the metric into the Einstein field equations, and discard all nonlinear terms that are not proportional to $\E_{ab}$ multiplied by itself.  We eventually obtain 
\begin{subequations} 
\label{M0_quad-quad_perturbation} 
\begin{align}  
p^{\rm qq}_{tt} &= -\frac{1}{15} r^4\, \EE^{\sf m} - \frac{2}{7} r^4\, \EE^{\sf q} 
+ ( c_4 - \tfrac{2}{3} )r^4\, \EE^{\sf h}, \\
p^{\rm qq}_{rr} &=-\frac{1}{5} r^4\, \EE^{\sf m} - \frac{4}{7} r^4\, \EE^{\sf q} 
+ c_4 r^4\, \EE^{\sf h}, \\
p^{\rm qq} &=-\frac{1}{14} r^4\, \EE^{\sf q}  + c_4 r^4\, \EE^{\sf h}. 
\end{align} 
\end{subequations} 
Except for $c_4$, all numerical coefficients are determined by the requirement that the metric be a solution to the vacuum field equations to second order in the curvature expansion. The value of $c_4$ remains arbitrary, because $p_{tt} = p_{rr} = p = c_4 r^4\, \EE^{\sf h}$ is a solution to the linearized equations. The constant, however, can be pinned down by demanding that the tidal moment $\E_{abcd}$ be compatible with its definition of Eq.~(\ref{EL_def}), not just to first order in the curvature expansion, but also to second order. To achieve this determination, we use the metric to calculate the second derivatives of the Weyl tensor evaluated on the world line (at $r = 0$), we take tetrad projections, with legs of the tetrad aligned with the Cartesian directions of Eq.~(\ref{cart_vs_spher}), and we symmetrize all indices and remove all traces. The computation returns $\E_{abcd}$ provided that 
\begin{equation} 
c_4 = 1. 
\label{M0_c4} 
\end{equation} 

The calculations described in this subsection can be simplified, without loss of generality, by imagining (again momentarily, for the purpose of the computation) that $\E_{ab}$ is axisymmetric. The justification is virtually identical to our previous defense of the stratagem: the perturbation is linear in the bilinear moments $\EE$, $\EE_{ab}$, and $\EE_{abcd}$, the angular functions of Eq.~(\ref{quad-quad_potentials}) contain the required number of spherical-harmonic components, and all nonlinearities associated with the quadrupole-quadrupole coupling are accounted for by Eq.~(\ref{EqEq}).    

\subsection{Metric construction: Quadrupole-octupole coupling} 
\label{subsec:M0_quad-oct} 

We continue with the construction of the metric to second order in the curvature expansion, and this time we place our attention on terms that are generated by the coupling between $\E_{ab}$ and $\E_{abc}$. The relevant angular functions were introduced in Eqs.~(\ref{quad-quad_potentials}), and these must now be multiplied by $r^5$ to produce candidate contributions to the metric perturbation. The bilinear terms occur at dipole ($\ell = 1$), octupole ($\ell = 3$), and triakontadipole $(\ell = 5)$ orders.  

For $\ell = 3$ and $\ell = 5$ we exploit the same coordinate conditions as for the first-order perturbation. For $\ell = 1$ we have the freedom to impose one additional condition. Several choices were considered. An option is to adopt the Zerilli gauge \cite{zerilli:70}, for which $p = 0$. Another option is to set $p_{tt} = p_{rr}$. Yet another choice is a nondiagonal gauge defined by $p_{tt} = p_{rr} = p$ and $g_{rA} \neq 0$. We adopt the nondiagonal gauge, for two reasons that will become clear when the calculation is generalized to the tidal deformation of a compact body in Sec.~\ref{sec:GR}. First, this gauge produces the simplest metric perturbation. Second, the metric is compatible with the standard post-Newtonian form when it is expanded in powers of $M/r$; this criterion will become important in Sec.~\ref{sec:PN}. 

To account for the quadrupole-octupole coupling, the metric of Eq.~(\ref{M0_linear_metric}) is updated to 
\begin{subequations}
\label{M0_quad-oct_metric} 
\begin{align}  
g_{tt} &= -1 + p^{\rm ls}_{tt} + p^{\rm qo}_{tt}, \\ 
g_{rr} &= 1 + p^{\rm ls}_{rr} + p^{\rm qo}_{rr}, \\ 
g_{rA} &= r \partial_A p^{\rm qo}_r, \\
g_{AB} &= r^2 \Omega_{AB} (1 + p^{\rm ls}  + p^{\rm qo}), 
\end{align} 
\end{subequations} 
where the ``qo'' label stands for ``quadrupole-octupole''. The perturbations $p^{\rm qo}_{tt}$, $p^{\rm qo}_{rr}$, $p^{\rm qo}_r$, and $p^{\rm qo}$ are superpositions of terms $r^5 \EE^{\sf d}$, $r^5 \EE^{\sf o}$, and $r^5 \EE^{\sf t}$, with unknown numerical coefficients. The additional factor of $r$ in $g_{rA}$ comes from the fact that this is a mixed radial-angular component of the metric, and $\partial_A$ stands for partial differentiation with respect to $\theta^A$.  

We insert the metric within the Einstein field equations, and eliminate all nonlinear terms that are not proportional to $\E_{ab}$ multiplied by $\E_{abc}$.  We eventually obtain
\begin{subequations} 
\label{M0_quad-oct_perturbation} 
\begin{align}  
p^{\rm qo}_{tt} &= -\frac{2}{35} r^5\, \EE^{\sf d} - \frac{2}{9} r^5\, \EE^{\sf o} 
+ (c_5 - \tfrac{7}{15}) r^5\, \EE^{\sf t}, \\ 
p^{\rm qo}_{rr} &=-\frac{2}{35} r^5\, \EE^{\sf d} - \frac{1}{9} r^5\, \EE^{\sf o} 
+ c_5 r^5\, \EE^{\sf t}, \\
p^{\rm qo}_{r} &= \frac{1}{5} r^5\, \EE^{\sf d}, \\
p^{\rm qo} &=-\frac{2}{35} r^5\, \EE^{\sf d} + \frac{1}{18} r^5\, \EE^{\sf o} 
+ c_5 r^5\, \EE^{\sf t}. 
\end{align} 
\end{subequations} 
Again we find that the calculation leaves one numerical coefficient undetermined, because $p_{tt} = p_{rr} = p = c_5 r^5\, \EE^{\sf t}$ is a solution to the linearized equations. The correct value for the constant is obtained by making sure that the tidal moment $\E_{abcde}$ is compatible with its definition of Eq.~(\ref{EL_def}), not just to first order in the curvature expansion, but also to second order. We find that this requires 
\begin{equation} 
c_5 = \frac{11}{15}. 
\label{M0_c5} 
\end{equation} 
With this, the metric is completely determined. 

The calculations carried out in this subsection can again be simplified by imagining (momentarily, as always) that $\E_{ab}$ and $\E_{abc}$ are both axisymmetric. The justification is the same as the one presented at the end of Sec.~\ref{subsec:M0_quad-quad}. 

\subsection{Metric construction: Time derivatives} 
\label{subsec:M0_time} 

To complete the task we must now incorporate the time-derivative terms in the metric. For this purpose we keep the perturbation to first order in the curvature expansion, and update Eq.~(\ref{M0_linear_metric}) to
\begin{subequations}
\label{M0_time_metric} 
\begin{align}  
g_{tt} &= -1 + p^{\rm ls}_{tt} + p^{\rm ld}_{tt}, \\ 
g_{tr} &= p^{\rm ld}_{tr}, \\ 
g_{rr} &= 1 + p^{\rm ls}_{rr} + p^{\rm ld}_{rr}, \\ 
g_{AB} &= r^2 \Omega_{AB} (1 + p^{\rm ls}  + p^{\rm ld}), 
\end{align} 
\end{subequations} 
where the ``ld'' label stands for ``linear and dynamic''. This expression for the metric reflects the same coordinate conditions as in Sec.~\ref{subsec:M0_linear-static}, augmented with $g_{tA} = 0$; we are still working with the $M=0$ implementation of the Regge-Wheeler gauge.

We can expect that for a multipole of order $\ell$, the new perturbation will consist of terms proportional to $r^{\ell + 1} \dot{\E}_L \Omega^L$ and $r^{\ell + 2} \ddot{\E}_L \Omega^L$, in which overdots indicate differentiation with respect to time. The numerical coefficients in front of these terms are then determined by the Einstein field equations. We obtain
\begin{subequations}
\label{M0_lin-dyn-perturbation} 
\begin{align}
p^{\rm ld}_{tt} &= -\frac{\ell+9}{(\ell-1)\ell(\ell+1)(2\ell+3)} r^{\ell + 2}\, \ddot{\E}_L \Omega^L, \\
p^{\rm ld}_{tr} &= -\frac{4}{(\ell-1)\ell(\ell+1)} r^{\ell + 1}\, \dot{\E}_L \Omega^L, \\
p^{\rm ld}_{rr} &= -\frac{\ell+9}{(\ell-1)\ell(\ell+1)(2\ell+3)} r^{\ell + 2}\, \ddot{\E}_L \Omega^L, \\
p^{\rm ld} &= -\frac{\ell^2+3\ell+6}{(\ell-1)\ell(\ell+1)(\ell+2)(2\ell+3)} r^{\ell + 2}\, \ddot{\E}_L \Omega^L.
\end{align}
\end{subequations}
We observe that an expected term proportional to $\dot{\E}_L$ actually vanishes in the diagonal components of the perturbation. On the other hand, this term is present in the $tr$ component, but an expected term proportional to $\ddot{\E}_L$ is absent. 

\subsection{Complete metric} 

Collecting results from the preceding subsections, we have that the metric of a vacuum region of spacetime in a neighborhood of a timelike geodesic $\gamma$ can be expressed as 
\begin{subequations} 
\label{M0_metric}
\begin{align} 
g_{tt} &= -1 - r^2\, \E^{\sf q} - \frac{1}{3} r^3\, \E^{\sf o} - \frac{1}{6} r^4\, \E^{\sf h}  - \frac{1}{10} r^5\, \E^{\sf t}
- \frac{1}{15} r^4\, \EE^{\sf m} - \frac{2}{7} r^4\, \EE^{\sf q}  + \frac{1}{3} r^4\, \EE^{\sf h} 
\nonumber \\ & \quad \mbox{} 
- \frac{2}{35} r^5\, \EE^{\sf d} - \frac{2}{9} r^5\, \EE^{\sf o}  + \frac{11}{15} r^5\, \EE^{\sf t}
- \frac{11}{42} r^4\, \ddot{\E}^{\sf q} - \frac{1}{18} r^5\, \ddot{\E}^{\sf o} + O(r^5, r^6), \\ 
g_{tr} &= -\frac{2}{3} r^3\, \dot{\E}^{\sf q} - \frac{1}{6} r^4\, \dot{\E}^{\sf o} + O(r^4, r^5), \\ 
g_{rr} &= 1 - r^2\, \E^{\sf q} - \frac{1}{3} r^3\, \E^{\sf o} - \frac{1}{6} r^4\, \E^{\sf h} - \frac{1}{10} r^5\, \E^{\sf t}
- \frac{1}{5} r^4\, \EE^{\sf m} - \frac{4}{7} r^4\, \EE^{\sf q} + r^4\, \EE^{\sf h} 
\nonumber \\ & \quad \mbox{} 
- \frac{2}{35} r^5\, \EE^{\sf d} - \frac{1}{9} r^5\, \EE^{\sf o}  + \frac{11}{15} r^5\, \EE^{\sf t}
- \frac{11}{42} r^4\, \ddot{\E}^{\sf q} - \frac{1}{18} r^5\, \ddot{\E}^{\sf o} + O(r^5, r^6), \\ 
g_{rA} &= r \biggl[ \frac{1}{5} r^5\, \partial_A \EE^{\sf d} + O(r^6) \biggr], \\
g_{AB} &= r^2 \Omega_{AB} \biggl[ 1 - r^2\, \E^{\sf q} - \frac{1}{3} r^3\, \E^{\sf o} - \frac{1}{6} r^4\, \E^{\sf h} 
- \frac{1}{10} r^5\, \E^{\sf t} -\frac{1}{14} r^4\, \EE^{\sf q}  + r^4\, \EE^{\sf h}, 
\nonumber \\ & \quad \mbox{}
- \frac{2}{35} r^5\, \EE^{\sf d} + \frac{1}{18} r^5\, \EE^{\sf o} + \frac{11}{15} r^5\, \EE^{\sf t}
- \frac{2}{21} r^4\, \ddot{\E}^{\sf q} - \frac{1}{45} r^5\, \ddot{\E}^{\sf o} + O(r^5, r^6) \biggr]. 
\end{align} 
\end{subequations} 
The metric is presented in coordinates $(t, r, \theta^A)$, with $t$ equal to proper time on $\gamma$, $r$ measuring the spatial distance to the world line, and $\theta^A$ representing a set of two polar angles. The metric is expressed as a curvature expansion in powers of $(r/{\cal R})^2$, with terms linear in the curvature constructed from $\E^{\sf q}$, $\E^{\sf o}$, $\E^{\sf h}$, and $\E^{\sf t}$, and terms quadratic in the curvature implicating $\EE^{\sf m}$, $\EE^{\sf d}$, $\EE^{\sf q}$, $\EE^{\sf o}$, $\EE^{\sf h}$, and $\EE^{\sf t}$. The metric is also expressed as a spatial-derivative expansion in powers of $r/{\cal L}$. In the terms linear in curvature, the contributions are of order $(r/{\cal R})^2 (r/{\cal L})^{\ell - 2}$, with each subsequent multipole adding an additional power of $r/{\cal L}$. In the quadratic terms, the monopole, quadrupole, and hexadecapole terms are proportional to $\E_{ab}$ multiplied by itself, and are of order $(r/{\cal R})^4 (r/{\cal L})^0$; the dipole, octupole, and triakontadipole terms are proportional to $\E_{ab}$ multiplied by $\E_{abc}$, and are of order $(r/{\cal R})^4 (r/{\cal L})$. Finally, the metric is expressed as a time-derivative expansion in powers of $(r/{\cal T})$; for a multipole of order $\ell$ this contributes terms that scale as $(r/{\cal R})^2 (r/{\cal L})^{\ell-2}  (r/{\cal T})$ and $(r/{\cal R})^2 (r/{\cal L})^{\ell-2} (r/{\cal T})^2$.   

The error terms in Eqs.~(\ref{M0_metric}) are short-hand forms for a more detailed description of the neglected contributions to the metric. For example, we have 
\begin{align} 
O(r^5, r^6) &:= O\bigl[ (r/{\cal R})^2 (r/{\cal L})^4 \bigr]
+ O\bigl[ (r/{\cal R})^4 (r/{\cal L})^2 \bigr] 
+ O\bigl[ (r/{\cal R})^6 \bigr]
\nonumber \\ & \quad \mbox{} 
+ O\bigl[(r/{\cal R})^2 (r/{\cal T})^3 \bigr]
+ O\bigl[(r/{\cal R})^2 (r/{\cal L}) (r/{\cal T})^3 \bigr]
+ O\bigl[(r/{\cal R})^2 (r/{\cal L})^2 (r/{\cal T})^2 \bigr]. 
\end{align} 
The contribution of order $r^5$ to the error budget comes from the fact that the $\ell = 2$ part of the metric perturbation does not incorporate a term involving the third derivative of $\E_{ab}(t)$. It would be a simple matter to remedy this situation and calculate this term --- it vanishes. We do not do so here (officially), because we shall not do so in Sec.~\ref{sec:GR}, when we generalize this calculation to the tidal deformation of a compact body. 

\section{Tidally deformed constant-density sphere in Newtonian gravity} 
\label{sec:constant-density} 

As another warmup exercise before the big task of Sec.~\ref{sec:GR}, in this section we examine the tidal deformation of a fluid body in Newtonian theory. Our goal is to obtain a description of the gravitational field that accounts for the linear and nonlinear tidal responses of the body, as well as its response to a time-varying tidal environment. To keep the technical difficulties to a minimum, we take the body to have a uniform density. This, to be sure, is a far cry from a realistic description of a star. But the simple model amply suffices for our purpose here, which is to illustrate the general structure of the Newtonian potential outside a tidally deformed body. This will serve as a useful guide in the relativistic generalization of Sec.~\ref{sec:GR}. 

\subsection{Governing equations} 

Our treatment is based on Euler's equation 
\begin{equation} 
\partial_t v_a + v^b \nabla_b v_a - \nabla_a (U - h) = 0 
\label{Euler} 
\end{equation} 
for the fluid variables, and Poisson's equation 
\begin{equation} 
\nabla^2 U = -4\pi G \rho 
\label{Poisson} 
\end{equation} 
for the gravitational potential. Here, $v_a$ is the fluid's velocity field, $\rho$ the mass density, and $h$ the specific enthalpy, defined in terms of the pressure $p$ by $dh = dp/\rho$; when the density is uniform we have that $h = p/\rho$. The body's surface is situated at $r = r_{\rm S}(\theta,\phi)$. The specific enthalpy and gravitational potential satisfy the surface conditions 
\begin{equation} 
h(r=r_{\rm S}) = 0, \qquad 
\bigl[ U \bigr] = 0, \qquad 
\bigl[ \nabla^a U \bigr] \nabla_a(r - r_{\rm S}) = 0, 
\label{surf_cond} 
\end{equation} 
where $[U] := U_{\rm out}(r=r_{\rm S}) - U_{\rm in}(r=r_{\rm S})$ is the difference between the exterior and interior potentials evaluated on the surface, while $[\nabla^a U]$ is the difference in the differentiated potentials.   

We shall demand that $M$, the body's total mass, stays the same as the body goes from its unperturbed state to a tidally deformed state. Because the mass is the integral of the (constant) density $\rho$ over the volume occupied by the body, this requirement translates to   
\begin{equation} 
R^3 = \frac{1}{4\pi} \oint r^3_{\rm S}(\theta,\phi)\, d\Omega, 
\label{mass_condition} 
\end{equation} 
where $R$ is the body's radius in the unperturbed state, and $d\Omega := \sin\theta\, d\theta d\phi$ is the element of solid angle on the body's surface.  

\subsection{Unperturbed body} 

As stated, we are considering a body of mass $M$ and radius $R$, consisting of a perfect fluid with a uniform mass density $\rho = 3M/(4\pi R^3)$. The body is spherical in the absence of a tidal field, and the Newtonian potential outside the body is given by 
\begin{equation} 
U_{\rm out} = \frac{GM}{r}. 
\end{equation}
The potential inside is 
\begin{equation} 
U_{\rm in} =\frac{GM}{2R} (3 - r^2/R^2). 
\end{equation} 
The potential is continuous and differentiable at the stellar surface, as required by Eq.~(\ref{surf_cond}). The velocity field vanishes, and Eq.~(\ref{Euler}) implies that $U_{\rm in} - h = \mbox{constant}$. Because $h$ vanishes on the surface by virtue of Eq.~(\ref{surf_cond}), we have that 
\begin{equation} 
h = \frac{GM}{2R} (1 - r^2/R^2). 
\end{equation} 
The preceding three equations provide a complete description of the unperturbed body. 

\subsection{Linear and static tides} 

We immerse the body within a tidal environment created by an external distribution of matter. The gravitational potential created by this remote matter is denoted $U^{\rm ext}$. For the time being we assume that $U^{\rm ext}$ is constant in time; the time dependence will be restored below. We also assume that $U^{\rm ext}$ varies spatially over a length scale that is very long compared with $R$. To reflect this we express it as the Taylor expansion 
\begin{equation} 
U^{\rm ext} = U_0 + g_a x^a - \frac{1}{2} \E_{ab}\, x^a x^b - \frac{1}{6} \E_{abc}\, x^a x^b x^c + \cdots, 
\end{equation} 
where $x^a$ is the position vector with respect to the body's center of mass, $U_0 := U^{\rm ext}(x^a = 0)$, $g_a := \partial_a U^{\rm ext}(x^a=0)$, and $\E_{ab}$, $\E_{abc}$ are the tidal moments defined by Eq.~(\ref{EL_Newton}). 

We choose to work in a noninertial frame of reference attached to the body's center of mass, which is accelerated in the field of the external matter. This forces the introduction of an inertial term in the gravitational potential, given by  
\begin{equation} 
U^{\rm inertial} = -a_a x^a, 
\end{equation} 
where $a_a$ is the acceleration vector. Adding this to $U^{\rm ext}$ and setting $a_a = g_a$, we obtain the tidal potential 
\begin{equation} 
U^{\rm tidal} = -\frac{1}{2} \E_{ab}\, x^a x^b - \frac{1}{6} \E_{abc}\, x^a x^b x^c
= -\frac{1}{2} r^2\, \E^{\sf q} - \frac{1}{6} r^3\, \E^{\sf o}.  
\label{Utidal} 
\end{equation} 
We discarded the irrelevant constant $U_0$ (because it produces no force), and in the second expression we replaced $x^a$ with $r\Omega^a(\theta^A)$, where $(r,\theta^A)$ are the spherical polar coordinates associated with the Cartesian coordinates. The expansion of the tidal potential begins at order $\ell = 2$, and for our purposes here it is truncated beyond $\ell = 3$. 

The body is deformed by forces created by $U^{\rm tidal}$, and we begin with a linear description of this deformation. A complete expression for the gravitational potential outside the body must include the body's response in addition to the tidal field described by Eq.~(\ref{Utidal}). The response is measured by multipole moments of the mass distribution. In a linear description, the quadrupole moment can be expected to be proportional to $\E_{ab}$, and the octupole moment will be proportional to $\E_{abc}$. Inserting appropriate scaling factors to ensure that the potential has the correct dimension, we write 
\begin{equation} 
U_{\rm out} = \frac{GM}{r} - \frac{1}{2} r^2\, \E^{\sf q} - k_2 \frac{R^5}{r^3}\, \E^{\sf q} 
- \frac{1}{6} r^3\, \E^{\sf o} - \frac{1}{3} k_3 \frac{R^7}{r^4}\, \E^{\sf o} 
\label{Uout_linear} 
\end{equation} 
for the exterior potential, where $k_2$ and $k_3$ --- the body's quadrupole and octupole Love numbers, respectively --- are numerical coefficients to be determined. It should be noted that each term in Eq.~(\ref{Uout_linear}) is a solution to Laplace's equation; the terms that grow with $r$ represent the tidal field, while the decaying terms represent the body's response. The exterior potential satisfies $\nabla^2 U_{\rm out} = 0$ by virtue of Eq.~(\ref{Poisson}) and the fact that the body's immediate neighborhood is empty of matter.  

The potential inside the body must be a solution to Eq.~(\ref{Poisson}) with a constant $\rho$. We assume that the density is not changed by the tidal deformation, and write the solution as 
\begin{equation} 
U_{\rm in} = \frac{GM}{2R} (3 - r^2/R^2) + b_2 r^2\, \E^{\sf q} + b_3 r^3\, \E^{\sf o}, 
\label{Uin_linear} 
\end{equation} 
where $b_2$ and $b_3$ are constants to be determined. The first term on the right-hand side of Eq.~(\ref{Uin_linear}) is the unperturbed potential, and it accounts for the right-hand side of Poisson's equation. The remaining terms are solutions to Laplace's equation; we exclude decaying terms because they are singular at the center of the body.  

Our (momentary) assumption that $U^{\rm ext}$ is time-independent implies that $v^a = 0$ within the perturbed body. Equation (\ref{Euler}) then implies that $U_{\rm in} - h$ continues to be a constant in the deformed state. The specific enthalpy is therefore given by 
\begin{equation} 
h = \frac{GM}{2R} (1 - r^2/R^2) + b_2 r^2\, \E^{\sf q} + b_3 r^3\, \E^{\sf o}.  
\label{h_linear} 
\end{equation} 
The body's surface is deformed by the tidal forces, and it is now described by $r = r_{\rm S}(\theta^A)$, with
\begin{equation} 
r_{\rm S} = R \biggl( 1 + d_2 \frac{R^3}{GM}\, \E^{\sf q} + d_3 \frac{R^4}{GM}\, \E^{\sf o} \biggr).  
\end{equation}  
The scaling factors involving $R$ and $GM$ are inserted to ensure that each term within brackets is properly dimensionless; the numerical factors $d_2$ and $d_3$ are to be determined.

The construction thus far has left a number of constants undetermined. They can all be obtained from the surface conditions of Eq.~(\ref{surf_cond}), which are linearized with respect to $\E_{ab}$ and $\E_{abc}$. Simple manipulations reveal that 
\begin{subequations}
\label{coeffs_linear} 
\begin{align}  
k_2 &= \frac{3}{4}, & b_2 &= -\frac{5}{4}, & d_2 &= -\frac{5}{4}, \\ 
k_3 &= \frac{3}{8}, & b_3 &= -\frac{7}{24}, & d_3 &= -\frac{7}{24}. 
\end{align} 
\end{subequations} 
The results for $k_2$ and $k_3$ match the general expression $k_\ell = 3/[4(\ell-1)]$ for the Love numbers of a constant-density sphere; see Ref.~\cite{poisson-will:14}, Eq.~(2.249). 

\subsection{Nonlinear tides: Quadrupole-quadrupole coupling} 

We continue the calculation with the inclusion of nonlinear terms in the description of the gravitational potential, specific enthalpy, and deformed surface. We begin with an account of the coupling between $\E_{ab}$ and itself. 

The gravitational potential satisfies Poisson's equation. With the perturbation of the density set to zero, both exterior and interior corrections to the potential are to be constructed from solutions to Laplace's equations, either growing or decaying. In spite of the linearity of Laplace's equation, the potentials will include nonlinear contributions involving $\EE^{\sf m}$, $\EE^{\sf q}$, and $\EE^{\sf h}$. The origin of the nonlinearity is in the fluid mechanics, which produces a nonlinear deformation of the surface; nonlinear terms therefore appear in the description of the body's response. 

We update the exterior potential to 
\begin{equation} 
U_{\rm out} = \frac{GM}{r} - \frac{1}{2} r^2\, \E^{\sf q} - k_2 \frac{R^5}{r^3}\, \E^{\sf q} 
- p_0 \frac{R^6}{GM} \frac{1}{r}\, \EE^{\sf m} 
- p_2 \frac{R^8}{GM} \frac{1}{r^3}\, \EE^{\sf q} 
- p_4 \frac{R^{10}}{GM} \frac{1}{r^5}\, \EE^{\sf h}. 
\label{Uout_qq} 
\end{equation} 
We include terms that are linear in $\E_{ab}$, both growing and decaying, as well as decaying terms that are bilinear in $\E_{ab}$. We exclude bilinear growing terms because they are not provided by the tidal potential of Eq.~(\ref{Utidal}), and we momentarily omit all terms associated with $\E_{abc}$. We insert scaling factors to ensure that each term in the potential has the correct dimension. The constants $p_2$ and $p_4$ can be interpreted as nonlinear Love numbers; they will be determined below. The third constant, $p_0$, will be shown to vanish. 

The internal potential is updated to 
\begin{equation} 
U_{\rm in} = \frac{GM}{2R} (3 - r^2/R^2) 
+ b_2 r^2\, \E^{\sf q} 
+ q_0 \frac{R^5}{GM}\, \EE^{\sf m} 
+ q_2 \frac{R^3}{GM} r^2\, \EE^{\sf q} 
+ q_4 \frac{R}{GM} r^4\, \EE^{\sf h}, 
\label{Uin_qq} 
\end{equation} 
with each new term a solution to Laplace's equation; the constants $q_0$, $q_2$, and $q_4$ are to be determined. Because $v^a$ continues to vanish in the deformed state, we still have that $U_{\rm in} - h$ is a constant, and we write the specific enthalpy as
\begin{equation} 
h = \frac{GM}{2R} (1 - r^2/R^2) 
+ b_2 r^2\, \E^{\sf q} 
+ h_0 \frac{R^5}{GM}\, \EE^{\sf m} 
+ q_2 \frac{R^3}{GM} r^2\, \EE^{\sf q} 
+ q_4 \frac{R}{GM} r^4\, \EE^{\sf h}. 
\label{h_qq} 
\end{equation} 
Notice that while the coefficients in front of $\EE^{\sf q}$ and $\EE^{\sf h}$ are the same as those for $U_{\rm in}$, the one in front of $\EE^{\sf m}$ is allowed to differ; the constant value of $U_{\rm in} - h$ may be different for the perturbed and unperturbed states. 

The surface equation becomes 
\begin{equation} 
r_{\rm S} = R \biggl[ 1 + d_2 \frac{R^3}{GM}\, \E^{\sf q} 
+ s_0 \frac{R^6}{(GM)^2}\, \EE^{\sf m} 
+ s_2 \frac{R^6}{(GM)^2}\, \EE^{\sf q} 
+ s_4 \frac{R^6}{(GM)^2}\, \EE^{\sf h} \biggr], 
\end{equation} 
where $s_0$, $s_2$, and $s_4$ shall also be determined. 

The numerical value of each constant is revealed by imposing the surface conditions of Eq.~(\ref{surf_cond}), as we did previously. We also need to impose Eq.~(\ref{mass_condition}), the requirement that the body's mass does not change as it goes from its unperturbed state to the deformed state. The computation yields
\begin{subequations}
\label{coeffs_qq} 
\begin{align}  
p_0 &= 0,                   & q_0 &= -\frac{5}{16}, & s_0 &= -\frac{5}{24}, & h_0 &= -\frac{25}{48}, \\ 
p_2 &= -\frac{75}{28}, & q_2 &= \frac{75}{56}, & s_2 &= \frac{75}{28}, \\ 
p_4 &= -\frac{75}{32}, & q_4 &= 0,                   & s_4 &= \frac{75}{32}. 
\end{align} 
\end{subequations} 
The fact that $p_0$ vanishes is intimately related to the preservation of the mass. And indeed, we can see that the monopole term in Eq.~(\ref{Uout_qq}) corresponds to a shift $\Delta M$ in the mass, given by 
\begin{equation} 
\frac{\Delta M}{M} = -p_0 \frac{R^6}{(GM)^2}\, \E_{ab} \E^{ab}. 
\end{equation} 
Setting $p_0 = 0$ ensures that this unphysical shift is properly eliminated. 

\subsection{Nonlinear tides: Quadrupole-octupole coupling} 
\label{subsec:N_quad-oct} 

Next we account for the coupling between $\E_{ab}$ and $\E_{abc}$. We ignore terms that are quadratic in $\E_{ab}$ --- they were dealt with in the preceding subsection ---- and we ignore terms that are quadratic in $\E_{abc}$, because they are of higher order in the spatial-derivative expansion. 

Following the same rules as previously, we write the exterior potential as
\begin{align} 
U_{\rm out} &= \frac{GM}{r} - \frac{1}{2} r^2\, \E^{\sf q} - k_2 \frac{R^5}{r^3}\, \E^{\sf q} 
- \frac{1}{6} r^3\, \E^{\sf o} - \frac{1}{3} k_3 \frac{R^7}{r^4}\, \E^{\sf o} 
\nonumber \\ & \quad \mbox{} 
- a_1 \frac{R^5}{GM} r\, \EE^{\sf d} 
- p_1 \frac{R^8}{GM} \frac{1}{r^2}\, \EE^{\sf d} 
- p_3 \frac{R^{10}}{GM} \frac{1}{r^4}\, \EE^{\sf o} 
- p_5 \frac{R^{12}}{GM} \frac{1}{r^6}\, \EE^{\sf t},  
\label{Uout_qo} 
\end{align} 
the interior potential as
\begin{equation} 
U_{\rm in} = \frac{GM}{2R} (3 - r^2/R^2) 
+ b_2 r^2\, \E^{\sf q} 
+ b_3 r^3\, \E^{\sf o} 
+ q_1 \frac{R^5}{GM} r\, \EE^{\sf d} 
+ q_3 \frac{R^3}{GM} r^3\, \EE^{\sf o} 
+ q_5 \frac{R}{GM} r^5\, \EE^{\sf t}, 
\label{Uin_qo} 
\end{equation} 
the specific enthalpy as 
\begin{equation} 
h = \frac{GM}{2R} (1 - r^2/R^2) 
+ b_2 r^2\, \E^{\sf q} 
+ b_3 r^3\, \E^{\sf o} 
+ q_1 \frac{R^5}{GM} r\, \EE^{\sf d} 
+ q_3 \frac{R^3}{GM} r^3\, \EE^{\sf o} 
+ q_5 \frac{R}{GM} r^5\, \EE^{\sf t}, 
\label{h_qo} 
\end{equation} 
and the surface equation as 
\begin{equation} 
r_{\rm S} = R \biggl[ 1 + d_2 \frac{R^3}{GM}\, \E^{\sf q}
+ d_3 \frac{R^4}{GM}\, \E^{\sf o}
+ s_1 \frac{R^7}{(GM)^2}\, \EE^{\sf d} 
+ s_3 \frac{R^7}{(GM)^2}\, \EE^{\sf o} 
+ s_5 \frac{R^7}{(GM)^2}\, \EE^{\sf t} \biggr].  
\end{equation} 
The constants $p_3$ and $p_5$ can be interpreted as another set of nonlinear Love numbers; the reason for excluding $p_1$ from this list will be made clear presently. The $a_1$ term in $U_{\rm out}$ actually violates the rules put in place in the preceding subsection: it is a growing term that is not supplied by the tidal potential of Eq.~(\ref{Utidal}). The need for this contribution will also be explained below. 

With the exception of $p_1$, the constants can all be determined from Eq.~(\ref{surf_cond}); the mass-preservation condition of Eq.~(\ref{mass_condition}) is not required here. We obtain 
\begin{subequations}
\label{coeffs_qo} 
\begin{align}  
a_1 &= \frac{1}{4},        & q_1 &= -\frac{7}{16} - p_1,   & s_1 &= -\frac{3}{16} - p_1, \\ 
p_3 &= -\frac{35}{32},  & q_3 &= \frac{35}{96},             & s_3 &= \frac{385}{288}, \\ 
p_5 &= -\frac{35}{32},  & q_5 &= 0,                               & s_5 &= \frac{35}{24}. 
\end{align} 
\end{subequations} 
The numerical value of $p_1$ remains arbitrary. This has to do with the freedom to shift the origin of the coordinate system away from the body's center of mass --- such a shift produces a decaying dipole term in the external potential. Indeed, a straightforward calculation reveals that the position of the center of mass, defined by $R^a := M^{-1} \int \rho x^a\, dV$, is given by  
\begin{equation} 
R^a = -p_1 \frac{R^8}{(GM)^2}\, \EE_a = -p_1 \frac{R^8}{(GM)^2}\, \E_{abc} \E^{bc}. 
\end{equation} 
Setting $p_1 = 0$ places the origin of the coordinates firmly at the center of mass, and eliminates the decaying dipole from the gravitational potential. We shall henceforth enforce this convention. 

\subsection{Tidal acceleration} 

Given that $p_1$ remains arbitrary when solving for the deformed potentials and surface, a solution could not have been obtained without the insertion of another dipole term, the growing one proportional to $a_1$. And indeed, we found that the numerical value of this coefficient is unambiguous. 

To understand the meaning of this, we recall that our frame of reference is noninertial, and attached to the body's accelerated center of mass. This implies that the gravitational potential includes an inertial term $-a_a x^a$, where $a_a$ is the body's acceleration; this has the structure of a growing dipole. The inertial potential was previously cancelled out against the $g_a x^a$ term in the external potential, the one created by the remote distribution of matter. What we have here is an additional growing dipole, one that originates from the nonlinear coupling between $\E_{ab}$ and $\E_{abc}$. To eliminate it from the potential, as we did previously for $g_a x^a$, we must make an adjustment to the acceleration vector: {\it the new dipole term is a manifestation of the fact that the body is accelerated by the tidal field}. A comparison between $U^{\rm inertial}$ and the growing dipole term in $U_{\rm out}$ reveals that the tidal acceleration is given by 
\begin{equation} 
g^{\rm tidal}_a = a_1 \frac{R^5}{GM}\, \EE_a = a_1 \frac{R^5}{GM}\, \E_{abc} \E^{bc}. 
\label{atidal1} 
\end{equation} 
The body's complete acceleration is then $a_a = g_a + g_a^{\rm tidal}$, and the growing dipole is thus eliminated from the gravitational potential. 

To better understand the origin of the tidal acceleration, we reflect on the motion of a nonspherical body in Newtonian gravity, as discussed in Sec.~1.6 of Poisson and Will \cite{poisson-will:14}. According to their Eq.~(1.194), the net force on a body of mass $M$ with a mass quadrupole moment $Q^{ab}$ moving in an external gravitational potential $U^{\rm ext}$ is given by $M(g_a + g^{\rm tidal})$, where 
\begin{equation} 
Mg^{\rm tidal}_a = \frac{1}{2} Q^{bc}\, \partial_{abc} U^{\rm ext} = -\frac{1}{2} Q^{bc}\, \E_{abc}. 
\end{equation} 
Our body, however, is spherical when unperturbed, and its quadrupole moment is created by the tidal deformation. It is related to the tidal moment $\E_{ab}$ by [see Eq.~(2.267) of Poisson and Will]
\begin{equation} 
G Q_{ab} = -\frac{2}{3} k_2 R^5\, \E_{ab}, 
\end{equation} 
where $k_2$ is the body's quadrupolar Love number. Inserting this within the expression for the tidal acceleration, we find that it becomes 
\begin{equation} 
g_a^{\rm tidal} = \frac{1}{3} k_2 \frac{R^5}{GM}\, \E_{abc} \E^{bc}. 
\label{atidal2} 
\end{equation} 
This agrees with Eq.~(\ref{atidal1}) provided that $a_1 = k_2/3$. 

In the case of a constant-density sphere, we found that $k_2 = 3/4$, and it follows from this that $a_1$ should be equal to $1/4$. This is precisely the value found in our previous calculation, as displayed in Eqs.~(\ref{coeffs_qo}). The meaning of the growing dipole is therefore fully clarified: it is an inertial term in the gravitational potential, associated with the body's acceleration in the tidal field created by the external matter. 

\subsection{Linear and dynamic tides} 

We return to the linear problem, and finally account for the time variation of the tidal environment. A static field produces a static deformation, and in such a situation the fluid's velocity field vanishes. Time-varying tidal forces, however, produce mass currents within the body, and we can expect that $v^a$ will be proportional to $\dot{\E}_L$. Neglecting terms quadratic in the velocity, we find from Euler's equation (\ref{Euler}) that $U$ and $h$ acquire new terms proportional to $\ddot{\E}_L$. Our updated expressions for the gravitational potential are then 
\begin{equation} 
U_{\rm out} = \frac{GM}{r} - \frac{1}{2} r^2\, \E^{\sf q} - k_2 \frac{R^5}{r^3}\, \E^{\sf q} 
- \ddot{k}_2 \frac{R^8}{GM} \frac{1}{r^3}\, \ddot{\E}^{\sf q} 
- \frac{1}{6} r^3\, \E^{\sf o} - \frac{1}{3} k_3 \frac{R^7}{r^4}\, \E^{\sf o} 
- \frac{1}{3} \ddot{k}_3 \frac{R^{10}}{GM} \frac{1}{r^4}\, \ddot{\E}^{\sf o} 
\end{equation} 
and 
\begin{equation} 
U_{\rm in} = \frac{GM}{2R} (3-r^2/R^2) + b_2 r^2\, \E^{\sf q} 
+ \ddot{b}_2 \frac{R^3}{GM} r^2\, \ddot{\E}^{\sf q} 
+ b_3 r^3\, \E^{\sf o} 
+ \ddot{b}_3 \frac{R^3}{GM} r^3\, \ddot{\E}^{\sf o},
\end{equation} 
where $\ddot{k}_2$, $\ddot{k}_3$ are new Love numbers associated with the time variation, and $\ddot{b}_2$, $\ddot{b}_3$ are corresponding constants for the interior potential. The new contributions to the potential come with a proportionality factor that compensates dimensionally for the time derivatives; the only available time scale is $(R^3/GM)^{1/2}$. 

The specific enthalpy becomes 
\begin{equation} 
h = \frac{GM}{2R} (1-r^2/R^2) + c_2 r^2\, \E^{\sf q} 
+ \ddot{c}_2 \frac{R^3}{GM} r^2\, \ddot{\E}^{\sf q} 
+ c_3 r^3\, \E^{\sf o} 
+ \ddot{c}_3 \frac{R^3}{GM} r^3\, \ddot{\E}^{\sf o},
\end{equation} 
and we observe that the coefficients $\ddot{c}_2$, $\ddot{c}_3$ associated with the time-derivative terms are not equal to the corresponding ones in $U_{\rm in}$; $U_{\rm in} - h$ is no longer a constant when we account for the time variation. 

We write the surface equation as 
\begin{equation} 
r_{\rm S} = R \biggl( 1 + d_2 \frac{R^3}{GM}\, \E^{\sf q} 
+ \ddot{d}_2 \frac{R^6}{(GM)^2}\, \ddot{\E}^{\sf q} 
+ d_3 \frac{R^4}{GM}\, \E^{\sf o} 
+ \ddot{d}_3 \frac{R^7}{(GM)^2}\, \ddot{\E}^{\sf o} \biggr), 
\end{equation} 
and the velocity field as 
\begin{equation} 
v_a = \nabla_a \biggl( \frac{1}{2} d_2 \frac{R^3}{GM} r^2\, \dot{\E}^{\sf q} 
+ \frac{1}{3} d_3 \frac{R^3}{GM} r^3\, \dot{\E}^{\sf o} \biggr). 
\end{equation} 
That the velocity vector is a gradient field is dictated by Euler's equation (still neglecting terms quadratic in the velocity). The fact that the coefficients are $d_2$ and $d_3$ is dictated by the requirement that $v_a(r=R)\Omega^a$, the radial component of the velocity evaluated at the surface, be equal to $\dot{r}_{\rm S}$. Notice that $\nabla_a v^a = 0$, as demanded by the continuity equation when $\rho = \mbox{constant}$.    

As usual the numerical value of all coefficients is determined by enforcing Eq.~(\ref{surf_cond}). We arrive at 
\begin{subequations}
\label{coeffs_time} 
\begin{align}  
\ddot{k}_2 &= -\frac{15}{16}, & \ddot{b}_2 &= \frac{15}{16}, 
& \ddot{c}_2 &= \frac{25}{16}, & \ddot{d}_2 &= \frac{25}{16}, \\ 
\ddot{k}_3 &= -\frac{7}{32}, & \ddot{b}_3 &= \frac{7}{96}, 
& \ddot{c}_2 &= \frac{49}{288}, & \ddot{d}_3 &= \frac{49}{288}.  
\end{align} 
\end{subequations} 

\subsection{Complete potential} 

To sum up, we have found that the gravitational potential outside a tidally deformed body in Newtonian theory is given by 
\begin{align} 
U_{\rm out} &= \frac{GM}{r} - \frac{1}{2} r^2\, \E^{\sf q} - k_2 \frac{R^5}{r^3}\, \E^{\sf q} 
- \frac{1}{6} r^3\, \E^{\sf o} - \frac{1}{3} k_3 \frac{R^7}{r^4}\, \E^{\sf o} 
\nonumber \\ & \quad \mbox{} 
- a_1 \frac{R^5}{GM} r\, \EE^{\sf d} 
- p_2 \frac{R^8}{GM} \frac{1}{r^3}\, \EE^{\sf q} 
- p_3 \frac{R^{10}}{GM} \frac{1}{r^4}\, \EE^{\sf o} 
- p_4 \frac{R^{10}}{GM} \frac{1}{r^5}\, \EE^{\sf h}  
- p_5 \frac{R^{12}}{GM} \frac{1}{r^6}\, \EE^{\sf t}
\nonumber \\ & \quad \mbox{} 
- \ddot{k}_2 \frac{R^8}{GM} \frac{1}{r^3}\, \ddot{\E}^{\sf q} 
- \frac{1}{3} \ddot{k}_3 \frac{R^{10}}{GM} \frac{1}{r^4}\, \ddot{\E}^{\sf o}.  
\label{Ucomplete} 
\end{align} 
The description of the tidal environment is truncated beyond $\ell = 3$, and the nonlinear response includes the coupling between $\E_{ab}$ and itself (the quadrupole and hexadecapole terms), as well as the coupling between $\E_{ab}$ and $\E_{abc}$ (the dipole, octupole, and triakontadipole terms); we neglect all terms that are quadratic in $\E_{abc}$. The constants $k_2$, $k_3$ are linear Love numbers; $p_2$, $p_3$, $p_4$, and $p_5$ are nonlinear Love numbers; and $\ddot{k}_2$, $\ddot{k}_3$ are time-variation Love numbers. We eliminated the decaying dipole contribution to the potential; as we have seen, $p_1$ can always be set to zero by placing the origin of the coordinate system at the body's center of mass. We kept the growing dipole, and recognize that $a_1 = k_2/3$ provides a dimensionless measure of the body's tidal acceleration.  

The numerical values of all constants were determined for a constant-density sphere --- refer back to Eqs.~(\ref{coeffs_linear}), (\ref{coeffs_qq}), (\ref{coeffs_qo}), and (\ref{coeffs_time}). The validity of Eq.~(\ref{Ucomplete}), however, extends well beyond this simple-minded model: it applies to any fluid body with any distribution of mass, provided that the numerical value of each Love number is adjusted appropriately. Their calculation for a generic body would require the specification of an equation of state. 

The exterior potential includes a quadrupole term given by $\frac{3}{2} Q_{ab} \Omega^a \Omega^b/r^3$. Comparison with Eq.~(\ref{Ucomplete}) reveals that the quadrupole moment of the mass distribution is given by 
\begin{equation} 
Q_{ab} = -\frac{2}{3} k_2 R^5\, \E_{ab} - \frac{2}{3} p_2 \frac{R^8}{GM}\, \E_{c\langle a} \E^c_{\ b\rangle} 
- \frac{2}{3} \ddot{k}_2 \frac{R^8}{GM}\, \ddot{\E}_{ab}. 
\label{quadmoment_Newtonian} 
\end{equation} 
We observe that $Q_{ab}$ consists of a linear-static contribution proportional to $k_2$, a nonlinear contribution proportional to $p_2$, and a linear-dynamic contribution proportional to $\ddot{k}_2$. It is this observation that provides an operational meaning to the Love numbers. Similarly, the potential includes an octupole term given by $\frac{5}{2} Q_{abc} \Omega^a \Omega^b \Omega^c/r^4$, and comparison with Eq.~(\ref{Ucomplete}) reveals that 
\begin{equation} 
Q_{abc} = -\frac{2}{5} k_3 R^7\, \E_{abc} - \frac{2}{5} p_3 \frac{R^{10}}{GM}\, \E_{d\langle ab} \E^d_{\ c\rangle} 
- \frac{2}{5} \ddot{k}_3 \frac{R^{10}}{GM}\, \ddot{\E}_{abc}. 
\label{octmoment_Newtonian} 
\end{equation} 
This consists of a linear-static contribution proportional to $k_3$, a nonlinear contribution proportional to $p_3$, and a linear-dynamic contribution proportional to $\ddot{k}_3$.

\section{Tidally deformed body in general relativity} 
\label{sec:GR} 

In this section we work toward our main goal, to obtain the metric of a tidally deformed compact body in general relativity. Our precise objectives are the same as those specified in Sec.~\ref{subsec:M0_goals}: We wish to express the metric in simultaneous curvature, spatial-derivative, and time-derivative expansions. We shall follow the general strategy adopted in Sec.~\ref{sec:worldline}.  

\subsection{The task} 
\label{subsec:task}

We begin with a more precise statement of our goals. We wish to construct the metric of a tidally deformed body of mass $M$ and radius $R$; this can be a material body or a black hole. The metric shall be defined in a neighborhood around the body, and we exclude the body's interior ($r < R$) from our considerations; the metric shall be a solution to the Einstein field equations in vacuum. The body's tidal environment is characterized by tidal multipole moments $\E_L$. In the context of this section, these are freely-specifiable functions of time that are not determined by the field equations. In the limit $M \to 0$, in which the body disappears and leaves behind a timelike geodesic $\gamma$, the tidal moments become related to the Weyl curvature tensor by Eq.~(\ref{EL_def}); in this limit, the metric shall reduce to the one displayed in Eq.~(\ref{M0_metric}). 

The metric shall be expressed as a curvature expansion in powers of $(r/{\cal R})^2 \ll 1$, where $r$ is the distance to the body, and ${\cal R}$ is a characteristic length scale associated with the strength of the tidal field. If $M'$ is a characteristic mass for the external matter responsible for the tidal environment, and $a$ is a measure of the distance between body and external matter, then ${\cal R} \sim (a^3/M')^{1/2}$. The curvature expansion shall be truncated beyond second order, and it shall account for the coupling between $\E_{ab}$ and itself, and the coupling between $\E_{ab}$ and $\E_{abc}$. 

The metric shall also be expressed as a spatial-derivative expansion in powers of $r/{\cal L} \ll 1$, where ${\cal L}$ is a characteristic length scale associated with inhomogeneities in the tidal field. In the example mentioned previously, ${\cal L} \sim a$. Terms linear in the curvature shall be constructed from $\E^{\sf q}$, $\E^{\sf o}$, $\E^{\sf h}$, and $\E^{\sf t}$; for these, the contributions from the spatial-derivative expansion are of order $(r/{\cal R})^2 (r/{\cal L})^{\ell - 2}$, with each subsequent multipole adding an additional factor of $r/{\cal L}$. Terms quadratic in the curvature shall be built from $\EE^{\sf m}$, $\EE^{\sf d}$, $\EE^{\sf q}$, $\EE^{\sf o}$, $\EE^{\sf h}$, and $\EE^{\sf t}$. For these, the monopole, quadrupole, and hexadecapole contributions are proportional to $\E_{ab}$ multiplied by itself, and are of order $(r/{\cal R})^4 (r/{\cal L})^0$; the dipole, octupole, and triakontadipole contributions are proportional to $\E_{ab}$ multiplied by $\E_{abc}$, and are of order $(r/{\cal R})^4 (r/{\cal L})$. 

Finally, the metric shall be expressed as a time-derivative expansion in powers of $r/{\cal T} \ll 1$, where ${\cal T}$ is a characteristic time scale associated with temporal variations of the tidal field. For our canonical example, ${\cal T} \sim (a^3/M')^{1/2}$, and it is of the same order of magnitude as ${\cal R}$. This expansion makes use of time derivatives of the tidal moments $\E_L$; the first derivative $\dot{\E}_L$ produces terms in the metric that scale as 
$(r/{\cal R})^2 (r/{\cal L})^{\ell - 2} (r/{\cal T})$, and terms of order $(r/{\cal R})^2 (r/{\cal L})^{\ell - 2} (r/{\cal T})^2$
come from $\ddot{\E}_L$. We shall truncate the time-derivative expansion beyond the second derivative.   

Formally, the metric shall be expanded through orders $r^4$ and $r^5$, with an error term 
\begin{align} 
O(r^5, r^6) &:= O\bigl[ (r/{\cal R})^2 (r/{\cal L})^4 \bigr]
+ O\bigl[ (r/{\cal R})^4 (r/{\cal L})^2 \bigr] 
+ O\bigl[ (r/{\cal R})^6 \bigr]
\nonumber \\ & \quad \mbox{} 
+ O\bigl[(r/{\cal R})^2 (r/{\cal T})^3 \bigr]
+ O\bigl[(r/{\cal R})^2 (r/{\cal L}) (r/{\cal T})^3 \bigr]
+ O\bigl[(r/{\cal R})^2 (r/{\cal L})^2 (r/{\cal T})^2 \bigr]. 
\label{error} 
\end{align} 
The contribution of order $r^5$ to the error comes from the fact that the $\ell = 2$ part of the metric perturbation does not incorporate a term involving the third derivative of $\E_{ab}(t)$. The metric, however, shall incorporate relativistic corrections of all orders in $M/r$; we make no compromise with the relativistic expansion. 
 
\subsection{Unperturbed body} 

By virtue of Birkhoff's theorem, the metric outside any spherically symmetric body of mass $M$ is given by the Schwarzschild solution
\begin{equation} 
ds^2 = -f\, dt^2 + f^{-1}\, dr^2 + r^2\, d\Omega^2, 
\end{equation} 
where 
\begin{equation} 
f := 1 - \frac{2M}{r} 
\end{equation} 
and $d\Omega^2 := \Omega_{AB}\, d\theta^A d\theta^B$ is the metric on the unit two-sphere; we recall that $\theta^A = (\theta, \phi)$ and $\Omega_{AB} = \mbox{diag}[1,\sin^2\theta]$. The metric is valid outside the body, where $r > R$. 

In the case of a material body, the exterior metric is matched at $r=R$ to an interior metric obtained by integrating the Einstein field equations together with the equation of hydrostatic equilibrium. In the case of a black hole, $R = 2M$, and the interior metric continues to be given by the Schwarzschild solution. An extension across $r = 2M$ can be accomplished by transforming the metric to the advanced-time coordinate 
\begin{equation} 
v = t + r + 2M \ln\biggl( \frac{r}{2M} - 1 \biggr). 
\label{advanced-time} 
\end{equation}  

\subsection{Linear and static tides} 
\label{subsec:metric_linear-static} 

We begin our construction of the perturbed metric with a linearized description of the tidal deformation. For the time being we imagine that the tidal environment is stationary; the time dependence will be incorporated at a later stage, in Secs.~\ref{subsec:first_time} and \ref{subsec:second_time}.  

The metric of a spherical body of mass $M$, tidally deformed by an external source of gravity, was previously obtained by Binnington and Poisson \cite{binnington-poisson:09}. The metric is presented in $(t, r, \theta^A)$ coordinates, and the tidal perturbation is calculated in the Regge-Wheeler gauge \cite{regge-wheeler:57} (refer to Appendix~\ref{app:RW} for definitions). It is expressed as 
\begin{subequations} 
\label{metric_linear} 
\begin{align} 
g_{tt} &= -f + p_{tt}^{\rm ls}, \\ 
g_{rr} &= f^{-1} + p_{rr}^{\rm ls}, \\ 
g_{AB} &= r^2 \Omega_{AB}(1 + p^{\rm ls}), 
\end{align}
\end{subequations} 
where 
$p^{\rm ls}_{tt}$, $p^{\rm ls}_{rr}$, $p^{\rm ls}$ (with the label standing for ``linear and static'') are the components of the metric perturbation. This expression can be compared with Eq.~(\ref{M0_linear_metric}), which applies in the limit $M \to 0$. 

For a tidal field of multipole order $\ell$, the perturbation is given by 
\begin{subequations} 
\label{pert_linear_ell} 
\begin{align} 
p^{\rm ls}_{tt} &=-\frac{2}{(\ell-1)\ell}\, f^2 r^\ell \Bigl[ A_\ell + 2 K_\ell (M/r)^{2\ell+1} B_\ell \Bigr]\, 
\E_L \Omega^L, \\ 
p^{\rm ls}_{rr} &= -\frac{2}{(\ell-1)\ell}\, r^\ell \Bigl[ A_\ell + 2 K_\ell (M/r)^{2\ell+1} B_\ell \Bigr]\, 
\E_L \Omega^L, \\ 
p^{\rm ls} &= -\frac{2}{(\ell-1)\ell}\, r^\ell \Bigl[ C_\ell + 2 K_\ell (M/r)^{2\ell+1} D_\ell \Bigr]\,
\E_L \Omega^L,
\end{align} 
\end{subequations} 
where 
\begin{equation} 
\E_L \Omega^L := \E_{a_1 a_2 \cdots a_\ell}\, \Omega^{a_1} \Omega^{a_2} \cdots \Omega^{a_\ell} 
\end{equation} 
and 
\begin{subequations} 
\label{ABCD} 
\begin{align} 
A_\ell &:= F(-\ell, -\ell+2; -2\ell; 2M/r), \\ 
B_\ell &:= F(\ell+1, \ell+3; 2\ell+2; 2M/r), \\ 
C_\ell &:= \frac{\ell+1}{\ell-1} F(-\ell, -\ell; -2\ell; 2M/r) 
- \frac{2}{\ell-1} F(-\ell-1, -\ell; -2\ell; 2M/r), \\ 
D_\ell &:= \frac{\ell}{\ell+2} F(\ell+1, \ell+1; 2\ell+2; 2M/r) 
+ \frac{2}{\ell+2} F(\ell, \ell+1; 2\ell+2; 2M/r),  
\end{align} 
\end{subequations} 
with $F(a,b;c;z)$ standing for the hypergeometric function. The radial functions admit the asymptotic expansions 
\begin{equation} 
A_\ell = 1 + O(M/r), \qquad 
B_\ell = 1 + O(M/r), \qquad 
C_\ell = 1 + O(M/r), \qquad 
D_\ell = 1 + O(M/r)
\end{equation}  
when $r \gg M$. The dimensionless constants $K_\ell$ are rescaled Love numbers. If we define an effective potential $U_{\rm eff}$ via $g_{tt} = -1 + 2U_{\rm eff}$, then 
\begin{equation} 
U_{\rm eff} = \frac{M}{r} - \frac{1}{(\ell-1)\ell}\, r^\ell \Bigl\{ 
\bigl[ 1 + O(M/r) \bigr] + 2 K_\ell (M/r)^{2\ell+1} \bigl[1 + O(M/r) \bigr] \Bigr\} \, \E_L \Omega^L. 
\label{U_eff} 
\end{equation} 
This matches the expression for the Newtonian potential when $r \gg M$ --- see, for example, Eq.~(2.269) of Poisson and Will \cite{poisson-will:14}\footnote{Note that Poisson and Will define $\E_L$ as $-\partial_L U^{\rm ext}$ instead of adopting the definition of Eq.~(\ref{EL_Newton}). This difference of definition gives rise to the different overall numerical factor in their Eq.~(2.269).} ---  provided that 
\begin{equation} 
K_\ell M^{2\ell+1} = k_\ell R^{2\ell+1}, 
\label{K_vs_k} 
\end{equation} 
where $k_\ell$ are the primitive and scalefree Love numbers.  

The functions $A_\ell$ and $C_\ell$ are terminating polynomials in $2M/r$. The functions $B_\ell$ and $D_\ell$ contain terms proportional to $\ln f$, and an expansion in powers of $2M/r$ does not terminate. In the case of a black hole, the perturbation must be smooth across the deformed horizon, and this requires $K_\ell = 0$ --- the Love numbers of a black hole are zero. The surface $r = 2M$ has no relevance in the case of material bodies (it occurs inside the body, where the metric does not apply), and for these we have that $K_\ell \neq 0$.  

The functions $A_\ell$, $B_\ell$, $C_\ell$, $D_\ell$ can be given an alternative representation in terms of Legendre functions; the details are provided in Appendix~\ref{app:Legendre}. An explicit listing of the functions for $\ell = \{2, 3, 4, 5\}$ is given in Appendix~\ref{app:ABCD}. 

\subsection{Nonlinear tides: Integration of the field equations} 
\label{subsec:integration} 

When we work at second order in the curvature expansion, the coupling between $\E_{ab}$ and itself produces terms in the metric of multipolar order $\ell = \{0, 2, 4\}$, involving $\EE^{\sf m}$, $\EE^{\sf q}$, and $\EE^{\sf h}$, respectively. On the other hand, the coupling between $\E_{ab}$ and $\E_{abc}$ produces terms of multipolar order $\ell = \{1, 3, 5\}$, implicating $\EE^{\sf d}$, $\EE^{\sf o}$, and $\EE^{\sf t}$, respectively.   

With the coordinate conditions described below, the metric is expressed as 
\begin{subequations} 
\label{metric_bilinear} 
\begin{align} 
g_{tt} &= -f +p_{tt}^{\rm ls}+ p_{tt}^{\rm qq} + p_{tt}^{\rm qo}, \\ 
g_{rr} &= f^{-1} + p_{rr}^{\rm ls} + p_{rr}^{\rm qq} + p_{rr}^{\rm qo}, \\ 
g_{rA} &= r\, p_{rA}^{\rm qo}, \\ 
g_{AB} &= r^2 \Omega_{AB}(1 + p^{\rm ls} + p^{\rm qq} + p^{\rm qo}). 
\end{align}
\end{subequations} 
The linear and static (``ls'') perturbations are listed in Eq.~(\ref{pert_linear}), the quadrupole-quadrupole (``qq'') terms are written as  
\begin{subequations} 
\label{pert_qq} 
\begin{align} 
p_{tt}^{\rm qq} &= 
h^0_{tt}(r)\, \EE^{\sf m} + h^2_{tt}(r)\, \EE^{\sf q} + h^4_{tt}(r)\, \EE^{\sf h} + O(r^6), \\ 
p_{rr}^{\rm qq} &= 
h^0_{rr}(r)\, \EE^{\sf m} + h^2_{rr}(r)\, \EE^{\sf q} + h^4_{rr}(r)\, \EE^{\sf h} + O(r^6), \\ 
p^{\rm qq} &= h^2(r)\, \EE^{\sf q} + h^4(r)\, \EE^{\sf h} + O(r^6), 
\end{align} 
\end{subequations} 
and the quadrupole-octupole (``qo'') terms are written as 
\begin{subequations} 
\label{pert_qo} 
\begin{align} 
p_{tt}^{\rm qo} &= 
h^1_{tt}(r)\, \EE^{\sf d} + h^3_{tt}(r)\, \EE^{\sf o} + h^5_{tt}(r)\, \EE^{\sf t} + O(r^6), \\ 
p_{rr}^{\rm qo} &= 
h^1_{rr}(r)\, \EE^{\sf d} + h^3_{rr}(r)\, \EE^{\sf o} + h^5_{rr}(r)\, \EE^{\sf t}+ O(r^6), \\ 
p_{rA}^{\rm qo} &= j^1(r)\, \partial_A \EE^{\sf d} + O(r^6), \\ 
p^{\rm qo} &= 
h^1(r)\, \EE^{\sf d} + h^3(r)\, \EE^{\sf o} + h^5(r)\, \EE^{\sf t} + O(r^6),  
\end{align} 
\end{subequations} 
where $\partial_A$ denotes a partial derivative with respect to $\theta^A$. Notice that $p_{rA}^{\rm qo}$ includes a single term with $\ell = 1$, and that $p^{\rm qq}$ excludes a term with $\ell = 0$.

For $\ell = \{2,3,4,5\}$ we continue to exploit the Regge-Wheeler gauge, and this choice is reflected in the expressions of Eqs.~(\ref{pert_qq}) and (\ref{pert_qo}). Letting $G_{\alpha\beta}$ denote the Einstein tensor, we find that the tracefree part of the angular components of the field equations, $G_\stf{AB} = 0$, provide an algebraic relation between $h^\ell_{tt}$ and $h^\ell_{rr}$. The components $G_{rA} = 0$ then allow us to express $dh^\ell/dr$ in terms of $h^\ell_{rr}$ and its derivative; from this we obtain $d^2h^\ell/dr^2$. Making the substitutions within $f^{-1} G_{tt} + f G_{rr} = 0$, we obtain a second-order differential equation for $h^\ell_{rr}$,   
\begin{equation} 
r^2 f \frac{d^2 h^\ell_{rr}}{dr^2} + 2(r+M) \frac{d h^\ell_{rr}}{dr} - \ell(\ell+1) h^\ell_{rr} + S^\ell = 0, 
\label{diff_eq} 
\end{equation} 
where $S^\ell(r)$ is a source term constructed from the linear perturbation. From the solution to this equation we immediately obtain $h^\ell_{tt}$, and with $dh^\ell/dr$ and $d^2h^\ell/dr^2$ determined previously, $G_{tt} = 0$ provides an algebraic equation for $h^\ell$. At this stage the problem is solved, and we find that the remaining field equations are redundant. 

A particular solution to Eq.~(\ref{diff_eq}) is 
\begin{equation} 
h^{\ell\ {\rm part}}_{rr} = \frac{1}{2\ell+1} \biggl( r^{-(\ell+1)} B_\ell \int r^\ell A_\ell\, f^2 S^\ell\, dr 
- r^\ell A_\ell \int r^{-(\ell+1)} B_\ell\, f^2 S^\ell\, dr \biggr), 
\label{diff_eq_part} 
\end{equation} 
where $A_\ell$ and $B_\ell$ are the functions defined by Eq.~(\ref{ABCD}). This result relies on the fact that $r^2 f^3 W = -(2\ell+1)$, where $W := g_1 (dg_2/dr) - g_2 (dg_1/dr)$ is the Wronskian of the functions $g_1 := r^\ell A_\ell$ and $g_2 := r^{-(\ell+1)} B_\ell$, the linearly independent solutions to the homogeneous differential equation; the equation implies that $r^2 f^3 W$ is a constant, and evaluation of the left-hand side in the limit $r \to \infty$ returns $-(2\ell+1)$. The general solution to Eq.~(\ref{diff_eq}) is then 
\begin{equation} 
h^\ell_{rr} = c_1\, r^\ell A_\ell + c_2\, r^{-(\ell+1)} B_\ell + h^{\ell\ {\rm part}}_{rr}, 
\end{equation} 
and it comes with two constants of integration, $c_1$ and $c_2$. For the source terms encountered in practice in these computations, we find that the first integral in Eq.~(\ref{diff_eq_part}) can be evaluated in a straightforward manner. For the second integral it is advantageous to integrate by parts when dealing with terms proportional to $(\ln f)^2$.  

The Regge-Wheeler gauge is not defined for $\ell = 0$. In this case we have the additional freedom to set to zero the angular components of the bilinear perturbation; this choice is reflected in Eq.~(\ref{pert_qq}). (Recall that we adopted the same gauge back in Sec.~\ref{subsec:M0_quad-quad}.) The component $G_{tt} = 0$ of the field equations gives rise to a first-order differential equation for $h^0_{rr}$, and $G_{rr} = 0$ produces another first-order differential equation for $h^0_{tt}$. Integration is straightforward, and the solution comes with two constants of integrations. All remaining field equations are redundantly satisfied. 

The Regge-Wheeler gauge is also not defined for $\ell = 1$. As was discussed back in Sec.~\ref{subsec:M0_quad-oct}, several choices of gauge were examined, and the one that produced the simplest results for the radial functions is the 
nondiagonal gauge defined by 
\begin{equation} 
h^1_{tt} = f h^1, \qquad 
h^1_{rr} = f^{-1} h^1, \qquad 
j^1 \neq 0. 
\label{Ngauge_def} 
\end{equation} 
In this gauge, the components $G_{rA} = 0$ of the field equations give $j^1$ algebraically in terms of $h^1$ and its derivative, and $G_{rr} = 0$ returns a first-order differential equation for $h_1$,  
\begin{equation} 
r f \frac{d h^1}{dr} + 2 \biggl(1 - \frac{M}{r} \biggr) h^1 + S^1 = 0, 
\end{equation} 
where $S^1(r)$ is a source term constructed from the linear perturbation. The general solution is 
\begin{equation} 
h^1 = -\frac{1}{r^2 f} \biggl( c_3 + \int r S^1\, dr \biggr), 
\label{h1_sol} 
\end{equation} 
where $c_3$ is a constant of integration. We again find it advantageous, when evaluating the integral, to integrate by parts when dealing with terms proportional to $(\ln f)^2$. With $h^1$ in hand, $j^1$ is obtained immediately, and all remaining field equations are redundantly satisfied. 

All manipulations described in this subsection rely on a spherical-harmonic decomposition of the Einstein field equations --- this implicates scalar, vector, and tensor harmonics. As in Sec.~\ref{sec:worldline}, the calculations are simplified by imagining (momentarily, for these immediate purposes) that all tidal tensors are axisymmetric, so that the angular functions are those given by Eqs.~(\ref{axi_potentials}), (\ref{axi_quad_quad_potentials}), and (\ref{axi_quad_oct_potentials}).  

\subsection{Nonlinear tides: Quadrupole-quadrupole coupling} 
\label{subsec:metric_quad-quad} 

We follow the methods outlined in the preceding subsection, and obtain general solutions for the bilinear perturbations that result from the coupling between $\E_{ab}$ and itself. We describe separately the solutions for $\ell = 0$, $\ell = 2$, and $\ell = 4$. 

\subsubsection{Monopole sector: $\ell = 0$} 

The nonvanishing radial functions are 
\begin{subequations} 
\label{radialf_L0} 
\begin{align} 
h^0_{tt} &= h^0_{tt}[\BH] + S_0 M^4 f + (K_2)^2 M^4\, \gothm^0_{tt} + K_2 M^4\, \gothn^0_{tt}, \\ 
h^0_{rr} &= h^0_{rr}[\BH] + (K_2)^2 M^4\, \gothm^0_{rr} + K_2 M^4\, \gothn^0_{rr},
\end{align} 
\end{subequations} 
where 
\begin{subequations}
\label{radialf_L0_BH} 
\begin{align} 
h^0_{tt}[\BH] &= T_0 \frac{M^5}{r} - \frac{1}{15} (r-2M)^3(r+2M), \\ 
h^0_{rr}[\BH] &= T_0 \frac{M^5}{r f^2} - \frac{1}{15} r^2 (r+2M)(3r-2M)
\end{align} 
\end{subequations} 
is the piece of the solution that survives when the compact object is a black hole, so that $K_2 = 0$. The remaining radial functions are listed in Appendix~\ref{app:MNO}.

The solution comes with two constants of integration. The first is $S_0$, and it corresponds to a residual gauge freedom that consists of a rescaling of the time coordinate; this constant can always be set to zero without loss of generality. The second is $T_0$, and as we shall show in Sec.~\ref{subsec:redefinitions}, it corresponds to a freedom to redefine $M$ by a shift proportional to $M^5 \E_{ab} \E^{ab}$; this constant also can be set equal to zero (or to any other preferred value) without loss of generality. 

The radial functions possess the asymptotic behavior 
\begin{subequations} 
\begin{align}  
h^0_{tt}[\BH] &\sim -\frac{1}{15} r^4, & 
h^0_{rr}[\BH] &\sim -\frac{1}{5} r^4, \\ 
\gothm^0_{tt} &\sim -\frac{4}{15}(M/r)^6, &  
\gothm^0_{rr} &\sim \frac{6}{5} (M/r)^6, \\  
\gothn^0_{tt} &\sim -\frac{32}{15} (M/r), & 
\gothn^0_{rr} &\sim -\frac{6}{5} (M/r)^2   
\end{align}
\end{subequations} 
when $r \gg M$. The solution displayed here agrees with the metric of Eq.~(\ref{M0_metric}) when $M \to 0$. 

\subsubsection{Quadrupole sector: $\ell = 2$} 

The radial functions are 
\begin{subequations} 
\label{radialf_L2} 
\begin{align} 
h^2_{tt} &= h^2_{tt}[\BH] - P_2 \frac{M^7}{r^3} f^2 B_2
+ (K_2)^2 M^4\, \gothm^2_{tt} + K_2 M^4\, \gothn^2_{tt}, \\ 
h^2_{rr} &= h^2_{rr}[\BH] - P_2 \frac{M^7}{r^3} B_2
+ (K_2)^2 M^4\, \gothm^2_{rr} + K_2 M^4\, \gothn^2_{rr}, \\
h^2 &= h^2[\BH] - P_2 \frac{M^7}{r^3} D_2
+ (K_2)^2 M^4\, \gothm^2 + K_2 M^4\, \gothn^2,
\end{align} 
\end{subequations} 
where 
\begin{subequations}
\label{radialf_L2_BH} 
\begin{align} 
h^2_{tt}[\BH] &= -T_2 M^2 r^2 f^2 A_2 - \frac{2}{7} r^3 (r+3M) f^2, \\ 
h^2_{rr}[\BH] &= -T_2 M^2 r^2 A_2 - \frac{4}{7} r^4 - \frac{2}{7} M r^3, \\ 
h^2[\BH] &= -T_2 M^2 r^2 C_2 - \frac{1}{14} r^4 - \frac{8}{7} M^2 r^2 + 4M^4
\end{align} 
\end{subequations} 
is the solution for a black hole, for which $K_2 = 0$. The remaining radial functions are listed in Appendix~\ref{app:MNO}.  

The solution comes with two constants of integration, $T_2$ and $P_2$. As we shall show in Sec.~\ref{subsec:redefinitions}, $T_2$ corresponds to a freedom to redefine the tidal tensor $\E_{ab}$ by a shift proportional to $M^2 \EE_{ab}$; it can always be set to zero without loss of generality. The second constant, $P_2$, is physically meaningful: it is a (rescaled) nonlinear Love number.  

The radial functions have the asymptotic behavior 
\begin{subequations} 
\begin{align}  
h^2_{tt}[\BH] &\sim -\frac{2}{7} r^4, & 
h^2_{rr}[\BH] &\sim -\frac{4}{7} r^4, & 
h^2[\BH] &\sim -\frac{1}{14} r^4, \\ 
\gothm^2_{tt} &\sim -\frac{8}{7}(M/r)^6, &  
\gothm^2_{rr} &\sim -\frac{16}{7} (M/r)^6, & 
\gothm^2 &\sim \frac{8}{7} (M/r)^6, \\ 
\gothn^2_{tt} &\sim - \frac{8}{7} (M/r), & 
\gothn^2_{rr} &\sim -\frac{16}{7} (M/r), & 
\gothn^2 &\sim 4 (M/r)  
\end{align}
\end{subequations} 
when $r \gg M$. The solution displayed here agrees with the metric of Eq.~(\ref{M0_metric}) when $M \to 0$. A comparison with the Newtonian potential of Eq.~(\ref{Ucomplete}) reveals that $P_2$ is related to $p_2$ by 
\begin{equation} 
P_2 M^8 = 2 p_2 R^8,  
\label{P2_scaling} 
\end{equation} 
where $p_2$ is the (primitive and scalefree) nonlinear Love number introduced in Sec.~\ref{sec:constant-density}. 

\subsubsection{Hexadecapole sector: $\ell = 4$} 

The radial functions are 
\begin{subequations} 
\label{radialf_4} 
\begin{align} 
h^4_{tt} &= h^4_{tt}[\BH] - P_4 \frac{M^9}{r^5} f^2 B_4
+ (K_2)^2 M^4\, \gothm^4_{tt} + K_2 M^4\, \gothn^4_{tt}, \\ 
h^4_{rr} &= h^4_{rr}[\BH] - P_4 \frac{M^9}{r^5} B_4
+ (K_2)^2 M^4\, \gothm^4_{rr} + K_2 M^4\, \gothn^4_{rr}, \\
h^4 &= h^4[\BH] - P_4 \frac{M^9}{r^5} D_4
+ (K_2)^2 M^4\, \gothm^4 + K_2 M^4\, \gothn^4,
\end{align} 
\end{subequations} 
where 
\begin{subequations}
\label{radialf_L4_BH} 
\begin{align} 
h^4_{tt}[\BH] &= -T_4 r^4 f^2 A_4 
+ \biggl( \frac{1}{3} r^4 + \frac{3}{2} M r^3 - \frac{61}{42} M^2 r^2 \biggr)f^2, \\ 
h^4_{rr}[\BH] &= -T_4 r^4 A_4 
+ r^4 + \frac{1}{6} M r^3 - \frac{61}{42} M^2 r^2, \\ 
h^4[\BH] &= -T_4 r^4 C_4 
+ r^4 + \frac{13}{18} M r^3 - \frac{37}{7} M^2 r^2 + \frac{710}{189} M^4 
\end{align} 
\end{subequations} 
is the solution for a black hole. The remaining radial functions are listed in Appendix~\ref{app:MNO}. 

There are two constants of integration. The first, $T_4$, is associated with a freedom to redefine $\E_{abcd}$ by a shift proportional to $\EE_{abcd}$. This freedom cannot be exercised, because the metric constructed here must agree with Eq.~(\ref{M0_metric}) in the limit $M \to 0$. Comparison dictates that 
\begin{equation} 
T_4 = 0. 
\end{equation} 
The second constant, $P_4$, is a (rescaled) nonlinear Love number. 

With $T_4$ properly set to zero, we find that the radial functions have the asymptotic behavior 
\begin{subequations} 
\begin{align}  
h^4_{tt}[\BH] &\sim \frac{1}{3} r^4, & 
h^4_{rr}[\BH] &\sim r^4, & 
h^4[\BH] &\sim r^4, \\ 
\gothm^4_{tt} &\sim -2 (M/r)^6, &  
\gothm^4_{rr} &\sim \frac{2}{3} (M/r)^6, & 
\gothm^4 &\sim \frac{11}{9} (M/r)^6, \\ 
\gothn^4_{tt} &\sim -2 (M/r), & 
\gothn^4_{rr} &\sim \frac{2}{3} (M/r), & 
\gothn^4 &\sim \frac{7}{3} (M/r)  
\end{align}
\end{subequations} 
when $r \gg M$. A comparison with the Newtonian potential of Eq.~(\ref{Ucomplete}) reveals that $P_4$ is related to $p_4$ by 
\begin{equation} 
P_4 M^{10} = 2 p_4 R^{10}, 
\label{P4_scaling} 
\end{equation} 
where $p_4$ is the (primitive and scalefree) nonlinear Love number introduced in Sec.~\ref{sec:constant-density}.   

\subsection{Nonlinear tides: Quadrupole-octupole coupling} 
\label{subsec:metric_quad-oct} 

We proceed with a presentation of the general solution for the bilinear perturbations that result from the coupling between $\E_{ab}$ and $\E_{abc}$. We list separately the solutions for $\ell = 1$, $\ell = 3$, and $\ell = 5$. 

\subsubsection{Dipole sector: $\ell = 1$} 

The nonvanishing radial functions are $h^1$, and $j^1$, and the choice of gauge implies that $h^1_{tt} = f h^1$ and  $h^1_{rr} = f^{-1} h^1$. They are given by 
\begin{subequations} 
\label{radialf_L1} 
\begin{align} 
h^1 &= h^1[\BH] + S_1 \frac{M^7}{r^2 f} + K_2 K_3 M^5\, \gothm^1_h + K_2 M^5\, \gothn^1_h
+ K_3 M^5\, \gotho^1_h, \\  
j^1 &= j^1[\BH] + S_1 \frac{M^8}{r^3 f^2} + K_2 K_3 M^5\, \gothm^1_j + K_2 M^5\, \gothn^1_j
+ K_3 M^5\, \gotho^1_j, 
\end{align} 
\end{subequations} 
where 
\begin{subequations}
\label{radialf_L1_BH} 
\begin{align} 
h^1[\BH] &= -\frac{2}{35} r^5 + \frac{6}{35} M r^4 - \frac{72}{175} M^3 r^2 + \frac{64}{175} M^4 r, \\ 
j^1[\BH] &= \frac{1}{6} r^5 - \frac{16}{35} M r^4 - \frac{2}{35} M^2 r^3 + \frac{88}{175} M^3 r^2 
- \frac{32}{175} M^4 r 
\end{align} 
\end{subequations} 
is the piece of the solution that survives when the compact object is a black hole, so that $K_2 = K_3 = 0$. The remaining radial functions are listed in Appendix~\ref{app:MNO}.

The solution comes with $S_1$ as a constant of integration. This corresponds to a residual gauge freedom that consists of a translation of the spatial origin of the coordinates. This constant can always be set to zero, or to any other preferred value, without loss of generality; this freedom is analogous to the one encountered back in Sec.~\ref{subsec:N_quad-oct}, in the case of the Newtonian potential. 

The radial functions have the asymptotic behavior 
\begin{subequations} 
\begin{align}  
h^1[\BH] &\sim -\frac{2}{35} r^5, & 
j^1[\BH] &\sim \frac{1}{6} r^5, \\ 
\gothm^1_h &\sim -\frac{8}{35}(M/r)^7, &  
\gothm^1_j &\sim -\frac{8}{5} (M/r)^7, \\  
\gothn^1_h &\sim -\frac{2}{3} (r/M), & 
\gothn^1_j &\sim -\frac{4}{15}, \\ 
\gotho^1_h &\sim -\frac{208}{175} (M/r)^3, & 
\gotho^1_j &\sim -\frac{4}{5} (M/r)^2   
\end{align}
\end{subequations} 
when $r \gg M$. The solution displayed here agrees with the metric of Eq.~(\ref{M0_metric}) when $M \to 0$. 

\subsubsection{Octupole sector: $\ell = 3$} 

The radial functions are 
\begin{subequations} 
\label{radialf_L3} 
\begin{align} 
h^3_{tt} &= h^3_{tt}[\BH] - P_3 \frac{M^9}{r^4}f^2 B_3 
+ K_2 K_3 M^5\, \gothm^3_{tt} + K_2 M^5\, \gothn^3_{tt} + K_3 M^5\, \gotho^3_{tt}, \\  
h^3_{rr} &= h^3_{rr}[\BH] - P_3 \frac{M^9}{r^4} B_3 
+ K_2 K_3 M^5\, \gothm^3_{rr} + K_2 M^5\, \gothn^3_{rr} + K_3 M^5\, \gotho^3_{rr}, \\  
h^3 &= h^3[\BH] - P_3 \frac{M^9}{r^4} D_3 
+ K_2 K_3 M^5\, \gothm^3 + K_2 M^5\, \gothn^3 + K_3 M^5\, \gotho^3,   
\end{align} 
\end{subequations} 
where 
\begin{subequations}
\label{radialf_L3_BH} 
\begin{align} 
h^3_{tt}[\BH] &= -T_3 M^2 r^3 f^2 A_3 + \biggl( -\frac{2}{9} r^5 - \frac{4}{9} M r^4 - \frac{2}{9} M^2 r^3
+ \frac{53}{45} M^3 r^2 \biggr) f^2, \\ 
h^3_{rr}[\BH] &= -T_3 M^2 r^3 A_3 - \frac{1}{9} r^5 - \frac{7}{9} M r^4 + \frac{53}{45} M^3 r^2, \\ 
h^3[\BH] &= -T_3 M^2 r^3 C_3 + \frac{1}{18} r^5 - \frac{13}{18} M r^4 - \frac{1}{9} M^2 r^3
+ \frac{14}{5} M^3 r^2 - \frac{28}{15} M^5
\end{align} 
\end{subequations} 
is the solution for a black hole, for which $K_2 = K_3 = 0$. The remaining radial functions are listed in Appendix~\ref{app:MNO}. 

The solution comes with two constants of integration, $T_3$ and $P_3$. As we shall show in Sec.~\ref{subsec:redefinitions}, $T_3$ corresponds to a freedom to redefine the tidal tensor $\E_{abc}$ by a shift proportional to $M^2 \EE_{abc}$; it can be set equal to zero without loss of generality. The second constant, $P_3$, is a physically meaningful (rescaled) nonlinear Love number.  

The radial functions admit the asymptotic behavior 
\begin{subequations} 
\begin{align}  
h^3[\BH]_{tt} &\sim -\frac{2}{9} r^5, & 
h^3[\BH]_{rr} &\sim -\frac{1}{9} r^5, & 
h^3[\BH] &\sim \frac{1}{18} r^5, \\ 
\gothm^3_{tt} &\sim -\frac{8}{9}(M/r)^7, &  
\gothm^3_{rr} &\sim -\frac{4}{9} (M/r)^7, &  
\gothm^3 &\sim \frac{2}{3} (M/r)^7, \\ 
\gothn^3_{tt} &\sim -\frac{4}{9}, &  
\gothn^3_{rr} &\sim -\frac{2}{9}, &  
\gothn^3 &\sim \frac{10}{9}, \\   
\gotho^3_{tt} &\sim -\frac{4}{9} (M/r)^2, &  
\gotho^3_{rr} &\sim -\frac{2}{9} (M/r)^2, &  
\gotho^3 &\sim \frac{8}{9} (M/r)^2 
\end{align}
\end{subequations} 
when $r \gg M$. The solution displayed here agrees with the metric of Eq.~(\ref{M0_metric}) when $M \to 0$. A comparison with the Newtonian potential of Eq.~(\ref{Ucomplete}) reveals that $P_3$ is related to $p_3$ by 
\begin{equation} 
P_3 M^{10} = 2 p_3 R^{10},  
\label{P3_scaling} 
\end{equation} 
where $p_3$ is the (primitive and scalefree) nonlinear Love number introduced in Sec.~\ref{sec:constant-density}.  

\subsubsection{Triakontadipole sector: $\ell = 5$} 

The radial functions are 
\begin{subequations} 
\label{radialf_L5} 
\begin{align} 
h^5_{tt} &= h^5_{tt}[\BH] - P_5 \frac{M^{11}}{r^6}f^2 B_5 
+ K_2 K_3 M^5\, \gothm^5_{tt} + K_2 M^5\, \gothn^5_{tt} + K_3 M^5\, \gotho^5_{tt}, \\  
h^5_{rr} &= h^5_{rr}[\BH] - P_5 \frac{M^{11}}{r^6} B_5 
+ K_2 K_3 M^5\, \gothm^5_{rr} + K_2 M^5\, \gothn^5_{rr} + K_3 M^5\, \gotho^5_{rr}, \\  
h^5 &= h^5[\BH] - P_5 \frac{M^{11}}{r^6} D_5 
+ K_2 K_3 M^5\, \gothm^5 + K_2 M^5\, \gothn^5 + K_3 M^5\, \gotho^5,   
\end{align} 
\end{subequations} 
where 
\begin{subequations}
\label{radialf_L5_BH} 
\begin{align} 
h^5_{tt}[\BH] &= -T_5 r^5 f^2 A_5 + \biggl( \frac{4}{15} r^5 + \frac{19}{30} M r^4 - \frac{29}{15} M^2 r^3
+ \frac{13}{15} M^3 r^2 \biggr) f^2, \\ 
h^5_{rr}[\BH] &= -T_5 r^5 A_5 + \frac{11}{15} r^5 - \frac{23}{30} M r^4 - M^2 r^3 + \frac{13}{15} M^3 r^2, \\ 
h^5[\BH] &= -T_5 r^5 C_5 + \frac{11}{15} r^5 - \frac{14}{15} M r^4 - \frac{46}{15} M^2 r^3
+ \frac{23}{5} M^3 r^2 - \frac{38}{35} M^5
\end{align} 
\end{subequations} 
is the solution for a black hole. The remaining radial functions are listed in Appendix~\ref{app:MNO}. 

There are two constants of integration. The first, $T_5$, is associated with a freedom to shift the tidal tensor $\E_{abcde}$ by something proportional to $\EE_{abcde}$. This freedom cannot be exercised, because the metric constructed here must agree with Eq.~(\ref{M0_metric}) in the limit $M \to 0$. Comparison dictates that 
\begin{equation} 
T_5 = 0. 
\end{equation} 
The second constant, $P_5$, is a physically meaningful (rescaled) nonlinear Love number. 

With $T_5$ properly set to zero, we find that the radial functions have the asymptotic behavior 
\begin{subequations} 
\begin{align}  
h^5_{tt}[\BH] &\sim \frac{4}{15} r^5, & 
h^5_{rr}[\BH] &\sim \frac{11}{15} r^5, & 
h^5[\BH] &\sim \frac{11}{15} r^5, \\ 
\gothm^5_{tt} &\sim -\frac{4}{3}(M/r)^7, &  
\gothm^5_{rr} &\sim \frac{8}{15} (M/r)^7, & 
\gothm^5 &\sim \frac{4}{5} (M/r)^7, \\ 
\gothn^5_{tt} &\sim -\frac{2}{3}, & 
\gothn^5_{rr} &\sim \frac{4}{15}, & 
\gothn^5 &\sim \frac{23}{30}, \\ 
\gotho^5_{tt} &\sim -\frac{2}{3} (M/r)^2, & 
\gotho^5_{rr} &\sim \frac{4}{15} (M/r)^2, & 
\gotho^5 &\sim \frac{11}{15} (M/r)^2 
\end{align}
\end{subequations} 
when $r \gg M$. A comparison with the Newtonian potential of Eq.~(\ref{Ucomplete}) reveals that $P_5$ is related to $p_5$ by 
\begin{equation} 
P_5 M^{12} = 2 p_5 R^{12}, 
\label{P5_scaling} 
\end{equation} 
where $p_5$ is the (primitive and scalefree) nonlinear Love number introduced in Sec.~\ref{sec:constant-density}.   

\subsection{Dynamic tides: First time derivative} 
\label{subsec:first_time}

We return to a linearized description of the tidal deformation, but now account for the temporal variation of the tidal moments $\E_L$. We begin with a consideration of the first time derivative of these moments. 

We update the metric of Eq.~(\ref{metric_linear}) to 
\begin{subequations} 
\label{metric_first-time} 
\begin{align} 
g_{tt} &= -f + p_{tt}^{\rm ls} +p_{tt}^{\rm first}, \\ 
g_{tr} &= p_{tr}^{\rm first}, \\ 
g_{rr} &= f^{-1} + p_{rr}^{\rm ls} + p_{rr}^{\rm first}, \\ 
g_{AB} &= r^2 \Omega_{AB}(1 + p^{\rm ls} + p^{\rm first}), 
\end{align}
\end{subequations} 
where 
$p^{\rm ls}_{tt}$, $p^{\rm ls}_{rr}$, $p^{\rm ls}$ are given by Eq.~(\ref{pert_linear_ell}), and $p^{\rm first}_{tt}$, $p^{\rm first}_{rr}$, $p^{\rm first}$ are new contributions proportional to $\dot{\E}_L(t)$, in which an overdot indicates differentiation with respect to time. The form of the metric indicates that we continue to work in the Regge-Wheeler gauge. 

For each multipole of order $\ell$ we write the new perturbation as 
\begin{subequations}
\label{pert_first_ell} 
\begin{align} 
p^{\rm first}_{tt} &= -\frac{2}{(\ell-1)\ell} f^2 r^\ell \, M \dot{e}^\ell_{tt}(r)\, \dot{\E}_L \Omega^L, \\    
p^{\rm first}_{tr} &= -\frac{4}{(\ell-1)\ell(\ell+1)} \frac{r^{\ell+1}}{f}\, \dot{e}^\ell_{tr}(t)\, \dot{\E}_L \Omega^L, \\
p^{\rm first}_{rr} &= -\frac{2}{(\ell-1)\ell} r^\ell\, M \dot{e}^\ell_{rr}(r)\, \dot{\E}_L \Omega^L, \\    
p^{\rm first} &= -\frac{2}{(\ell-1)\ell} r^\ell\, M \dot{e}^\ell(r)\, \dot{\E}_L \Omega^L, 
\end{align} 
\end{subequations} 
where $\dot{e}^\ell_{tt}$, $\dot{e}^\ell_{tr}$, $\dot{e}^\ell_{rr}$, and $\dot{e}^\ell$ are radial functions that shall be determined by the field equations. Notice that we adorn these functions with an overdot, as a reminder of the fact that they are associated with time-derivative terms in the metric; in this case the overdot does not mean that the functions are themselves differentiated with respect to $t$.  

The field equations $G_{tr} = 0$ and $G_{tA} = 0$ implicate $\dot{e}^\ell_{tr}$ and none of the other radial functions. The first equation provides an algebraic expression for it, and the second equation is then redundantly satisfied. We thus obtain 
\begin{align} 
\dot{e}^\ell_{tr} &= \biggl[ 
-\frac{2(\ell+1)}{(\ell-1)\ell} \biggl( 1 - \frac{3M}{2r} \biggr) F(-\ell,-\ell;-2\ell;2M/r) 
+ \frac{\ell^2+\ell+2}{(\ell-1)\ell} F(-\ell-1,-\ell;-2\ell;2M/r) \biggr] 
\nonumber \\ & \quad \mbox{} 
- 2 \frac{\ell+1}{\ell} K_\ell \biggl( \frac{M}{r} \biggr)^{2\ell+1} \biggl[ 
\frac{2\ell}{(\ell+1)(\ell+2)} \biggl( 1 - \frac{3M}{2r} \biggr) F(\ell+1,\ell+1;2\ell+2;2M/r) 
\nonumber \\ & \quad \mbox{} 
+ \frac{\ell^2+\ell+2}{(\ell+1)(\ell+2)} F(\ell,\ell+1;2\ell+2;2M/r) \biggr]. 
\label{edot_tr} 
\end{align} 
The functions within square brackets both admit the asymptotic behavior $1 + O(M/r)$ when $r \gg M$. An explicit listing is provided in Appendix~\ref{app:edot_tr}. 

The remaining field equations implicate $\dot{e}^\ell_{tt}$, $\dot{e}^\ell_{rr}$, and $\dot{e}^\ell$, and they are all identical to the equations for a static perturbation; there are no additional source terms that might have come from the static tidal field. The solutions are therefore the same as in Eq.~(\ref{pert_linear_ell}), but with new constants of integration. We have 
\begin{subequations} 
\label{edot} 
\begin{align} 
\dot{e}^\ell_{tt} = \dot{e}^\ell_{rr} &= \dot{T}_\ell\, A_\ell + 2 \dot{K}_\ell (M/r)^{2\ell+1}\, B_\ell, \\ 
\dot{e}^\ell &= \dot{T}_\ell\, C_\ell + 2 \dot{K}_\ell (M/r)^{2\ell+1}\, D_\ell,
\end{align} 
\end{subequations} 
where $\dot{T}_\ell$ and $\dot{K}_\ell$ are the new constants. We shall see in Sec.~\ref{subsec:redefinitions} that $\dot{T}_\ell$ corresponds to a freedom to redefine the tidal moments $\E_L$ by a shift proportional to $M \dot{\E}_L$; it can be set equal to zero without loss of generality. The second constant, $\dot{K}_\ell$, is physically meaningful: it is a (rescaled) Love number associated with the temporal variation of the tidal environment. 

The factor of $M$ inserted in front of $\dot{e}^\ell_{tt}$, $\dot{e}^\ell_{rr}$, and $\dot{e}^\ell$ in Eq.~(\ref{pert_first_ell}) ensures that the corresponding components of the metric perturbation vanish in the $M \to 0$ limit. The only surviving member is the $tr$ component, and with $\dot{e}^\ell_{tr} \to 1$ implied by Eq.~(\ref{edot_tr}), we recover the results of Eq.~(\ref{M0_lin-dyn-perturbation}). 

The Newtonian potential of Eq.~(\ref{Ucomplete}) does not feature terms proportional to the first time derivative of the tidal moments.\footnote{An extension of the Newtonian model to include viscosity would reveal the existence of such terms. This observation motivates the conjecture that $\dot{K}_\ell$ is associated with dissipative effects, and might therefore vanish in the absence of dissipation. The conjecture is supported by the explicit calculation of $\dot{K}_\ell$ for a black hole in Sec.~\ref{subsec:BHLove}; the nonzero result can be associated with dissipation taking place at the event horizon.} This property is also linked to the factor of $M$ in front of $\dot{e}^\ell_{tt}$, which makes this contribution to the metric appear beyond Newtonian order in a post-Newtonian expansion. The absence of such a term in the Newtonian potential prevents us from assigning a definite scaling for the Love numbers $\dot{K}_\ell$. In an analogy with Eq.~(\ref{K_vs_k}), we may surmise that they are related to primitive and scalefree Love numbers $\dot{k}_\ell$ by 
\begin{equation} 
\dot{K}_\ell M^{2\ell+1} = \dot{k}_\ell R^{2\ell+1}. 
\label{Kdot_vs_kdot} 
\end{equation} 
It is possible, however, that $\dot{K}_\ell$ comes with additional factors of $M/R$.  

\subsection{Linear and dynamic tides: Second time derivative} 
\label{subsec:second_time} 

Next we incorporate second time-derivative terms in the metric perturbation. As before, we continue to work in the Regge-Wheeler gauge. The metric is expressed as 
\begin{subequations} 
\label{metric_second-time} 
\begin{align} 
g_{tt} &= -f + p_{tt}^{\rm ls} +p_{tt}^{\rm first} + p_{tt}^{\rm second}, \\ 
g_{tr} &= p_{tr}^{\rm first} + p_{tr}^{\rm second}, \\ 
g_{rr} &= f^{-1} + p_{rr}^{\rm ls} + p_{rr}^{\rm first} + p_{rr}^{\rm second}, \\ 
g_{AB} &= r^2 \Omega_{AB}(1 + p^{\rm ls} + p^{\rm first} + p^{\rm second}), 
\end{align}
\end{subequations} 
and we write 
\begin{subequations}
\label{pert_second_ell} 
\begin{align} 
p^{\rm second}_{tt} &= -\frac{2}{(\ell-1)\ell} f^2 r^\ell \, M^2 \ddot{e}^\ell_{tt}(r)\, \ddot{\E}_L \Omega^L, \\    
p^{\rm second}_{tr} &= -\frac{4}{(\ell-1)\ell(\ell+1)} \frac{r^{\ell+1}}{f}\, M \ddot{e}^\ell_{tr}(t)\, \ddot{\E}_L \Omega^L, \\
p^{\rm second}_{rr} &= -\frac{2}{(\ell-1)\ell} r^\ell\, M^2 \ddot{e}^\ell_{rr}(r)\, \ddot{\E}_L \Omega^L, \\    
p^{\rm second} &= -\frac{2}{(\ell-1)\ell} r^\ell\, M^2 \ddot{e}^\ell (r)\, \ddot{\E}_L \Omega^L 
\end{align} 
\end{subequations} 
for a tidal multipole of order $\ell$, where $\ddot{e}^\ell_{tt}$, $\ddot{e}^\ell_{tr}$, $\ddot{e}^\ell_{rr}$, and $\ddot{e}^\ell$ are new radial functions. 

When we insert the metric within the Einstein field equations, we again find that the equations for $\ddot{e}^\ell_{tr}$ are decoupled from the equations for the remaining variables. The function is obtained algebraically from $G_{tr} = 0$, and $G_{tA} = 0$ is then redundant. The solution takes the form of 
\begin{equation} 
\ddot{e}^\ell_{tr} = \dot{T}_\ell\, \gothf_\ell + \dot{K}_\ell\, \gothg_\ell, 
\label{eddot_tr} 
\end{equation} 
where the exact form of $\gothf_\ell$ and $\gothg_\ell$ for $\ell = \{2, 3, 4, 5\}$ is given in Appendix~\ref{app:ABCDFG}. 

The remaining field equations implicate the other radial functions. These equations are no longer homogeneous: there are now source terms that come from the static and first time-derivative pieces of the metric perturbation. The tracefree part of the angular equations, $G_\stf{AB} = 0$, implies immediately that 
\begin{equation} 
\ddot{e}^\ell_{tt} = \ddot{e}^\ell_{rr}. 
\end{equation} 
The equations $G_{rr} = 0$ and $G_{rA} = 0$ allow us to express $d \ddot{e}^\ell/dr$ in terms of $\ddot{e}_{rr}^\ell$, and from this we also obtain $d^2 \ddot{e}^\ell/dr^2$. Making the substitutions in a linear combination of $G_{tt} = 0$ and $G_{rr} = 0$, we obtain the second-order differential equation 
\begin{equation} 
r^2 f \frac{d^2 \ddot{e}^\ell_{rr}}{dr^2} + \bigl[ 2(\ell+1) r - 2(2\ell-1) M \bigr] \frac{d \ddot{e}^\ell_{rr}}{dr} 
- 2(\ell-2) \ell \frac{M}{r} \ddot{e}^\ell_{rr} + S^\ell = 0, 
\label{diff_eq_ddot} 
\end{equation} 
with $S^\ell$ a source term constructed from the static and first time-derivative perturbations. The equation can be cast in the form of Eq.~(\ref{diff_eq}) by introducing $h^\ell_{rr} = r^\ell \ddot{e}^\ell_{rr}$ as a new dependent variable. Once a solution is found, $G_{tt} = 0$ provides an algebraic equation for $\ddot{e}^\ell$, and all remaining field equations are redundantly satisfied. At this stage the problem is completely solved. 

The integration techniques described in Sec.~\ref{subsec:integration} can be recycled  to find the general solution to Eq.~(\ref{diff_eq_ddot}). The details are slightly different, because here we use $A_\ell$ and $r^{-(2\ell+1)} B_\ell$ as a basis of linearly independent solutions to the homogeneous equation, instead of the basis $r^\ell A_\ell$ and $r^{-(\ell+1)} B_\ell$ associated with Eq.~(\ref{diff_eq}). Apart from this slight change, however, the calculations are virtually identical, and there is no need to describe the details here. 

The general solution to Eq.~(\ref{diff_eq_ddot}) takes the form of 
\begin{subequations} 
\label{eddot} 
\begin{align} 
\ddot{e}^\ell_{tt} = \ddot{e}^\ell_{rr} &= \ddot{T}_\ell\, A_\ell + 2 \ddot{K}_\ell (M/r)^{2\ell+1}\, B_\ell
+ \gotha_\ell + K_\ell\, \gothb_{\ell}, \\ 
\ddot{e}^\ell &= \ddot{T}_\ell\, C_\ell + 2 \ddot{K}_\ell (M/r)^{2\ell+1}\, D_\ell 
+ \gothc_\ell + K_\ell\, \gothd_{\ell}, 
\end{align} 
\end{subequations} 
where $\ddot{T}_\ell$ and $\ddot{K}_\ell$ are two constants of integration, and where the functions $\gotha_\ell$, $\gothb_\ell$, $\gothc_\ell$, $\gothd_\ell$ for $\ell = \{2, 3, 4, 5\}$ are displayed in Appendix~\ref{app:ABCDFG}. As we shall show in Sec.~\ref{subsec:redefinitions}, $\ddot{T}_\ell$ corresponds to a freedom to redefine the tidal moments $\E_L$ by a shift proportional to $M^2 \ddot{\E}_L$; it can be set equal to zero without loss of generality. The second constant, $\ddot{K}_\ell$, is another (rescaled) Love number associated with the time variation of the tidal environment.   

For $\ell = 2$ the radial functions possess the asymptotic behavior 
\begin{equation} 
\gotha_2 \sim \frac{11}{42} (r/M)^2, \qquad 
\gothb_2 \sim (M/r)^3, \qquad 
\gothc_2 \sim \frac{2}{21} (r/M)^2, \qquad 
\gothd_2 \sim -(M/r)^3 
\end{equation} 
and 
\begin{equation} 
\gothf_2\sim 1, \qquad 
\gothg_2 \sim - 3 (M/r)^5 
\end{equation} 
when $r \gg M$. Inserting these within Eqs.~(\ref{eddot_tr}) and (\ref{eddot}), and making the substitution in Eq.~(\ref{pert_second_ell}), we recover the perturbation of Eq.~(\ref{M0_lin-dyn-perturbation}) in the limit $M \to 0$.  A comparison with the Newtonian potential of Eq.~(\ref{Ucomplete}) reveals that the time-variation Love number scales as 
\begin{equation} 
\ddot{K}_2 M^8 = \ddot{k}_2 R^8, 
\label{Kddot2_scaling}  
\end{equation} 
where $\ddot{k}_2$ is the primitive and scalefree Love number. 

For $\ell = 3$ the radial functions behave as 
\begin{equation} 
\gotha_3 \sim \frac{1}{6} (r/M)^2, \qquad 
\gothb_3 \sim \frac{1}{3} (M/r)^5, \qquad 
\gothc_3 \sim \frac{1}{15} (r/M)^2, \qquad 
\gothd_3 \sim -\frac{1}{3} (M/r)^5 
\end{equation} 
and 
\begin{equation} 
\gothf_3\sim 1, \qquad 
\gothg_3 \sim - \frac{8}{3} (M/r)^7 
\end{equation} 
when $r \gg M$. Inserting these within Eqs.~(\ref{eddot_tr}) and (\ref{eddot}), and making the substitution in Eq.~(\ref{pert_second_ell}), we recover the perturbation of Eq.~(\ref{M0_lin-dyn-perturbation}) in the limit $M \to 0$. Comparison with the Newtonian potential of Eq.~(\ref{Ucomplete}) reveals that the Love number scales as 
\begin{equation} 
\ddot{K}_3 M^{10} = \ddot{k}_3 R^{10}, 
\label{Kddot3_scaling}  
\end{equation} 
where $\ddot{k}_3$ is the primitive and scalefree Love number.  

For $\ell = 4$ we have 
\begin{equation} 
\gotha_4 \sim \frac{13}{110} (r/M)^2, \qquad 
\gothb_4 \sim \frac{1}{7} (M/r)^7, \qquad 
\gothc_4 \sim \frac{17}{330} (r/M)^2, \qquad 
\gothd_4 \sim -\frac{4}{21} (M/r)^7 
\end{equation} 
and 
\begin{equation} 
\gothf_4\sim 1, \qquad 
\gothg_4 \sim - \frac{5}{2} (M/r)^9
\end{equation} 
when $r \gg M$. Comparison with the Newtonian potential of Eq.~(\ref{Ucomplete}), generalized so as to include a $\ell=4$ contribution, reveals that the Love number scales as 
\begin{equation} 
\ddot{K}_4 M^{12} = \ddot{k}_4 R^{12}, 
\label{Kddot4_scaling}  
\end{equation} 
where $\ddot{k}_4$ is the primitive and scalefree Love number.  

Finally, for $\ell = 5$ the asymptotic behavior is 
\begin{equation} 
\gotha_5 \sim \frac{7}{78} (r/M)^2, \qquad 
\gothb_5 \sim \frac{1}{15} (M/r)^9, \qquad 
\gothc_5 \sim \frac{23}{546} (r/M)^2, \qquad 
\gothd_5 \sim -\frac{2}{15} (M/r)^9 
\end{equation} 
and 
\begin{equation} 
\gothf_5\sim 1, \qquad 
\gothg_5 \sim - \frac{12}{5} (M/r)^{11}
\end{equation} 
when $r \gg M$. Comparison with the Newtonian potential of Eq.~(\ref{Ucomplete}), generalized so as to include a $\ell=5$ contribution, reveals that 
\begin{equation} 
\ddot{K}_5 M^{14} = \ddot{k}_5 R^{14}, 
\label{Kddot5_scaling}  
\end{equation} 
where $\ddot{k}_5$ is the primitive and scalefree Love number.  

\subsection{Summary: Complete metric} 
\label{subsec:metric_summary} 

Collecting results, we have found that the metric of a tidally deformed compact body can be expressed as  
\begin{subequations} 
\label{metric_complete} 
\begin{align} 
g_{tt} &= -f + p_{tt}^{\rm ls}+ p_{tt}^{\rm qq} + p_{tt}^{\rm qo} +p_{tt}^{\rm first} + p_{tt}^{\rm second}, \\ 
g_{tr} &= p_{tr}^{\rm first} + p_{tr}^{\rm second}, \\ 
g_{tt} &= f^{-1} + p_{rr}^{\rm ls}+ p_{rr}^{\rm qq} + p_{rr}^{\rm qo} +p_{rr}^{\rm first} + p_{rr}^{\rm second},\\ 
g_{rA} &= r\, p_{rA}^{\rm qo}, \\ 
g_{AB} &= r^2 \Omega_{AB}(1 + p^{\rm ls}+ p^{\rm qq} + p^{\rm qo} +p^{\rm first} + p^{\rm second}),
\end{align}
\end{subequations} 
where $f := 1-2M/r$. The linear and static perturbations are 
\begin{subequations} 
\label{pert_linear} 
\begin{align} 
p^{\rm ls}_{tt} &= -f^2 r^2 \Bigl[ A_2 + 2 K_2 (M/r)^{5} B_2 \Bigr]\, \E^{\sf q} 
- \frac{1}{3} f^2 r^3 \Bigl[ A_3 + 2 K_3 (M/r)^{7} B_3 \Bigr]\, \E^{\sf o} 
\nonumber \\ & \quad \mbox{} 
- \frac{1}{6} f^2 r^4 \Bigl[ A_4 + 2 K_4 (M/r)^{9} B_4 \Bigr]\, \E^{\sf h} 
- \frac{1}{10} f^2 r^5 \Bigl[ A_5 + 2 K_5 (M/r)^{11} B_5 \Bigr]\, \E^{\sf t} + O(r^6), \\ 
p^{\rm ls}_{rr} &= -r^2 \Bigl[ A_2 + 2 K_2 (M/r)^{5} B_2 \Bigr]\, \E^{\sf q} 
- \frac{1}{3} r^3 \Bigl[ A_3 + 2 K_3 (M/r)^{7} B_3 \Bigr]\, \E^{\sf o} 
\nonumber \\ & \quad \mbox{} 
- \frac{1}{6} r^4 \Bigl[ A_4 + 2 K_4 (M/r)^{9} B_4 \Bigr]\, \E^{\sf h} 
- \frac{1}{10} r^5 \Bigl[ A_5 + 2 K_5 (M/r)^{11} B_5 \Bigr]\, \E^{\sf t} + O(r^6), \\ 
p^{\rm ls}  &= -r^2 \Bigl[ C_2 + 2 K_2 (M/r)^{5} D_2 \Bigr]\, \E^{\sf q} 
- \frac{1}{3} r^3 \Bigl[ C_3 + 2 K_3 (M/r)^{7} D_3 \Bigr]\, \E^{\sf o} 
\nonumber \\ & \quad \mbox{} 
- \frac{1}{6} r^4 \Bigl[ C_4 + 2 K_4 (M/r)^{9} D_4 \Bigr]\, \E^{\sf h} 
- \frac{1}{10} r^5 \Bigl[ C_5 + 2 K_5 (M/r)^{11} D_5 \Bigr]\, \E^{\sf t} + O(r^6),  
\end{align} 
\end{subequations} 
with $A_\ell$, $B_\ell$, $C_\ell$, and $D_\ell$ given explicitly by Eqs.~(\ref{ABCD2})--(\ref{ABCD5}). 

The quadrupole-quadrupole perturbations are those displayed in Eq.~(\ref{pert_qq}), 
\begin{subequations} 
\label{pert_qq_repeat} 
\begin{align} 
p_{tt}^{\rm qq} &= 
h^0_{tt}\, \EE^{\sf m} + h^2_{tt}\, \EE^{\sf q} + h^4_{tt}\, \EE^{\sf h} + O(r^6), \\ 
p_{rr}^{\rm qq} &= 
h^0_{rr}\, \EE^{\sf m} + h^2_{rr}\, \EE^{\sf q} + h^4_{rr}\, \EE^{\sf h} + O(r^6), \\ 
p^{\rm qq} &= h^2\, \EE^{\sf q} + h^4\, \EE^{\sf h} + O(r^6), 
\end{align} 
\end{subequations} 
and the quadrupole-octupole perturbations are copied from Eq.~(\ref{pert_qo}), 
\begin{subequations} 
\label{pert_qo_repeat} 
\begin{align} 
p_{tt}^{\rm qo} &= 
h^1_{tt}\, \EE^{\sf d} + h^3_{tt}\, \EE^{\sf o} + h^5_{tt}\, \EE^{\sf t} + O(r^6), \\ 
p_{rr}^{\rm qo} &= 
h^1_{rr}\, \EE^{\sf d} + h^3_{rr}\, \EE^{\sf o} + h^5_{rr}\, \EE^{\sf t}+ O(r^6), \\ 
p_{rA}^{\rm qo} &= j^1\, \partial_A \EE^{\sf d} + O(r^6), \\ 
p^{\rm qo} &= 
h^1\, \EE^{\sf d} + h^3\, \EE^{\sf o} + h^5\, \EE^{\sf t} + O(r^6). 
\end{align} 
\end{subequations} 
The radial functions that accompany these perturbations are listed in Secs.~\ref{subsec:metric_quad-quad} and \ref{subsec:metric_quad-oct}, as well as Appendix~\ref{app:MNO}. 
 
The perturbations associated with the first time derivative of the tidal moments were obtained in Sec.~\ref{subsec:first_time}. We have 
\begin{subequations} 
\label{pert_first} 
\begin{align} 
p^{\rm first}_{tt} &= -f^2 r^2 \bigl[ \dot{T}_2\, A_2 + 2 \dot{K}_2 (M/r)^5\, B_2 \bigr] M \dot{\E}^{\sf q}
- \frac{1}{3} f^2 r^3 \bigl[ \dot{T}_3\, A_3 + 2 \dot{K}_3 (M/r)^7\, B_3 \bigr] M \dot{\E}^{\sf o}
\nonumber \\ & \quad \mbox{} 
- \frac{1}{6} f^2 r^4 \bigl[ \dot{T}_4\, A_4 + 2 \dot{K}_4 (M/r)^9\, B_4 \bigr] M \dot{\E}^{\sf h}
- \frac{1}{10} f^2 r^5 \bigl[ \dot{T}_5\, A_5 + 2 \dot{K}_5 (M/r)^{11}\, B_5 \bigr] M \dot{\E}^{\sf t}
+ O(r^6), \\ 
p^{\rm first}_{tr} &= -\frac{2}{3} f^{-1} r^3\, \dot{e}^2_{tr}\, \dot{\E}^{\sf q} 
- \frac{1}{6} f^{-1} r^4\, \dot{e}^3_{tr}\, \dot{\E}^{\sf o} 
- \frac{1}{15} f^{-1} r^5\, \dot{e}^4_{tr}\, \dot{\E}^{\sf h}
- \frac{1}{30} f^{-1} r^6\, \dot{e}^5_{tr}\, \dot{\E}^{\sf t}
+ O(r^7), \\  
p^{\rm first}_{rr} &= -r^2 \bigl[ \dot{T}_2\, A_2 + 2 \dot{K}_2 (M/r)^5\, B_2 \bigr] M \dot{\E}^{\sf q}
- \frac{1}{3} r^3 \bigl[ \dot{T}_3\, A_3 + 2 \dot{K}_3 (M/r)^7\, B_3 \bigr] M \dot{\E}^{\sf o}
\nonumber \\ & \quad \mbox{} 
- \frac{1}{6} r^4 \bigl[ \dot{T}_4\, A_4 + 2 \dot{K}_4 (M/r)^9\, B_4 \bigr] M \dot{\E}^{\sf h}
- \frac{1}{10} r^5 \bigl[ \dot{T}_5\, A_5 + 2 \dot{K}_5 (M/r)^{11}\, B_5 \bigr] M \dot{\E}^{\sf t}
+ O(r^6), \\ 
p^{\rm first} &= -r^2 \bigl[ \dot{T}_2\, C_2 + 2 \dot{K}_2 (M/r)^5\, D_2 \bigr] M \dot{\E}^{\sf q}
- \frac{1}{3} r^3 \bigl[ \dot{T}_3\, C_3 + 2 \dot{K}_3 (M/r)^7\, D_3 \bigr] M \dot{\E}^{\sf o}
\nonumber \\ & \quad \mbox{} 
- \frac{1}{6} r^4 \bigl[ \dot{T}_4\, C_4 + 2 \dot{K}_4 (M/r)^9\, D_4 \bigr] M \dot{\E}^{\sf h}
- \frac{1}{10} r^5 \bigl[ \dot{T}_5\, C_5 + 2 \dot{K}_5 (M/r)^{11}\, D_5 \bigr] M \dot{\E}^{\sf t}
+ O(r^6), 
\end{align} 
\end{subequations} 
with the radial functions $\dot{e}^\ell_{tr}$ listed explicitly in Eq.~(\ref{edot_tr_explicit}). The perturbations associated with the second time derivative of the tidal moments were calculated in Sec.~\ref{subsec:second_time}. We found that 
\begin{subequations} 
\label{pert_second} 
\begin{align} 
p^{\rm second}_{tt} &= -f^2 r^2 \bigl[ \ddot{T}_2\, A_2 + 2 \ddot{K}_2 (M/r)^5\, B_2 
+ \gotha_2 + K_2\, \gothb_2\bigr] M^2 \ddot{\E}^{\sf q}
- \frac{1}{3} f^2 r^3 \bigl[ \ddot{T}_3\, A_3 + 2 \ddot{K}_3 (M/r)^7\, B_3 
+ \gotha_3 + K_3\, \gothb_3 \bigr] M^2 \ddot{\E}^{\sf o}   
\nonumber \\ & \quad \mbox{} 
- \frac{1}{6} f^2 r^4 \bigl[ \ddot{T}_4\, A_4 + 2 \ddot{K}_4 (M/r)^9\, B_4 
+ \gotha_4 + K_4\, \gothb_4 \bigr] M^2 \ddot{\E}^{\sf h}
- \frac{1}{10} f^2 r^5 \bigl[ \ddot{T}_5\, A_5 + 2 \ddot{K}_5 (M/r)^{11}\, B_5 
+ \gotha_5 + K_5\, \gothb_5 \bigr] M^2 \ddot{\E}^{\sf t} 
\nonumber \\ & \quad \mbox{} 
+ O(r^6), \\ 
p^{\rm second}_{tr} &= 
-\frac{2}{3} f^{-1} r^3 \bigl( \dot{T}_2\, \gothf_2 + \dot{K}_2\, \gothg_2 \bigr) M \ddot{\E}^{\sf q} 
- \frac{1}{6} f^{-1} r^4 \bigl( \dot{T}_3\, \gothf_3 + \dot{K}_3\, \gothg_3 \bigr) M \ddot{\E}^{\sf o} 
\nonumber \\ & \quad \mbox{} 
- \frac{1}{15} f^{-1} r^5 \bigl( \dot{T}_4\, \gothf_4 + \dot{K}_4\, \gothg_4 \bigr) M \ddot{\E}^{\sf h} 
- \frac{1}{30} f^{-1} r^6 \bigl( \dot{T}_5\, \gothf_5 + \dot{K}_5\, \gothg_5 \bigr) M \ddot{\E}^{\sf t} 
+ O(r^7), \\ 
p^{\rm second}_{rr} &= -r^2 \bigl[ \ddot{T}_2\, A_2 + 2 \ddot{K}_2 (M/r)^5\, B_2 
+ \gotha_2 + K_2\, \gothb_2\bigr] M^2 \ddot{\E}^{\sf q}
- \frac{1}{3} r^3 \bigl[ \ddot{T}_3\, A_3 + 2 \ddot{K}_3 (M/r)^7\, B_3 
+ \gotha_3 + K_3\, \gothb_3 \bigr] M^2 \ddot{\E}^{\sf o} 
\nonumber \\ & \quad \mbox{} 
- \frac{1}{6} r^4 \bigl[ \ddot{T}_4\, A_4 + 2 \ddot{K}_4 (M/r)^9\, B_4 
+ \gotha_4 + K_4\, \gothb_4 \bigr] M^2 \ddot{\E}^{\sf h}
- \frac{1}{10} r^5 \bigl[ \ddot{T}_5\, A_5 + 2 \ddot{K}_5 (M/r)^{11}\, B_5 
+ \gotha_5 + K_5\, \gothb_5 \bigr] M^2 \ddot{\E}^{\sf t} 
\nonumber \\ & \quad \mbox{} 
+ O(r^6), \\  
p^{\rm second} &= -r^2 \bigl[ \ddot{T}_2\, C_2 + 2 \ddot{K}_2 (M/r)^5\, D_2 
+ \gothc_2 + K_2\, \gothd_2\bigr] M^2 \ddot{\E}^{\sf q}
- \frac{1}{3} r^3 \bigl[ \ddot{T}_3\, C_3 + 2 \ddot{K}_3 (M/r)^7\, D_3 
+ \gothc_3 + K_3\, \gothd_3 \bigr] M^2 \ddot{\E}^{\sf o} 
\nonumber \\ & \quad \mbox{} 
- \frac{1}{6} r^4 \bigl[ \ddot{T}_4\, C_4 + 2 \ddot{K}_4 (M/r)^9\, D_4 
+ \gothc_4 + K_4\, \gothd_4 \bigr] M^2 \ddot{\E}^{\sf h}
- \frac{1}{10} r^5 \bigl[ \ddot{T}_5\, C_5 + 2 \ddot{K}_5 (M/r)^{11}\, D_5 
+ \gothc_5 + K_5\, \gothd_5 \bigr] M^2 \ddot{\E}^{\sf t} 
\nonumber \\ & \quad \mbox{} 
+ O(r^6),
\end{align} 
\end{subequations} 
and the radial functions are displayed in Eqs.~(\ref{ABCD_L2}), (\ref{ABCD_L3}), (\ref{ABCD_L4}), and (\ref{ABCD_L5}).  

The metric depends on a number of integration constants. Two of these, $S_0$ and $S_1$, reflect a residual gauge freedom, and they can be set equal to zero (or to any other preferred value) without loss of generality. The constants $T_\ell$, $\dot{T}_\ell$, and $\ddot{T}_\ell$ are associated with the freedom to redefine the mass and tidal multipole moments, as described below in Sec.~\ref{subsec:redefinitions}. These can all be set to zero without loss of generality; two members of this set, $T_4$ and $T_5$, {\it must} be made to vanish, in order for the tidal moments to respect their definition when $M \to 0$. And finally, the constants $K_\ell$, $P_\ell$, $\dot{K}_\ell$, and $\ddot{K}_\ell$ are physically meaningful: they are rescaled versions of the primitive and scalefree Love numbers first introduced in Sec.~\ref{sec:constant-density}. The metric displayed here fulfills (indeed exceeds) the objectives specified back in Sec.~\ref{subsec:task}.  

\subsection{Love numbers for a black hole} 
\label{subsec:BHLove} 

The Love numbers $K_\ell$, $P_\ell$, $\dot{K}_\ell$, and $\ddot{K}_\ell$ provide a measure of the body's response to an applied tidal field. For a material body, the Love numbers are determined by joining the exterior metric of Eq.~(\ref{metric_complete}) to an interior metric at the body's deformed surface. For a black hole, the exterior metric extends to the interior, and it must be smooth across the deformed event horizon when presented in well-behaved coordinates. We shall show that this requirement implies that 
\begin{equation} 
K_\ell[\BH] = 0, \qquad P_\ell[\BH] = 0, \qquad \ddot{K}_\ell[\BH] = 0 
\label{BH_Love1} 
\end{equation} 
but that 
\begin{equation} 
\dot{K}_\ell[\BH] = -\frac{2^{2\ell}(\ell-2)!(\ell-1)!\ell! (\ell+2)!}{(2\ell-1)! (2\ell+1)!}. 
\label{BH_Love2} 
\end{equation} 
The expressions for $K_\ell[\BH]$ and $\dot{K}_\ell[\BH]$ can be deduced for all values of $\ell$. The nonlinear Love numbers $P_\ell$ are defined for $\ell = \{ 2, 3, 4, 5 \}$ only, and they all vanish for a black hole. Our vanishing result for $\ddot{K}_\ell[\BH]$ is a conjecture that was verified explicitly for $\ell = \{ 2, 3, 4, 5 \}$ only. It seems highly plausible that the pattern extends to all values of $\ell$, but we are not able to provide a proof of the conjecture. As we shall explain below, the zero value for $\ddot{K}_\ell[\BH]$ requires setting $\dot{T}_\ell = 0$.  

Our well-behaved coordinates shall be the Eddington-Finkelstein system $(v, r, \theta^A)$, where $v$ is the advanced-time coordinate of Eq.~(\ref{advanced-time}), 
\begin{equation} 
v = t + M \Delta, \qquad 
\Delta := r/M + 2\ln(r/2M) + 2\ln f. 
\end{equation} 
In these coordinates the components of the metric perturbation are
\begin{equation} 
p^{\rm EF}_{vv} = p_{tt}, \qquad 
p^{\rm EF}_{vr} = p_{tr} - f^{-1} p_{tt}, \qquad 
p^{\rm EF}_{rr} = p_{rr} - 2 f^{-1} p_{tr} + f^{-2} p_{tt};
\label{EF_perturbations} 
\end{equation} 
$p_{rA}$, $p_{AB}$ stay unchanged (recall that $p_{tA} = 0$ in our selected gauge). All components of the perturbation must be bounded at $r=2M$, and be free of terms proportional to $\ln f$.  

We first examine the linear and static perturbation of Sec.~\ref{subsec:metric_linear-static}. It is immediate that smoothness at the horizon requires $K_\ell = 0$. Next we examine the bilinear perturbations of Secs.~\ref{subsec:metric_quad-quad} and \ref{subsec:metric_quad-oct}; inspection of the results reveals that $P_\ell = 0$ is required to ensure smoothness across the horizon. 

To reach a correct conclusion about the time-derivative contributions of Secs.~\ref{subsec:first_time} and \ref{subsec:second_time}, we must ensure that the tidal moments $\E_L$ are properly expressed in terms of $v$ instead of $t$. Taking advantage of the time-derivative expansion, we write 
\begin{equation} 
\E_L(t) = \E_L(v - M\Delta) = \E_L(v) - M \Delta\, \dot{\E}_L(v) 
+ \frac{1}{2} M^2 \Delta^2\, \ddot{\E}_L(v) + \cdots, 
\end{equation} 
and expand its derivatives in a similar manner. The entirety of the linearized piece of the $vv$ component (say) of the perturbation, for each multipole $\ell$, is then 
\begin{align} 
p^{{\rm linear}}_{vv} &= -\frac{2}{(\ell-1)\ell} f^2 r^\ell \biggl[ 
e_{tt}\, \E_L(t)  + M\dot{e}_{tt}\, \dot{\E}_L(t) + M^2\ddot{e}_{tt}\, \ddot{\E}_L(t) \biggr] \Omega^L 
\nonumber \\ 
&= -\frac{2}{(\ell-1)\ell} f^2 r^\ell \biggl[ 
e_{tt}\, \E_L(v)  + M \bigl( \dot{e}_{tt} - \Delta e_{tt} \bigr)\, \dot{\E}_L(v) 
+ M^2 \bigl( \ddot{e}_{tt} - \Delta \dot{e}_{tt} + \tfrac{1}{2} \Delta^2 e_{tt} \bigr)\, \ddot{\E}_L(v) \biggr] \Omega^L. 
\end{align} 
The new radial functions in front of $\E_L(v)$, $\dot{\E}_L(v)$, and $\ddot{\E}_L(v)$ must be bounded at $r=2M$ and free of any occurrence of $\ln f$. Similar considerations apply to the other components of the metric perturbation. 

Let us first examine $\dot{e}_{tt} - \Delta e_{tt}$, with $\dot{e}_{tt}$ given by Eq.~(\ref{edot}), and $e_{tt}$ by Eq.~(\ref{pert_linear_ell}), which reduces to $e_{tt} = A_\ell$ when $K_\ell = 0$. We observe that $B_\ell$ contains terms proportional to $\ln f$, and we must choose $\dot{K}_\ell$ in such a way that these are cancelled out by the corresponding presence of $\ln f$ in $\Delta$. To achieve this we recall the definition of $A_\ell$ and $B_\ell$ from 
Eq.~(\ref{ABCD}), and make use of the identities of Eqs.~(\ref{id5}) and (\ref{id6}), 
\begin{subequations}
\begin{align}  
(2M/r)^{-\ell} F(-\ell,-\ell+2;-2\ell;2M/r) &= \frac{(\ell-2)!(\ell-1)!}{2(2\ell-1)!} (y+1)^2 P''_\ell(y), \\
(2M/r)^{\ell + 1} F(\ell+1, \ell+3; 2\ell+2; 2M/r) &= \frac{2(2\ell+1)!}{\ell!(\ell+2)!} (y+1)^2 Q''_\ell(y), 
\end{align} 
\end{subequations} 
where $y := r/M-1$, $P_\ell$ and $Q_\ell$ are Legendre functions, and a prime indicates differentiation with respect to the argument. These representations are helpful, because it is known that 
\begin{equation} 
Q_\ell(y) = \frac{1}{2} P_\ell(y)\, \ln\frac{y+1}{y-1} + \mbox{polynomial in $y$}. 
\end{equation} 
The logarithm is recognized as $-\ln f$, and differentiating twice with respect to $y$, we can collect all terms in $B_\ell$ that are proportional to $\ln f$. We obtain 
\begin{equation} 
B_\ell = -\frac{2(2\ell-1)!(2\ell+1)!}{(\ell-2)!(\ell-1)!\ell!(\ell+2)!} (2M/r)^{-(2\ell+1)} A_\ell\, \ln f  
+ \mbox{other terms}, 
\end{equation} 
where the ``other terms'' do not involve $\ln f$. Making the substitution in $\dot{e}_{tt} - \Delta e_{tt}$, we find that the assignment of Eq.~(\ref{BH_Love2}) eliminates all occurrences of $\ln f$ in the radial function. 

We proceed with a similar examination of $\ddot{e}_{tt} - \Delta \dot{e}_{tt} + \tfrac{1}{2} \Delta^2 e_{tt}$, with $\ddot{e}_{tt}$ given by Eq.~(\ref{eddot}) and the listings of Eqs.~(\ref{ABCD_L2})--(\ref{ABCD_L5}) for $\ell = \{2, 3, 4, 5\}$. Here the examination relies on the explicit expressions for these individual cases, and we verify that all $\ln f$ and $(\ln f)^2$ terms get eliminated when 
\begin{equation} 
\ddot{K}_\ell = \dot{T}_\ell\, \dot{K}_\ell, 
\label{Kddot_BH} 
\end{equation} 
with $\dot{K}_\ell$ given by Eq.~(\ref{BH_Love2}). Setting $\dot{T}_\ell$ to zero as we are free to do, we obtain the statement of Eq.~(\ref{BH_Love1}). We were not able to devise a formal proof that Eq.~(\ref{Kddot_BH}) applies to values of $\ell$ beyond $\ell = 5$. This would require a careful consideration of the ``other terms'' in $B_\ell$, and how they manifest themselves in $\gotha_\ell$ and $\gothb_\ell$. Based on the evidence provided, however, we can feel confident that the conjecture is true.    

All other components of the metric perturbation are examined in a similar way, and we find that the assignments of Eqs.~(\ref{BH_Love1}) and (\ref{BH_Love2}) produce a metric that is bounded and smooth across the black-hole horizon. A careful calculation is required, because the radial functions possess individual terms that diverge when $r = 2M$; these must be shown to cancel out when the perturbations of Eq.~(\ref{EF_perturbations}) are computed. We find that they do indeed.  

\subsection{Redefinitions} 
\label{subsec:redefinitions} 

The body's mass $M$ can be redefined according to 
\begin{equation} 
M = \bar{M}\bigl(1 + \lambda_0 \bar{M}^4 \E_{ab} \E^{ab} \bigr), 
\end{equation} 
where $\bar{M}$ is the new mass parameter, and $\lambda_0$ is a dimensionless constant. Making the substitution in the metric, it is easy to see that the shift in mass is equivalent to a resetting of $T_0$, according to 
\begin{equation} 
T_0 \to \bar{T}_0 = T_0 + 2\lambda_0. 
\end{equation}   
As was stated previously, it follows that the value of $T_0$ can be altered at will by a redefinition of the mass. The constant, in particular, can be set equal to zero, or to any other preferred value. 

Similarly, the quadrupole tidal moment $\E_{ab}$ can be redefined according to 
\begin{equation} 
\E_{ab} = \bar{\E}_{ab} + \lambda_2 M^2 \bar{\E}_{c\langle a} \bar{\E}^c_{\ b\rangle}, 
\end{equation} 
where $\bar{\E}_{ab}$ is the new moment. Making the substitution in the metric reveals that this shift is equivalent to a resetting of $T_2$ and $P_2$, 
\begin{equation} 
T_2 \to \bar{T}_2 = T_2 + \lambda_2, \qquad 
P_2 \to \bar{P}_2 = P_2 + 2 \lambda_2 K_2. 
\end{equation} 
The shift can be exploited to set $T_2 = 0$; once this is done, the Love number $P_2$ becomes invariant. 

The octupole tidal moment $\E_{abc}$ can be redefined according to 
\begin{equation} 
\E_{abc} = \bar{\E}_{abc} + \lambda_3 M^2 \bar{\E}_{d\langle ab} \bar{\E}^d_{\ c\rangle}, 
\end{equation} 
where $\bar{\E}_{abc}$ is the new moment. This shift is equivalent to the resetting 
\begin{equation} 
T_3 \to \bar{T}_3 = T_3 + \frac{1}{3} \lambda_3, \qquad 
P_3 \to \bar{P}_3 = P_3 + \frac{2}{3} \lambda_3 K_3. 
\end{equation} 
It allows us to set $T_3 = 0$; once this is done, the Love number $P_3$ becomes invariant. 

The tidal moments can be further redefined according to 
\begin{equation} 
\E_L = \bar{\E}_L + \dot{\lambda}_\ell\, M \dot{\bar{\E}}_L, 
\end{equation} 
where $\dot{\lambda}_\ell$ is a dimensionless constant, and $\bar{\E}_L$ are the new moments. The redefinition produces the changes 
\begin{equation} 
\dot{T}_\ell \to \dot{\bar{T}}_\ell = \dot{T}_\ell + \dot{\lambda}_\ell, \qquad 
\dot{K}_\ell \to \dot{\bar{K}}_\ell = \dot{K}_\ell + \dot{\lambda}_\ell K_\ell 
\end{equation} 
in the integration constants. It allows us to set $\dot{T}_\ell$ to zero, or to any other preferred value. With specific values selected for $\dot{T}_\ell$, the Love numbers $\dot{K}_\ell$ become invariant.   

Finally, the tidal moments can be redefined according to 
\begin{equation} 
\E_L = \bar{\E}_L + \ddot{\lambda}_\ell\, M^2 \ddot{\bar{\E}}_L.  
\end{equation} 
This produces the changes 
\begin{equation} 
\ddot{T}_\ell \to \ddot{\bar{T}}_\ell = \ddot{T}_\ell + \ddot{\lambda}_\ell, \qquad 
\ddot{K}_\ell \to \ddot{\bar{K}}_\ell = \ddot{K}_\ell + \ddot{\lambda}_\ell K_\ell 
\end{equation} 
in the integration constants. We may therefore set $\ddot{T}_\ell$ to zero, and once this freedom has been exercised, the Love numbers $\ddot{K}_\ell$ become invariant.   

\section{Compact body as a member of a post-Newtonian system} 
\label{sec:PN} 

\subsection{Two descriptions of the gravitational field and an overlapping domain of validity} 
\label{subsec:zones} 

The metric of a tidally deformed compact body was constructed in Sec.~\ref{sec:GR}, and expressed in terms of tidal moments $\E_L(t)$. As long as considerations are limited to a small neighborhood around the body, the tidal moments continue to be freely-specifiable functions of time; they are not determined by the Einstein field equations. To give a concrete appearance to the tidal moments, one must insert the metric within a larger spacetime that includes the external matter responsible for the tidal environment. In this section we do this in a post-Newtonian setting. While we describe the compact body's {\it strong internal gravity} in full general relativity, we approximate the {\it weak mutual gravity} between body and external matter as a post-Newtonian expansion truncated beyond the first post-Newtonian order ($1\pn$). The external matter might consist of a companion star in a bound orbit, and in this case the compact body would be a member of a binary system. The companion could be weakly self-gravitating (in which case it could also be described by a post-Newtonian approximation to general relativity), or it could be strongly self-gravitating (in which case it would be treated in full general relativity, just as our compact body). Alternatively, there might be a larger number of companions, the orbits might be unbounded, or the external matter might be a diffuse distribution. We shall not make an explicit choice; what matters is that the gravitational interaction between compact body and external matter is a weak one that is adequately described by post-Newtonian theory. 

\begin{figure} 
\includegraphics[width=0.9\linewidth]{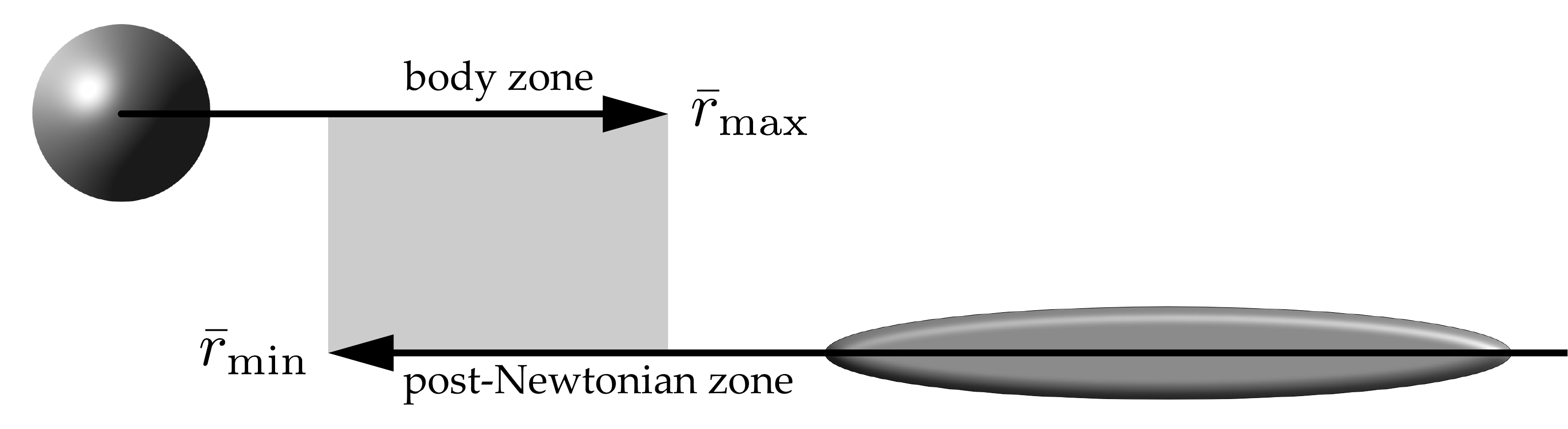}
\caption{A compact body (circle, left) is in a weak gravitational interaction with a distribution of external matter (ellipse, right). The body zone extends from the body out to $\bar{r} = \bar{r}_{\rm max}$; gravity is strong deep in the body zone, and it becomes progressively weaker as the boundary is approached. The post-Newtonian zone extends inward to $\bar{r} = \bar{r}_{\rm min}$; gravity is weak everywhere in this zone. The overlap zone (shaded box) corresponds to the interval $\bar{r}_{\rm min} < \bar{r} < \bar{r}_{\rm max}$.}  
\label{fig1} 
\end{figure} 

The construction detailed below requires the definition of three zones (see Fig.~\ref{fig1}). For these definitions we need the following scaling quantities: the mass $M$ of the compact body, its radius $R$, the mass scale $M'$ associated with the external matter, and the length scale $a$ associated with the distance between compact body and external matter. We assume that $a \gg M, R$. The {\it body zone} is the compact body's immediate neighborhood, our domain of consideration in Sec.~\ref{sec:GR}. If $\bar{r}$ is the spatial distance to the body, then the body zone is defined by $\bar{r} < \bar{r}_{\rm max}$, where $\bar{r}_{\rm max}$ is much larger than $M$ but much smaller than $a$. The gravity of the compact body is strong deep inside the body zone, and it is described in full general relativity; it becomes progressively weaker as $\bar{r}$ approaches $\bar{r}_{\rm max}$. The {\it post-Newtonian zone} is a region of spacetime in which gravity is weak everywhere and adequately described by a post-Newtonian expansion of the metric. The zone surrounds the compact body and contains the external matter, but it excises the compact body when its gravity is about to become too strong. In terms of $\bar{r}$, the post-Newtonian zone is defined by $\bar{r}_{\rm min} < \bar{r} < \lambda$, where $\bar{r}_{\rm min}$ is also much larger than $M$ and much smaller than $a$, and $\lambda \gg a$ is a characteristic wavelength for the gravitational radiation emitted by the system. We choose $\bar{r}_{\rm min}$ to be smaller than $\bar{r}_{\rm max}$, and the region $\bar{r}_{\rm min} < \bar{r} < \bar{r}_{\rm max}$ is an {\it overlap zone} in which both descriptions of the gravitational field (fully relativistic and post-Newtonian) are valid. 

The relativistic and post-Newtonian descriptions of the spacetime are both partial ones. The fully relativistic description is limited to the body zone, which excludes the external matter; as a result it leaves the tidal moments unspecified. On the other hand, the post-Newtonian description is limited to its own zone, which excludes the compact body; it also comes with unknowns, including the body's multipole structure and its motion as viewed in the post-Newtonian spacetime. To determine all these, the relativistic and post-Newtonian metrics are matched in the overlap zone. Post-Newtonian information determines the tidal moments of the body zone, and relativistic information determines the multipole structure of the post-Newtonian zone. The body's equations of motion are contained in the transformation between the coordinates employed in the post-Newtonian zone --- harmonic coordinates attached to the system's barycenter --- and those utilized in the body zone --- Schwarzschild coordinates extended with the Regge-Wheeler gauge. After matching, the partial descriptions become whole, and we obtain a complete metric for the spacetime. (Our considerations here exclude the wave zone $\bar{r} > \lambda$. It is known, however, how to extend the post-Newtonian metric to this region of the spacetime.)  

The following subsections are devoted to an implementation of this program. We rely on a working knowledge of post-Newtonian gravity (acquired, for example, by studying Chapters 8 and 9 of Poisson and Will \cite{poisson-will:14}), including the theory of transformations between coordinate systems (their Sec.~8.3). The methods exploited here originate from Damour, Soffel, and Xu \cite{damour-soffel-xu:91, damour-soffel-xu:92, damour-soffel-xu:93}, and they were implemented for weakly self-gravitating bodies of arbitrary structure by Racine and Flanagan \cite{racine-flanagan:05}. They were implemented for black holes in Refs.~\cite{taylor-poisson:08, poisson-corrigan:18}. This is a generalization to any type of compact object, either material body or black hole. 

\subsection{Post-Newtonian metric in the barycentric frame} 
\label{subsec:bary} 

We consider the post-Newtonian zone as defined in the preceding subsection, and install a metric $g_{\alpha\beta}$ in this region of spacetime. The metric is given the standard post-Newtonian form 
\begin{subequations} 
\label{PNmetric} 
\begin{align} 
g_{tt} &= -1 + 2U + 2(\Psi - U^2) + 2\pn, \\ 
g_{ta} &= -4 U_a + 2\pn, \\ 
g_{ab} &= (1 + 2U) \delta_{ab} + 2\pn, 
\end{align} 
\end{subequations} 
in terms of a Newtonian potential $U$, a post-Newtonian potential $\Psi$, and a vector potential $U_a$. The post-Newtonian potential is conventionally (and conveniently) expressed as 
\begin{equation} 
\Psi = \psi + \frac{1}{2} \partial_{tt} X,  
\label{Psi_def}
\end{equation} 
in terms of another potential $\psi$ and the superpotential $X$. The metric is presented in harmonic coordinates $(t, x^a)$, and the potentials are required to satisfy 
\begin{equation} 
\partial_t U + \partial_a U^a = 0.
\label{harmonic} 
\end{equation}      
In a region of spacetime empty of matter (away from the external matter), the Einstein field equations reduce to 
\begin{equation} 
\nabla^2 U = 0, \qquad 
\nabla^2 U_a = 0, \qquad 
\nabla^2 \psi = 0, \qquad 
\nabla^2 X = 2U, 
\label{PN_field-eqns} 
\end{equation} 
where $\nabla^2$ is the Laplacian operator in a flat, three-dimensional space.  

The coordinates $(t, x^a)$ are distinct from those employed in Sec.~\ref{sec:GR}. We imagine them to be anchored at the center of mass of the entire gravitating system\footnote{The post-Newtonian center of mass is defined, for example, in Sec.~9.3.6 of Poisson and Will. The precise definition is not important for our purposes here, and it plays no specific role in the subsequent developments. What is important is that the center of mass is a well-defined quantity, and that its motion is necessarily uniform when radiative effects can be neglected. The center of mass, therefore, can be chosen as the origin of an inertial frame of reference.}, which comprises the compact body and the external matter. This center of mass (which is distinct from the body's own center of mass) is designated as ``barycenter''; the frame of reference attached to the harmonic coordinates will be called the ``barycentric frame''. 

The matching procedure to be carried out below justifies the following post-Newtonian description of the compact body: It is skeletonized object with a multipole structure, situated at position $\bm{x} = \bm{z}(t)$ in the harmonic chart. The body comes with a mass $M$, a mass dipole moment 
\begin{equation} 
Q_a(t) = Q_a[\pn], 
\end{equation} 
and a mass quadrupole moment 
\begin{equation} 
Q_{ab}(t) = Q_{ab}[\n] + Q_{ab}[\pn].  
\end{equation} 
Here the quadrupole moment, a symmetric and tracefree tensor, is split into Newtonian and post-Newtonian contributions, and we indicate that the dipole moment is of post-Newtonian origin. (The need for a dipole moment will arise during the matching procedure.) The skeletonized object also possesses higher-order multipole moments, but to keep things simple we choose to truncate the multipole structure beyond the quadrupole order. This shall suffice for our purposes. 

To proceed we introduce the vector $\bm{s} := \bm{x} - \bm{z}(t)$, the separation from the compact object in the harmonic coordinates, its length $s := |\bm{s}|$ calculated with the Euclidean metric, and the unit vector $\bm{n} := \bm{s}/s$. The vector $\bm{v}(t) := d\bm{z}/dt$ is the body's velocity in the harmonic chart, and $\bm{a}(t) := d\bm{v}/dt$ is its acceleration. All tensor indices below are lowered and raised with the Euclidean metric. 

In accordance with the skeletonized description of the compact body introduced above, the potentials are written as 
\begin{subequations} 
\label{PNpotentials1} 
\begin{align} 
U &= \frac{M}{s} + \frac{1}{2} Q^{ab}[\n]\, \partial_{ab} \frac{1}{s} + U_{\rm ext},\\ 
U^a &= \biggl( \frac{M}{s} + \frac{1}{2} Q^{bc}[\n]\, \partial_{bc}\frac{1}{s} \biggr) v^a 
- \frac{1}{2} \dot{Q}^{ab}[\n]\, \partial_b \frac{1}{s} + U^a_{\rm ext}, \\ 
\psi &= \frac{M\mu}{s} - Q^a[\pn]\, \partial_a \frac{1}{s} 
+ \frac{1}{2} Q^{ab}[\pn] \partial_{ab}\, \frac{1}{s} + \psi_{\rm ext}, \\ 
X &= M s + \frac{1}{2} Q^{ab}[\n]\, \partial_{ab} s + X_{\rm ext}, 
\end{align} 
\end{subequations} 
where $U_{\rm ext}$, $U^a_{\rm ext}$, $\psi_{\rm ext}$, and $X_{\rm ext}$ are the potentials created by the external matter. In $U$ we recognize the mass and quadrupole terms, and these are multiplied by $v^a$ to create a piece of the vector potential; the remaining piece is required by the gauge condition of Eq.~(\ref{harmonic}) --- an overdot indicates differentiation with respect to time. We also have mass and quadrupole terms in $\psi$, with $\mu(t)$ a post-Newtonian adjustment to the mass parameter, but in addition we include a mass dipole, which will be required in the matching procedure. Finally, $X$ also comes with mass and quadrupole terms, to ensure that it properly satisfies $\nabla^2 X = 2U$. 

The potentials depend on a number of unknowns, namely the multipole moments and the position $\bm{z}(t)$ of the skeletonized object. As was pointed out previously, the multipole moments will be determined by matching, and the procedure will also deliver equations of motion for the position vector. The potentials are singular in the limit $s \to 0$. The singularity is only apparent, however, because $s=0$ is excluded from the post-Newtonian zone, in which $s \gg M$. 

More explicit expressions for the potentials are obtained by making use of the identities 
\begin{subequations} 
\label{iden1} 
\begin{align} 
\partial_a s &= n_a, \\ 
\partial_a n_b &= -\frac{1}{s} (n_a n_b - \delta_{ab}), \\ 
\partial_t s &= -v_n, \\ 
\partial_t n_a &= -\frac{1}{s} \bigl( v_a - v_n n_a \bigr), \\
\partial_t v_n &= a_n - \frac{1}{s} \bigl( v^2 - v_n^2 \bigr), \\
\partial_{tt\langle ab \rangle} s &= -\frac{v_n^2}{s^3} \Bigl( 15 n_a n_b - \delta_{ab} \Bigr) 
+ \frac{v^2}{s^3} \Bigl( 3n_a n_b - \tfrac{1}{3} \delta_{ab} \Bigr) 
+ \frac{6v_n}{s^3} \Bigl( v_a n_b + n_a v_b \Bigr) 
\nonumber \\ & \quad \mbox{} 
- \frac{2}{s^3}\, v_a v_b 
+ \frac{1}{s^2} \Bigl( a_a n_b + n_a a_b \Bigr) 
- \frac{a_n}{s^2} \Bigl( 3 n_a n_b - \tfrac{1}{3} \delta_{ab} \Bigr) 
\end{align} 
\end{subequations} 
and 
\begin{equation} 
\partial_a \frac{1}{s} = -\frac{1}{s^2} n_a, \qquad 
\partial_{ab} \frac{1}{s} = \frac{3}{s^3} n_\stf{ab}, \qquad 
\partial_{abc} \frac{1}{s} = -\frac{15}{s^4} n_\stf{abc}, 
\label{iden2} 
\end{equation} 
where $v_n := \bm{v} \cdot \bm{n}$, $a_n := \bm{a} \cdot \bm{n}$, and $v^2 := \bm{v} \cdot \bm{v}$. Making the substitutions, we find that 
\begin{subequations} 
\label{PNpotentials2} 
\begin{align} 
U &= \frac{1}{s^3} \biggl( \frac{3}{2} Q^{ab}[\n]\, n_a n_b \biggr) + \frac{M}{s} + U_{\rm ext}, \\ 
U^a &= \frac{1}{s^3} \biggl( \frac{3}{2} Q^{bc}[\n]\, v^a n_b n_c \biggr) 
+ \frac{1}{s^2} \biggl( \frac{1}{2} \dot{Q}^{ab}[\n]\, n_b \biggr) 
+ \frac{Mv^a}{s} + U^a_{\rm ext}, \\ 
\Psi &= \frac{1}{s^3} \biggl( \frac{3}{2} Q^{ab}[\pn]\, n_a n_b + 
\frac{3}{4} (v^2 - 5 v_n^2) Q^{ab}[\n]\, n_a n_b + 3 v_n Q^{ab}[\n]\, v_a n_b  
- \frac{1}{2} Q^{ab}[\n]\, v_a v_b \biggr) 
\nonumber \\ & \quad \mbox{} 
+ \frac{1}{s^2} \biggl( Q^a[\pn]\, n_a + \frac{1}{2} Q^{ab}[\n]\, a_a n_b - \frac{3}{4} a_n Q^{ab}[\n]\, n_a n_b 
+ \dot{Q}^{ab}[\n]\, v_a n_b - \frac{3}{2} v_n \dot{Q}^{ab}[\n]\, n_a n_b \biggr) 
\nonumber \\ & \quad \mbox{} 
+ \frac{1}{s} \biggl[ M \Bigl( \mu + \frac{1}{2} v^2 - \frac{1}{2} v_n^2 \Bigr) 
- \frac{1}{4} \ddot{Q}^{ab}[\n]\, n_a n_b \biggr] 
- \frac{1}{2} M a_n + \Psi_{\rm ext}, 
\end{align} 
\end{subequations} 
where $\Psi_{\rm ext} = \psi_{\rm ext} + \frac{1}{2} \partial_{tt} X_{\rm ext}$. 

\subsection{Transformation to the body frame}  
\label{subsec:transf} 

We wish to subject the metric of Eq.~(\ref{PNmetric}) to a coordinate transformation that brings the compact body to the spatial origin of the new coordinate system, denoted $(\bar{t}, \bar{x}^a)$. The reference frame attached to these new coordinates will be called the ``body frame''. The transformation from barycentric frame to body frame is required to preserve the post-Newtonian form of the metric, and the new metric will therefore be written in terms of transformed potentials $\bar{U}$, $\bar{U}^a$, and $\bar{\Psi}$. The details of this transformation are worked out in Sec.~8.3 of Poisson and Will \cite{poisson-will:14}, adapted from the original treatment by Racine and Flanagan \cite{racine-flanagan:05}. We shall provide very few of these details here; an extended discussion can be found in 
Refs.~\cite{taylor-poisson:08, poisson-corrigan:18}. 

The transformation $t = t(\bar{t}, \bm{\bar{x}})$, $x^a = x^a(\bar{t},\bm{\bar{x}})$ is characterized by a number of freely-specifiable functions of the new coordinates. The most important ones are a time-dilation function $A(\bar{t})$, a translation vector $\bm{z}(\bar{t})$, a frame-precession tensor $\epsilon_{abc} R^c(\bar{t})$, and a gauge function $\beta(\bar{t},\bm{\bar{x}})$ that will allow us to go from the harmonic gauge of the barycentric metric to the Regge-Wheeler gauge of the body-frame metric. As the notation suggests, the vector $\bm{z}$ is the assigned position of the compact body in the harmonic chart, except that it is re-expressed in terms of $\bar{t}$ instead of the original $t$; it comes with the associated velocity $\bm{v} = d\bm{z}/d\bar{t}$ and acceleration $\bm{a} = d\bm{v}/d\bar{t}$ vectors. The acceleration is decomposed as
\begin{equation} 
\bm{a} = \bm{a}[\n] + \bm{a}[\pn], 
\end{equation} 
into Newtonian and post-Newtonian pieces. 

As detailed in Sec.~8.3 of Poisson and Will, the transformation begins with the definition of ``hatted potentials'' $\hat{U}$, $\hat{U}^a$, and $\hat{\Psi}$, related to the barycentric potentials $U$, $U^a$, and $\Psi$ by equations of the form 
\begin{equation} 
\hat{U}(\bar{t}, \bar{\bm{x}}) := U\big( t = \bar{t}, \bm{x} = \bm{\bar{x}} + \bm{z}(\bar{t}) \bigr). 
\label{Uhat_def}
\end{equation} 
In words, a hatted potential is the original potential evaluated at the new time $\bar{t}$ and expressed in terms of the new spatial variables $\bar{x}^a$. For example, 
\begin{equation} 
\hat{U} = \frac{1}{\bar{r}^3} \biggl( \frac{3}{2} Q^{ab}[\n]\, \bar{n}_a \bar{n}_b \biggr) 
+ \frac{M}{\bar{r}} + \hat{U}_{\rm ext}, 
\end{equation} 
where $\bar{r} := |\bm{\bar{x}}|$ and $\bm{\bar{n}} := \bm{\bar{x}}/\bar{r}$; the potential depends on $\bar{t}$ through the quadrupole moment $Q^{ab}[\n]$, and through the external term $\hat{U}_{\rm ext}$. The hatted potentials are then boosted to the body frame, and this gives rise to the transformed potentials $\bar{U}$, $\bar{U}^a$, and $\bar{\Psi}$.

Throughout our developments we take advantage of the fact that all external potentials vary over a length scale $a$ that is very long compared with $\bar{r}$ --- we recall that $a$ is a characteristic distance to the external matter. This observation allows us to express each potential as a Taylor expansion about $\bm{\bar{x}} = \bm{0}$. For example, we write 
\begin{equation} 
\hat{U}_{\rm ext} = \hat{U}_{\rm ext} (\bar{t},\bm{0}) + \bar{r}\, g_a(\bar{t})\, \bar{n}^a  
+ \frac{1}{2} \bar{r}^2\, \partial_{\bar{a}\bar{b}} \hat{U}_{\rm ext}(\bar{t},\bm{0})\, \bar{n}^a \bar{n}^b  
+ \frac{1}{6} \bar{r}^3\, \partial_{\bar{a}\bar{b}\bar{c}} \hat{U}_{\rm ext}(\bar{t},\bm{0})\, \bar{n}^a \bar{n}^b \bar{n}^c 
+ O(\bar{r}^4),   
\label{Uext_taylor} 
\end{equation} 
where 
\begin{equation} 
g_a := \partial_{\bar{a}} \hat{U}_{\rm ext}(\bar{t},\bm{0})  
\label{g_def} 
\end{equation} 
is the external Newtonian field. The gauge function is expressed in a similar manner,\footnote{The $\beta$ of Eq.~(\ref{beta_taylor}) is not equal to the $\beta$ of Box~8.2 of Poisson and Will. The transformation described in Box~8.2 is designed to make the new coordinates $(\bar{t},\bm{\bar{x}})$ harmonic. We make no such requirement here, and indeed, the new coordinates will not be harmonic.}  
\begin{equation} 
\beta = \frac{1}{\bar{r}}\, \mbox{}_{-1} \beta_{ab}(\bar{t})\, \bar{n}^a \bar{n}^b 
+ \mbox{}_0 \beta(\bar{t}) 
+ \bar{r}\, \mbox{}_1 \beta_a(\bar{t})\, \bar{n}^a  
+ \frac{1}{2} \bar{r}^2\, \mbox{}_2 \beta_{ab}(\bar{t})\, \bar{n}^a \bar{n}^b 
+ \frac{1}{6} \bar{r}^3\, \mbox{}_3 \beta_{abc}(\bar{t})\, \bar{n}^a \bar{n}^b \bar{n}^c 
+ O(\bar{r}^4).  
\label{beta_taylor} 
\end{equation} 
We observe that $\beta$ is allowed to possess a term that is singular in the formal limit $\bar{r} \to 0$; its specific form will be justified during the matching procedure. 

We plough through the details of Secs.~8.3.2 and 8.3.3 of Poisson and Will and find, after a long calculation, that the transformed potentials are given by 
\begin{subequations} 
\label{PNpotentials_transf} 
\begin{align} 
\bar{U} &= \frac{1}{\bar{r}^3} \Bigl( \mbox{}_{-3} \bar{U}_{ab}\, \bar{n}^a \bar{n}^b \Bigr)  
+ \frac{M}{\bar{r}} 
+ \mbox{}_0 \bar{U} 
+ \bar{r}\, \mbox{}_1 \bar{U}_a\, \bar{n}^a
+ \frac{1}{2} \bar{r}^2\, \mbox{}_2 \bar{U}_{ab}\, \bar{n}^a \bar{n}^b 
+ \frac{1}{6} \bar{r}^3\, \mbox{}_3 \bar{U}_{abc}\, \bar{n}^a \bar{n}^b \bar{n}^c 
+ O(\bar{r}^4), \\ 
\bar{U}_a &= \frac{1}{\bar{r}^2} \Bigl( \mbox{}_{-2} \bar{U}^{\flat}_{ab}\, \bar{n}^b 
+ \mbox{}_{-2} \bar{U}^\sharp_{bc}\, \bar{n}_a \bar{n}^b \bar{n}^c \Bigr)   
+ \mbox{}_0 \bar{U}_a   
+ \bar{r}\, \mbox{}_1 \bar{U}_{ab}\, \bar{n}^b 
+ \frac{1}{2} \bar{r}^2\, \mbox{}_2 \bar{U}_{abc}\, \bar{n}^b \bar{n}^c 
+ O(\bar{r}^3), \\
\bar{\Psi} &= \frac{1}{\bar{r}^3} \Bigl( \mbox{}_{-3} \bar{\Psi}_{ab}\, \bar{n}^a \bar{n}^b \Bigr)  
+ \frac{1}{\bar{r}^2} \Bigl( \mbox{}_{-2} \bar{\Psi}_{a}\, \bar{n}^a \Bigr) 
+ \frac{1}{\bar{r}} \Bigl( \mbox{}_{-1} \bar{\Psi}_{ab}\, \bar{n}^a \bar{n}^b 
+ \mbox{}_{-1} \bar{\Psi} \Bigr)  
\nonumber \\ & \quad \mbox{} 
+ \mbox{}_0 \bar{\Psi} 
+ \bar{r}\, \mbox{}_1 \bar{\Psi}_a\, \bar{n}^a
+ \frac{1}{2} \bar{r}^2\, \mbox{}_2 \bar{\Psi}_{ab}\, \bar{n}^a \bar{n}^b 
+ \frac{1}{6} \bar{r}^3\, \mbox{}_3 \bar{\Psi}_{abc}\, \bar{n}^a \bar{n}^b \bar{n}^c 
+ O(\bar{r}^4),
\end{align}
\end{subequations} 
where the expansion coefficients are given by\footnote{The vector $F^a = H^a - Av^a$, defined by Eq.~(8.59a) of Poisson and Will, is set equal to zero in all following expressions. The justification for this comes from the matching procedure: the vector would produce terms in the potentials that are too singular in the formal limit $\bar{r} \to 0$ to be accounted for by the body-zone potentials.} 
\begin{subequations} 
\begin{align} 
\mbox{}_{-3} \bar{U}_{ab} &= \frac{3}{2} Q_{ab}[\n], \\ 
\mbox{}_0 \bar{U} &= \hat{U}_{\rm ext} - \dot{A} + \frac{1}{2} v^2, \\ 
\mbox{}_1 \bar{U}_a &= g_a - a_a[\n], \\ 
\mbox{}_2 \bar{U}_{ab} &= \partial_{\bar{a}\bar{b}} \hat{U}_{\rm ext}, \\  
\mbox{}_3 \bar{U}_{abc} &= \partial_{\bar{a}\bar{b}\bar{c}} \hat{U}_{\rm ext}, 
\end{align}  
\end{subequations} 
\begin{subequations} 
\begin{align} 
4\, \mbox{}_{-2} \bar{U}^{\flat}_{ab} &= 2 \dot{Q}_{ab}[\n] + 2\, \mbox{}_{-1} \beta_{ab}, \\ 
4\, \mbox{}_{-2} \bar{U}^\sharp_{ab} &= -3\, \mbox{}_{-1} \beta_{ab}, \\ 
4\, \mbox{}_0 \bar{U}_a &= 4 \hat{P}_a + (\dot{A} - v^2) v_a - A a_a[\n]  
+ \epsilon_{abc} v^b R^c + \mbox{}_1 \beta_a, \\ 
4\, \mbox{}_1 \bar{U}_{ab} &= 4 \partial_{\bar{b}} \hat{P}_a 
+ \frac{3}{2} v_a a_b[\n] + \frac{1}{2} a_a[\n] v_b 
+ (\ddot{A} - 2 v_c a^c[\n]) \delta_{ab} 
- \epsilon_{abc} \dot{R}^c 
+ \mbox{}_2 \beta_{ab}, \\ 
4\, \mbox{}_2 \bar{U}_{abc} &= 4 \partial_{\bar{b}\bar{c}} \hat{P}_a  
+ 2\delta_{a(b} \dot{a}_{c)}[\n] - \delta_{bc} \dot{a}_a[\n] 
+ \mbox{}_3 \beta_{abc}, 
\end{align} 
\end{subequations} 
with $\hat{P}_a := \hat{U}^{\rm ext}_a - v_a \hat{U}^{\rm ext}$, and
\begin{subequations} 
\begin{align} 
\mbox{}_{-3} \bar{\Psi}_{ab} &= \frac{3}{2} Q_{ab}[\pn] + \frac{9}{2} (\dot{A}-v^2) Q_{ab}[\n] 
+ \frac{3}{2} v_{\langle a} Q_{b \rangle c}[\n] v^c + \frac{3}{2} A \dot{Q}_{ab}[\n] 
- 3 \epsilon_{jc(a} R^c Q^j_{\ b)}[\n], \\ 
\mbox{}_{-2} \bar{\Psi}_{a} &= Q_a[\pn] + 2Q_{ab}[\n] a^b[\n] - \dot{Q}_{ab}[\n] v^b, \\ 
\mbox{}_{-1} \bar{\Psi}_{ab} &= -\frac{1}{4} \ddot{Q}_{ab} - \mbox{}_{-1} \dot{\beta}_{ab}, \\ 
\mbox{}_{-1} \bar{\Psi} &= M(\mu + \dot{A} - 2v^2), \\
\mbox{}_0 \bar{\Psi} &= \hat{P} + \frac{1}{2} \dot{A}^2 + \frac{1}{4} v^4 
+ A \bigl( v_a a^a[\n] + \partial_{\bar{t}} \hat{U}_{\rm ext} \bigr) 
- \mbox{}_0 \dot{\beta}, 
\label{Psi0} \\ 
\mbox{}_1 \bar{\Psi}_a &= \partial_{\bar{a}} \hat{P} + v_a \partial_{\bar{t}} \hat{U}_{\rm ext} 
- a_a[\pn] + \Bigl( \dot{A} 
- \frac{1}{2} v^2 \Bigr) \bigl( a_a[\n] - g_a \bigr) - \frac{1}{2} v_a v^b g_b 
\nonumber \\ & \quad \mbox{}
- \Bigl( \ddot{A} - \frac{3}{2} v_b a^b[\n] \Bigr) v_a 
+ A \dot{g}_a + \epsilon_{abc} R^b g^c - \epsilon_{abc} v^b \dot{R}^c 
- \mbox{}_1 \dot{\beta}_a, \\ 
\mbox{}_2 \bar{\Psi}_{ab} &= \partial_{\bar{a}\bar{b}} \hat{P} 
- 2 \Bigl( \dot{A} - \frac{1}{2} v^2 \Bigr) \partial_{\bar{a}\bar{b}} \hat{U}_{\rm ext} 
- v^c v_{(a} \partial_{\bar{b})\bar{c}} \hat{U}_{\rm ext} 
+ 2 v_{(a} \bigl( \dot{g}_{b)} - \dot{a}_{b)}[\n] \bigr) 
+ \delta_{ab}\, v_c \dot{a}^c[\n] 
\nonumber \\ & \quad \mbox{}
- 2 a_{(a}[\n] \Bigl( g_{b)} - \frac{1}{2} a_{b)}[\n] \Bigr) + \delta_{ab}\, a_c[\n] g^c 
+ A \partial_{\bar{t}\bar{a}\bar{b}} \hat{U}_{\rm ext} 
- 2\epsilon_{jc(a} R^c \partial^{\bar{\jmath}}_{\ \bar{b})} \hat{U}_{\rm ext} 
- \mbox{}_2 \dot{\beta}_{ab}, \\ 
\mbox{}_3 \bar{\Psi}_{abc} &= 
\partial_{\bar{a}\bar{b}\bar{c}} \hat{P} 
+ 3 v_{(a} \partial_{\bar{t}\bar{b}\bar{c})} \hat{U}_{\rm ext} 
- 3 \Bigl( \dot{A} - \frac{1}{2} v^2 \Bigr) \partial_{\bar{a}\bar{b}\bar{c}} \hat{U}_{\rm ext} 
- \frac{3}{2} v_{(a} \partial_{\bar{b}\bar{c})\bar{d}} \hat{U}_{\rm ext}\, v^d 
\nonumber \\ & \quad \mbox{}
- 6 a_{(a}[\n] \partial_{\bar{b}\bar{c})} \hat{U}_{\rm ext} 
+ 3 \delta_{(ab} \partial_{\bar{c})\bar{d}} \hat{U}_{\rm ext}\, a^d[\n]  
+ A \partial_{\bar{t}\bar{a}\bar{b}\bar{c}} \hat{U}_{\rm ext} 
- 3\epsilon_{jd(a} R^d \partial^{\bar{\jmath}}_{\ \bar{b}\bar{c})} \hat{U}_{\rm ext} 
- \mbox{}_3 \dot{\beta}_{abc},
\end{align} 
\end{subequations} 
with $\hat{P} := \hat{\Psi}_{\rm ext} - 4 v_a \hat{U}^a_{\rm ext} + 2 v^2 \hat{U}_{\rm ext}$. It is understood
that all external potentials are evaluated at $\bar{\bm{x}} = \bm{0}$ after differentiation. 

\subsection{Body-zone potentials} 
\label{subsec:body-zone}

The metric of a tidally deformed compact body was obtained in Sec.~\ref{sec:GR}. The calculation was performed in full general relativity, and it is a simple matter to extract potentials $\bar{U}$, $\bar{U}^a$, and $\bar{\Psi}$ from this metric; we shall refer to these as body-zone potentials. Because the metric is presented in the rest frame of the compact body, we adopt the overbar notation, which reminds us that the potentials refer to the body frame introduced in Sec.~\ref{subsec:transf}. 

The first step in the extraction of the potentials is to introduce a system of Lorentzian-like coordinates $(\bar{t}, \bar{x}^a)$ to replace the spherical-like coordinates $(t, r, \theta^A)$ employed in Sec.~\ref{sec:GR}. The relation is 
\begin{equation} 
\bar{t} = t, \qquad 
\bar{x}^a = \bar{r}\, \Omega^a(\theta^A), \qquad 
\bar{r} = r - M, 
\label{S_to_L} 
\end{equation} 
where $\Omega^a = [\sin\theta\cos\phi, \sin\theta\sin\phi, \cos\theta]$ was previously introduced in Eq.~(\ref{Omega_def}). It is perhaps helpful to note that the $t$ of Sec.~\ref{sec:GR} is distinct from the barycentric time introduced in Sec.~\ref{subsec:bary}; the risk for confusion should be minimal. The shift of $r$ by $M$ makes the $(\bar{t},\bar{x}^a)$ coordinates harmonic in the unperturbed Schwarzschild spacetime, and this ensures that the Schwarzschild metric admits a post-Newtonian expansion compatible with Eq.~(\ref{PNmetric}). The coordinates are no longer harmonic in the perturbed spacetime, but the post-Newtonian expansion of the perturbed metric continues to be compatible with Eq.~(\ref{PNmetric}). This was one of the guiding criteria for the adoption of the Regge-Wheeler gauge in Sec.~\ref{sec:GR}, together with its nondiagonal extension for $\ell = 1$. 

With a metric originally expressed as 
\begin{subequations} 
\begin{align} 
g_{tt} &= -f + p_{tt}, \\ 
g_{tr} &= p_{tr}, \\ 
g_{rr} &= f^{-1} + p_{rr}, \\ 
g_{rA} &= r\, p_{rA}, \\ 
g_{AB} &= r^2 \Omega_{AB} (1 + p), 
\end{align} 
\end{subequations} 
where $f := 1 - 2M/r$, the transformation of Eq.~(\ref{S_to_L}) produces 
\begin{subequations}
\begin{align} 
g_{\bar{t}\bar{t}} &= -f + p_{tt}, \\ 
g_{\bar{t}\bar{a}} &= p_{tr}\, \Omega_a, \\ 
g_{\bar{a}\bar{b}} &= (r/\bar{r})^2 (1 + p) \delta_{ab} 
+ \bigl[ f^{-1} + p_{rr} - (r/\bar{r})^2 (1 + p) \bigr] \Omega_a \Omega_b 
+ (r/\bar{r}) \bigl( \Omega_a p_b + p_a \Omega_b \bigr), 
\end{align} 
\end{subequations} 
where $p_a := p_{rA} \Omega^A_a$ --- recall the notation introduced in Sec.~\ref{subsec:tranf_cart_sphe}. This is compatible with Eq.~(\ref{PNmetric}) provided that the $\Omega_a \Omega_b$ and $\Omega_a p_b + p_a \Omega_b$ terms in the spatial metric are of $2\pn$ order (or higher), and provided that the Newtonian potentials extracted from $g_{\bar{t}\bar{t}}$ and the $\delta_{ab}$ part of the spatial metric agree with each other. 

To carry out the post-Newtonian expansion of the metric we introduce an ordering parameter $\epsilon$, such that a term of order $\epsilon$ is declared to be Newtonian ($\n = 0\pn$), a term of order $\epsilon^2$ is post-Newtonian ($1\pn$), and so on; $\epsilon$ is set equal to $1$ at the end of the exercise. We take $M$ to be of order $\epsilon$.  
Because a tidal moment $\bar{\E}_L$ scales as $M'/a^{\ell+1}$, where $M'$ is a characteristic mass scale for the external matter and $a$ a characteristic distance to it, we expand it as 
\begin{equation} 
\bar{\E}_L =\bar{\E}_L[\n] + \bar{\E}_L[\pn] + 2\pn, 
\end{equation} 
with the first term of order $\epsilon$, and the second term of order $\epsilon^2$; we place an overbar on the tidal moments to again remind ourselves that they refer to the body zone. The bilinear moments are similarly expanded as 
\begin{equation} 
\bEE_L = \bEE_L[\pn] + \bEE_L[2\pn] + 3\pn, 
\end{equation} 
with $\bEE_L[\pn]$ constructed entirely from $\bar{\E}_L[\n]$, while $\bEE_L[2\pn]$ involves a combination of Newtonian and post-Newtonian tidal moments. Because the time scale of variation of the tidal moments is comparable to $(a^3/M')^{1/2}$, we have that the first derivative of $\bar{\E}_L$ begins at order $\epsilon^{3/2}$, while its second derivative begins at order $\epsilon^2$. We express this as 
\begin{equation} 
\dot{\bar{\E}}_L = \dot{\bar{\E}}_L[\pn] + \dot{\bar{\E}}_L[2\pn] + 3\pn, \qquad  
\ddot{\bar{\E}}_L = \ddot{\bar{\E}}_L[\pn] + \ddot{\bar{\E}}_L[2\pn] + 3\pn. 
\end{equation} 
We use the same post-Newtonian ordering for both quantities, in spite of the fact that they come with different powers of $\epsilon$. The reason is that the first derivative of a tidal moment will appear in the vector potential only, and that it is conventional, for this potential, to reset the post-Newtonian counter so that $\epsilon^{3/2} = 1\pn$.    

We incorporate the scalings of the Love numbers $K_\ell$, $P_\ell$, $\dot{K}_\ell$, and $\ddot{K}_\ell$ provided by Eqs.~(\ref{K_vs_k}), (\ref{P2_scaling}), (\ref{P4_scaling}), (\ref{P3_scaling}), (\ref{P5_scaling}), (\ref{Kdot_vs_kdot}), (\ref{Kddot2_scaling}), (\ref{Kddot3_scaling}), (\ref{Kddot4_scaling}), and (\ref{Kddot5_scaling}). They are written in terms of the primitive and scalefree $k_\ell$, $p_\ell$, $\dot{k}_\ell$, and $\ddot{k}_\ell$, respectively.  

The extraction also relies on making scaling choices for the constants of integration $S_0$, $S_1$, $T_\ell$, $\dot{T}_\ell$, and $\ddot{T}_\ell$. We recall that $S_0$ reflects a residual gauge freedom associated with a rescaling of the time coordinate, and we eliminate it by setting $S_0 = 0$. Similarly, $S_1$ reflects a residual gauge freedom associated with a translation of the spatial origin of the coordinates. A scaling of $S_1 M^8 = s_1 R^8$ would produce a Newtonian shift, and we choose to rule this out. We do, however, retain the possibility of a post-Newtonian shift by adopting the scaling $S_1 M^7 = s_1 R^7$, with $s_1$ postulated to be primitive and scalefree. As we saw back in Sec.~\ref{subsec:redefinitions}, the constant $T_0$ is associated with the freedom to redefine the mass parameter $M$ by a shift proportional to $M^5 \E_{ab} \E^{ab}$. An assumed scaling of $T_0 M^6 = t_0 R^6$ would produce a Newtonian adjustment to the mass, which we do not allow. We retain, however, the possibility of a post-Newtonian adjustment by setting $T_0 M^5 = t_0 R^5$, with $t_0$ assumed to be primitive and scalefree. We set all remaining $T_\ell$'s to zero, and also put $\dot{T}_\ell = \ddot{T}_\ell = 0$. 

With all these ingredients in place, we find that the metric of Eq.~(\ref{metric_complete}) gives us the potentials 
\begin{subequations} 
\label{bodyzone_potentials1} 
\begin{align} 
\bar{U} &= \frac{M}{\bar{r}} 
- \frac{1}{2} \bar{r}^2 \bigl[ 1 + 2 k_2 (R/\bar{r})^5 \bigr]\, \bar{\E}^{\sf q}[\n] 
- \frac{1}{6} \bar{r}^3 \bigl[ 1 + 2 k_3 (R/\bar{r})^7 \bigr]\, \bar{\E}^{\sf o}[\n] 
- \frac{1}{12} \bar{r}^4 \bigl[ 1 + 2 k_4 (R/\bar{r})^9 \bigr]\, \bar{\E}^{\sf h}[\n] 
\nonumber \\ & \quad \mbox{}
- \frac{1}{20} \bar{r}^5 \bigl[ 1 + 2 k_5 (R/\bar{r})^{11} \bigr]\, \bar{\E}^{\sf t}[\n] 
- \frac{1}{3} k_2 \frac{R^5}{M} \bar{r}\, \bEE^{\sf d}[\pn] 
- p_2 \frac{R^8}{M} \frac{1}{\bar{r}^3}\, \bEE^{\sf q}[\pn] 
- p_3 \frac{R^{10}}{M} \frac{1}{\bar{r}^4}\, \bEE^{\sf o}[\pn] 
\nonumber \\ & \quad \mbox{}
- p_4 \frac{R^{10}}{M} \frac{1}{\bar{r}^5}\, \bEE^{\sf h}[\pn]
- p_5 \frac{R^{12}}{M} \frac{1}{\bar{r}^6}\, \bEE^{\sf t}[\pn] 
- \ddot{k}_2 \frac{R^8}{M} \frac{1}{\bar{r}^3}\, \ddot{\bar{\E}}^{\sf q}[\pn] 
- \frac{1}{3} \ddot{k}_3 \frac{R^{10}}{M} \frac{1}{\bar{r}^4}\, \ddot{\bar{\E}}^{\sf o}[\pn] 
- \frac{1}{6} \ddot{k}_4 \frac{R^{12}}{M} \frac{1}{\bar{r}^5}\, \ddot{\bar{\E}}^{\sf h}[\pn] 
\nonumber \\ & \quad \mbox{}
- \frac{1}{10} \ddot{k}_5 \frac{R^{14}}{M} \frac{1}{\bar{r}^6}\, \ddot{\bar{\E}}^{\sf t}[\pn],  \\
\bar{U}_a &= \biggl\{ \frac{1}{6} \bar{r}^3 \Bigl[1 - 3k_2 (R/\bar{r})^5 \Bigr]\, \dot{\bar{\E}}^{\sf q}[\pn]
+ \frac{1}{24} \bar{r}^4 \Bigl[1 - \frac{8}{3} k_3 (R/\bar{r})^7 \Bigr]\, \dot{\bar{\E}}^{\sf o}[\pn]  
+ \frac{1}{60} \bar{r}^5 \Bigl[1 - \frac{5}{2} k_4 (R/\bar{r})^9 \Bigr]\, \dot{\bar{\E}}^{\sf h}[\pn]  
\nonumber \\ & \quad \mbox{}
+ \frac{1}{120} \bar{r}^6 \Bigl[1 - \frac{12}{5} k_5 (R/\bar{r})^{11} \Bigr]\, \dot{\bar{\E}}^{\sf t}[\pn] 
\biggr\}\, \Omega_a \\
\bar{\Psi} &= -\frac{1}{2} \bar{r}^2 \bigl[ 1 + 2 k_2 (R/\bar{r})^5 \bigr]\, \bar{\E}^{\sf q}[\pn] 
- \frac{1}{6} \bar{r}^3 \bigl[ 1 + 2 k_3 (R/\bar{r})^7 \bigr]\, \bar{\E}^{\sf o}[\pn] 
- \frac{1}{12} \bar{r}^4 \bigl[ 1 + 2 k_4 (R/\bar{r})^9 \bigr]\, \bar{\E}^{\sf h}[\pn] 
\nonumber \\ & \quad \mbox{}
- \frac{1}{20} \bar{r}^5 \bigl[ 1 + 2 k_5 (R/\bar{r})^{11} \bigr]\, \bar{\E}^{\sf t}[\pn] 
+ \biggl( \frac{1}{2} t_0 - \frac{14}{15} k_2 \biggr) \frac{R^5}{\bar{r}}\, \bEE^{\sf m}[\pn] 
- \frac{1}{3} k_2 \frac{R^5}{M} \bar{r}\, \bEE^{\sf d}[2\pn] 
\nonumber \\ & \quad \mbox{}
+ \biggl( -\frac{1}{2} s_1 + \frac{2}{35} k_3 \biggr) \frac{R^7}{\bar{r}^2}\, \bEE^{\sf d}[\pn] 
- p_2 \frac{R^8}{M} \frac{1}{\bar{r}^3}\, \bEE^{\sf q}[2\pn] 
- p_3 \frac{R^{10}}{M} \frac{1}{\bar{r}^4}\, \bEE^{\sf o}[2\pn] 
+ \frac{5}{12} \bar{r}^4 \, \bEE^{\sf h}[\pn] 
\nonumber \\ & \quad \mbox{}
- p_4 \frac{R^{10}}{M} \frac{1}{\bar{r}^5}\, \bEE^{\sf h}[2\pn]
+ \frac{3}{10} \bar{r}^5\, \bEE^{\sf t}[\pn] 
- p_5 \frac{R^{12}}{M} \frac{1}{\bar{r}^6}\, \bEE^{\sf t}[2\pn] 
- \biggl( \frac{11}{84} \bar{r}^4 + \frac{1}{2} k_2 \frac{R^5}{\bar{r}} \biggr)\, \ddot{\bar{\E}}^{\sf q}[\pn] 
\nonumber \\ & \quad \mbox{}
- \ddot{k}_2 \frac{R^8}{M} \frac{1}{\bar{r}^3}\, \ddot{\bar{\E}}^{\sf q}[2\pn] 
- \biggl( \frac{1}{36} \bar{r}^5 + \frac{1}{18} k_3 \frac{R^7}{\bar{r}^2} \biggr)\, \ddot{\bar{\E}}^{\sf o}[\pn]
- \frac{1}{3} \ddot{k}_3 \frac{R^{10}}{M} \frac{1}{\bar{r}^4}\, \ddot{\bar{\E}}^{\sf o}[2\pn] 
- \biggl( \frac{13}{1320} \bar{r}^6 + \frac{1}{84} k_4 \frac{R^9}{\bar{r}^3} \biggr)\, \ddot{\bar{\E}}^{\sf h}[\pn] 
\nonumber \\ & \quad \mbox{}
- \frac{1}{6} \ddot{k}_4 \frac{R^{12}}{M} \frac{1}{\bar{r}^5}\, \ddot{\bar{\E}}^{\sf h}[2\pn] 
- \biggl( \frac{7}{1560} \bar{r}^7 + \frac{1}{300} k_5 \frac{R^{11}}{\bar{r}^4} \biggr)\, \ddot{\bar{\E}}^{\sf t}[\pn] 
- \frac{1}{10} \ddot{k}_5 \frac{R^{14}}{M} \frac{1}{\bar{r}^6}\, \ddot{\bar{\E}}^{\sf t}[2\pn]. 
\end{align} 
\end{subequations} 
The square of $\bar{U}$, which enters the computation of $\bar{\Psi}$, is calculated with the help of Eqs.~(\ref{EqEq}) and (\ref{EqEo}). We observe that in $\bar{U}$, a bilinear tidal moment $\bEE_L[\pn]$, formally of post-Newtonian order, is divided by $M$ to produce a quantity of Newtonian order. The same phenomenon occurs with $\ddot{\bar{\E}}_L[\pn]$, and we witness a similar conversion in $\bar{\Psi}$, when a tidal moment that is formally of second post-Newtonian order is divided by $M$. The Newtonian potential can be seen to agree with the one of Eq.~(\ref{Ucomplete}), which was obtained entirely within Newtonian theory; the comparison requires the truncation of the multipole expansion beyond $\ell =3$.          

Inspection reveals that $\bar{\Psi}$ contains a term proportional to $\bar{r}^{-1}\, \bEE^{\sf m}[\pn]$. This monopole term corresponds to a post-Newtonian adjustment of the mass parameter, and we can eliminate it by setting 
\begin{equation} 
t_0 = \frac{28}{15} k_2. 
\label{t0} 
\end{equation} 
We observe also that the post-Newtonian potential includes a term proportional to $\bar{r}^{-2}\, \bEE^{\sf d}[\pn]$. This dipole term corresponds to a post-Newtonian shift of the coordinate system away from the body's center of mass; we eliminate it by setting 
\begin{equation} 
s_1 = \frac{4}{35} k_3. 
\label{s1} 
\end{equation} 
These choices give us body-zone potentials for which $M$ is the body's mass at both Newtonian and post-Newtonian orders, and for which the spatial origin of the coordinate system is anchored at the body's center of mass. 
 
Equation (\ref{bodyzone_potentials1}) gives us a description of the potentials that includes tidal and mass multipole moments (both linear and bilinear) up to $\ell = 5$. The barycentric potentials of Eqs.~(\ref{PNpotentials1}), on the other hand, feature mass multipole moments up to $\ell = 2$ only, because for the sake of simplicity we chose, back in Sec.~\ref{subsec:bary}, to truncate the multipole structure beyond the quadrupole order. To be consistent we shall apply the same truncation to Eq.~(\ref{bodyzone_potentials1}); we therefore eliminate all mass moments beyond $\ell = 2$, and all tidal moments beyond $\ell = 3$. This gives   
\begin{subequations} 
\label{bodyzone_potentials2} 
\begin{align} 
\bar{U} &= \frac{1}{\bar{r}^3} \biggl( -k_2 R^5\, \bar{\E}_{ab}[\n] 
- p_2 \frac{R^8}{M}\, \bEE_{ab}[\pn] 
- \ddot{k}_2 \frac{R^8}{M}\, \ddot{\bar{\E}}_{ab}[\pn] \biggr) \Omega^a \Omega^b 
+ \frac{M}{\bar{r}} 
\nonumber \\ & \quad \mbox{} 
+ \bar{r} \biggl( -\frac{1}{3} k_2 \frac{R^5}{M}\, \bEE_{a}[\pn] \biggr) \Omega^a 
- \frac{1}{2} \bar{r}^2\, \bar{\E}_{ab}[\n]\, \Omega^a \Omega^b 
- \frac{1}{6} \bar{r}^3\, \bar{\E}_{abc}[\n]\, \Omega^a \Omega^b \Omega^c
+ O(\bar{r}^4), \\ 
\bar{U}_a &= \frac{1}{\bar{r}^2} \biggl( -\frac{1}{2} k_2 R^5\, \dot{\bar{\E}}_{bc}[\pn] \biggr) \Omega_a \Omega^b \Omega^c + O(\bar{r}^3), \\ 
\bar{\Psi} &= \frac{1}{\bar{r}^3} \biggl( -k_2 R^5\, \bar{\E}_{ab}[\pn] 
- p_2 \frac{R^8}{M}\, \bEE_{ab}[2\pn] 
- \ddot{k}_2 \frac{R^8}{M}\, \ddot{\bar{\E}}_{ab}[2\pn] \biggr) \Omega^a \Omega^b 
\nonumber \\ & \quad \mbox{} 
+ \frac{1}{\bar{r}} \biggl( -\frac{1}{2} k_2 R^5\, \ddot{\bar{\E}}_{ab}[\pn] \biggr) \Omega^a \Omega^b 
+ \bar{r} \biggl( -\frac{1}{3} k_2 \frac{R^5}{M}\, \bEE_{a}[2\pn] \biggr) \Omega^a 
\nonumber \\ & \quad \mbox{} 
- \frac{1}{2} \bar{r}^2\, \bar{\E}_{ab}[\pn]\, \Omega^a \Omega^b 
- \frac{1}{6} \bar{r}^3\, \bar{\E}_{abc}[\pn]\, \Omega^a \Omega^b \Omega^c
+ O(\bar{r}^4). 
\end{align} 
\end{subequations}  

\subsection{Matching} 
\label{subsec:match} 
 
The potentials of Eqs.~(\ref{PNpotentials_transf}) were obtained by subjecting the barycentric potentials of Eqs.~(\ref{PNpotentials2}) to a coordinate transformation from the barycentric frame (in which the compact body moves in the gravitational field of the external matter) to the body frame (in which it remains at rest at the spatial origin of the new coordinates). On the other hand, the potentials of Eqs.~(\ref{bodyzone_potentials2}) were extracted from the fully relativistic metric of a tidally deformed compact body, as computed in Sec.~\ref{sec:GR}. Each set of potentials describes the same gravitational field in the same region of spacetime --- the overlap zone introduced in Sec.~\ref{subsec:zones} --- in the same coordinate system; the potentials must therefore agree. A term by term comparison between the sets reveals (i) the relation between the mass quadrupole moment $Q_{ab}$ of the barycentric potentials and the tidal moments $\bar{\E}_{ab}$ of the body-zone potentials, (ii) an expression for the mass dipole moment $Q_a[\pn]$ of the barycentric post-Newtonian potential, (iii) an expression for $\mu$, the post-Newtonian adjustment to the body's mass parameter, (iv) the relation between the tidal moments and the barycentric potentials $U_{\rm ext}$, $U^a_{\rm ext}$, and $\Psi_{\rm ext}$ created by the external matter, (v) the equations of motion for the vector $\bm{z}(t)$, which points to the body's position in the harmonic chart of the barycentric potentials, and (vi) all other details of the coordinate transformation, including expressions for the time-dilation function $A$, the frame-precession tensor $\epsilon_{abc} R^c$, and the gauge function $\beta$. 

The match relies on the fact that $\Omega^a = \bar{n}^a = \bar{x}^a/\bar{r}$. Regarding the mass quadrupole moment, we find 
\begin{subequations} 
\label{Qab} 
\begin{align} 
Q_{ab}[\n] &= -\frac{2}{3} k_2 R^5\, \bar{\E}_{ab}[\n] 
- \frac{2}{3} p_2 \frac{R^8}{M}\, \bEE_{ab}[\pn] 
- \frac{2}{3} \ddot{k}_2 \frac{R^8}{M}\, \ddot{\bar{\E}}_{ab}[\pn], \\ 
Q_{ab}[\pn] &= -\frac{2}{3} k_2 R^5\, \bar{\E}_{ab}[\pn] 
- \frac{2}{3} p_2 \frac{R^8}{M}\, \bEE_{ab}[2\pn] 
- \frac{2}{3} \ddot{k}_2 \frac{R^8}{M}\, \ddot{\bar{\E}}_{ab}[2\pn]
\nonumber \\ & \quad \mbox{}
- 3\bigl( \hat{U}_{\rm ext} - \tfrac{1}{2} v^2 \bigr) Q_{ab}[\n] 
- v_{\langle a} Q_{b\rangle c}[\n] v^c 
- A \dot{Q}_{ab}[\n] + 2\epsilon_{jc(a} R^c Q^j_{\ b)}[\n]. 
\end{align} 
\end{subequations}
The Newtonian piece of the quadrupole moment can be compared with the expression of Eq.~(\ref{quadmoment_Newtonian}), which was obtained in pure Newtonian theory. 

Regarding the mass dipole moment, we get 
\begin{equation} 
Q_a[\pn] = -2 Q_{ab}[\n] a^b[\n] + \dot{Q}_{ab}[\n] v^b. 
\label{Qa} 
\end{equation} 
The post-Newtonian adjustment to the mass parameter is 
\begin{equation} 
\mu = -\hat{U}_{\rm ext} + \frac{3}{2} v^2. 
\end{equation} 
In this expression, and in Eq.~(\ref{Qab}) above, it is understood that the external potential is evaluated at $\bm{\bar{x}} = \bm{0}$, the body's position. 

Regarding the relation between tidal moments and external potentials, we obtain 
\begin{subequations} 
\label{Eab_vs_potentials} 
\begin{align} 
\bar{\E}_{ab}[\n] &= -\partial_{\bar{a}\bar{b}} \hat{U}_{\rm ext}, \\ 
\bar{\E}_{ab}[\pn] &= -\partial_\stf{\bar{a}\bar{b}} \hat{\Psi}_{\rm ext} 
- 4 \partial_{\bar{t} \langle \bar{a}} \hat{U}^{\rm ext}_{b\rangle} 
+ 4 v_c\, \partial_{\bar{a}\bar{b}} \hat{U}^c_{\rm ext} 
+ 2(v^2 - \hat{U}_{\rm ext}) \bar{\E}_{ab}[\n] 
- v_{\langle a} \bar{\E}_{b\rangle c}[\n] v^c
\nonumber \\ & \quad \mbox{}
+ 2 v_{\langle a} \dot{g}_{b\rangle} 
+ 3 \bigl( g_{\langle a} + \bar{g}^{\rm tidal}_{\langle a}[\n] \bigr) 
\bigl( g_{b \rangle} - \bar{g}^{\rm tidal}_{b \rangle}[\n] \bigr) 
+ A \dot{\bar{\E}}_{ab}[\pn] - 2 \epsilon_{jc(a} R^c \bar{\E}^j_{\ b)}[\n] 
\end{align} 
\end{subequations} 
and 
\begin{subequations} 
\label{Eabc_vs_potentials} 
\begin{align} 
\bar{\E}_{abc}[\n] &= -\partial_{\bar{a}\bar{b}\bar{c}} \hat{U}_{\rm ext}, \\ 
\bar{\E}_{abc}[\pn] &= -\partial_\stf{\bar{a}\bar{b}\bar{c}} \hat{\Psi}_{\rm ext} 
- 4 \partial_{\bar{t} \langle \bar{a}\bar{b}} \hat{U}^{\rm ext}_{c\rangle} 
+ 4 v_d\, \partial_{\bar{a}\bar{b}\bar{c}} \hat{U}^d_{\rm ext} 
+ (2v^2 - 3\hat{U}_{\rm ext}) \bar{\E}_{abc}[\n] 
- \frac{3}{2} v_{\langle a} \bar{\E}_{bc\rangle d}[\n] v^d
\nonumber \\ & \quad \mbox{}
- v_{\langle a} \dot{\bar{\E}}_{bc\rangle}[\pn] 
- 10 \bigl( g_{\langle a} + \bar{g}^{\rm tidal}_{\langle a}[\n] \bigr) \bar{\E}_{bc\rangle}[\n] 
+ A \dot{\bar{\E}}_{abc}[\pn] - 3 \epsilon_{jc(a} R^c \bar{\E}^j_{\ bc)}[\n].  
\end{align} 
\end{subequations} 
Here it is understood that the external potentials are evaluated at $\bm{\bar{x}} = \bm{0}$ after differentiation. 

Regarding the equations of motion, we find 
\begin{subequations} 
\label{acceleration} 
\begin{align} 
a_a[\n] &= g_a + \bar{g}_a^{\rm tidal}[\n], \\ 
a_a[\pn] &= \partial_{\bar{a}} \hat{\Psi}_{\rm ext} 
+ 4 \partial_{\bar{t}} \hat{U}^{\rm ext}_a 
- 4 v_b\, \partial_{\bar{a}} \hat{U}^b_{\rm ext} 
- 3 v_a\, \partial_{\bar{t}} \hat{U}_{\rm ext} 
+ v^2 \bigl( g_a - \bar{g}_a^{\rm tidal}[\n] \bigr) 
- \hat{U}_{\rm ext} \bigl( 4 g_a + 3 \bar{g}_a^{\rm tidal}[\n] \bigr) 
\nonumber \\ & \quad \mbox{}
- \frac{1}{2} v_a v^b \bigl( 2g_b + \bar{g}_b^{\rm tidal}[\n] \bigr)   
+ \bar{g}_a^{\rm tidal}[\pn] 
- A \dot{\bar{g}}^{\rm tidal}_a[\n] 
+ \epsilon_{jca} R^c \bar{g}^j_{\rm tidal}[\n], 
\end{align} 
\end{subequations} 
where 
\begin{subequations} 
\label{gtidal} 
\begin{align} 
g_a &= \partial_{\bar{a}} \hat{U}_{\rm ext}, \\ 
\bar{g}^{\rm tidal}_a[\n] &= \frac{1}{3} k_2 \frac{R^5}{M}\, \bEE_a[\pn], \\ 
\bar{g}^{\rm tidal}_a[\pn] &= \frac{1}{3} k_2 \frac{R^5}{M}\, \bEE_a[2\pn]. 
\end{align} 
\end{subequations} 
Again it is understood that the external potentials are evaluated at $\bm{\bar{x}} = \bm{0}$ after differentiation. The expression for the Newtonian piece of the tidal acceleration can be compared with Eq.~(\ref{atidal2}), which was obtained in a purely Newtonian treatment. 

Regarding the details of the coordinate transformation, we find that the time-dilation function and frame-precession tensor are determined by integrating 
\begin{equation}  
\dot{A} = \hat{U}_{\rm ext} + \frac{1}{2} v^2 
\label{Adot} 
\end{equation} 
and 
\begin{equation} 
\epsilon_{abc} \dot{R}^c = -4 \partial_{[\bar{a}} \hat{U}^{\rm ext}_{b]} 
+ \bigl( 3 g_{[a} - \bar{g}^{\rm tidal}_{[a}[\n] \bigr) v_{b]},  
\label{Rdot} 
\end{equation} 
respectively. The gauge function $\beta$ is constructed piece by piece, using the expansion of Eq.~(\ref{beta_taylor}). The singular term is given by 
\begin{equation} 
\mbox{}_{-1} \beta_{ab} = -\dot{Q}_{ab}[\n], 
\end{equation} 
and it allows us to convert the vector potential from the harmonic gauge of the barycentric frame to the Regge-Wheeler gauge of the body frame. The regular terms are given by 
\begin{subequations} 
\begin{align} 
\mbox{}_1 \beta_a &= -4\hat{U}^{\rm ext}_a + \bigl( 3 \hat{U}_{\rm ext} + \tfrac{1}{2} v^2 \bigr) v_a 
+ A \bigl( g_a + \bar{g}^{\rm tidal}_a[\n] \bigr) 
+ \epsilon_{abc} R^b v^c, \\  
\mbox{}_2 \beta_{ab} &= - 4\partial_{(\bar{a}} \hat{U}^{\rm ext}_{b)} 
+ 2 \bigl( g_{(a} - \bar{g}^{\rm tidal}_{(a}[\n] \bigr) v_{b)} 
+ \delta_{ab} \bigl( \partial_{\bar{c}} \hat{U}^c_{\rm ext} + v^c \bar{g}^{\rm tidal}_c \bigl), \\ 
\mbox{}_3 \beta_{abc} &= -4 \partial_{\langle \bar{a}\bar{b}} \bigl( \hat{U}^{\rm ext}_{c\rangle} 
- v_{c\rangle} \hat{U}_{\rm ext} \bigr) + \frac{1}{5} \bigl( \delta_{ab}\, \dot{g}_c 
+ \delta_{ac}\, \dot{g}_b + \delta_{bc}\, \dot{g}_a \bigr). 
\end{align} 
\end{subequations} 
An expression for $\mbox{}_0 \beta$ can be obtained by integrating $\mbox{}_0 \bar{\Psi} = 0$, as given by Eq.~(\ref{Psi0}); this expression, however, is not required in any of the calculations summarized here. 

The term by term matching between the transformed barycentric potentials and the extracted body-zone potentials reveals some apparent inconsistencies. For example, a comparison between the $\bar{r}^{-2}$ terms in the vector potentials reveals that 
\begin{equation} 
\dot{Q}_{ab}[\n] = -\frac{2}{3} k_2 R^5 \dot{\bar{\E}}_{ab}[\n], 
\end{equation} 
and this expression is incomplete in view of Eq.~(\ref{Qab}). The source of inconsistency is the fact that the fully relativistic metric of Sec.~\ref{sec:GR} does not account for the time derivative of the bilinear tidal moments, nor for the third time derivative of the linear moments; such terms are beyond the accuracy goals described in Sec.~\ref{subsec:task}. It seems most plausible that a more complete construction that includes these missing terms would eliminate the inconsistency. As another example along the same lines, a comparison between the $\bar{r}^{-1}$ terms in the post-Newtonian potential reveals that 
\begin{equation} 
\ddot{Q}_{ab}[\n] = -\frac{2}{3} k_2 R^5 \ddot{\bar{\E}}_{ab}[\n]. 
\end{equation} 
An expression consistent with Eq.~(\ref{Qab}) would have required a relativistic metric that includes the second time derivative of the bilinear moments, and the fourth time derivative of the linear moments. 

Another source of apparent inconsistency is our omission of all gravitomagnetic tidal moments ${\cal B}_L$ from the description of the body's tidal environment. This, however, has no bearing on the results displayed in this subsection, nor on results obtained below. 

\subsection{Barycentric tidal moments} 
\label{subsec:bary_moments} 

The mass quadrupole moment $Q_{ab}(t)$ is native to the barycentric frame, and in Eq.~(\ref{Qab}) it is expressed in terms of the tidal moment $\bar{\E}_{ab}(\bar{t})$, which is native to the body frame. The dichotomy of natural homes gives rise to the awkward presence of $A$ and $\epsilon_{abc} R^c$ in this expression. 

Following Racine and Flanagan \cite{racine-flanagan:05}, we introduce the barycentric tidal moments
\begin{subequations} 
\label{M-transf} 
\begin{align} 
\E_{ab}(t) &:= {\cal M}_a^{\ j}(\bar{t}) {\cal M}_b^{\ k}(\bar{t})\, \bar{\E}_{jk}(\bar{t}), \\ 
\E_{abc}(t) &:= {\cal M}_a^{\ j}(\bar{t}) {\cal M}_b^{\ k}(\bar{t}) {\cal M}_c^{\ n}(\bar{t})\, \bar{\E}_{jkn}(\bar{t}), 
\end{align} 
\end{subequations} 
where 
\begin{equation} 
t = \bar{t} + A(\bar{t}) + 2\pn, \qquad 
{\cal M}_a^{\ j}(\bar{t}) = \delta_a^{\ j} + \epsilon^{\ \ j}_{ca} R^c(\bar{t}) + 2\pn. 
\end{equation} 
The transformation accounts for the difference in time coordinates, and for the relative precession between frames. Its inverse is 
\begin{subequations} 
\begin{align} 
\bar{\E}_{ab}(\bar{t}) &:= {\cal N}_a^{\ j}(t) {\cal N}_b^{\ k}(t)\, \E_{jk}(t), \\ 
\bar{\E}_{abc}(\bar{t}) &:= {\cal N}_a^{\ j}(t) {\cal N}_b^{\ k}(t) {\cal N}_c^{\ n}(t)\, \E_{jkn}(t), 
\end{align} 
\end{subequations} 
with
\begin{equation} 
\bar{t} = t - A(t) + 2\pn, \qquad 
{\cal N}_a^{\ j}(t) = \delta_a^{\ j} - \epsilon^{\ \ j}_{ca} R^c(t) + 2\pn. 
\end{equation} 
The transformation is also applied to the bilinear moments. It follows, for example, that 
\begin{subequations} 
\begin{align} 
\EE_a(t) &= {\cal M}_a^{\ j}(\bar{t})\, \bEE_j(\bar{t}), \\ 
\EE_{ab}(t) &= {\cal M}_a^{\ j}(\bar{t}) {\cal M}_b^{\ k}(\bar{t})\, \bEE_{jk}(\bar{t}). 
\end{align} 
\end{subequations} 

We apply the transformation to Eqs.~(\ref{Qab}), (\ref{Eab_vs_potentials}), (\ref{Eabc_vs_potentials}), (\ref{acceleration}), and (\ref{gtidal}). Simultaneously, we ``unhat'' the external potentials by inverting Eq.~(\ref{Uhat_def}) and its analogues. Some care must be exercised when the hatted potentials are differentiated, because Eq.~(\ref{Uhat_def}) implies that $\partial_{\bar{t}} \hat{U} = \partial_t U + v^a \partial_a U$ and $\partial_{\bar{a}} \hat{U} = \partial_a U$. 

The end result of this exercise is   
\begin{subequations} 
\label{Qab_bary} 
\begin{align} 
Q_{ab}[\n] &= -\frac{2}{3} k_2 R^5\, \E_{ab}[\n] 
- \frac{2}{3} p_2 \frac{R^8}{M}\, \EE_{ab}[\pn] 
- \frac{2}{3} \ddot{k}_2 \frac{R^8}{M}\, \ddot{\E}_{ab}[\pn], \\ 
Q_{ab}[\pn] &= -\frac{2}{3} k_2 R^5\, \E_{ab}[\pn] 
- \frac{2}{3} p_2 \frac{R^8}{M}\, \EE_{ab}[2\pn] 
- \frac{2}{3} \ddot{k}_2 \frac{R^8}{M}\, \bigl( \ddot{\E}_{ab}[2\pn] + {\cal F}_{ab} \bigr) 
\nonumber \\ & \quad \mbox{}
- 3\bigl( U_{\rm ext} - \tfrac{1}{2} v^2 \bigr) Q_{ab}[\n] 
- v_{\langle a} Q_{b\rangle c}[\n] v^c 
\end{align} 
\end{subequations}
with 
\begin{subequations} 
\label{Eab_bary} 
\begin{align} 
\E_{ab}[\n] &= -\partial_{ab} U_{\rm ext}, \\ 
\E_{ab}[\pn] &= -\partial_\stf{ab} \Psi_{\rm ext} 
- 4 \partial_{t \langle a} U^{\rm ext}_{b\rangle} 
- 4 v^c \bigl( \partial_{c\langle a} U^{\rm ext}_{b\rangle} - \partial_{ab} U^{\rm ext}_c \bigr) 
+ 2(v^2 - U_{\rm ext}) \E_{ab}[\n] 
\nonumber \\ & \quad \mbox{}
- v_{\langle a} \E_{b\rangle c}[\n] v^c
+ 2 v_{\langle a} \dot{g}_{b\rangle} 
+ 3 \bigl( g_{\langle a} + g^{\rm tidal}_{\langle a}[\n] \bigr) 
\bigl( g_{b \rangle} - g^{\rm tidal}_{b \rangle}[\n] \bigr) 
\end{align} 
\end{subequations} 
and 
\begin{subequations} 
\label{Eabc_bary} 
\begin{align} 
\E_{abc}[\n] &= -\partial_{abc} U_{\rm ext}, \\ 
\E_{abc}[\pn] &= -\partial_\stf{abc} \Psi_{\rm ext} 
- 4 \partial_{t \langle ab} U^{\rm ext}_{c\rangle} 
- 4 v^d \bigl( \partial_{d\langle ab} U^{\rm ext}_{c\rangle} - \partial_{abc} U^{\rm ext}_d \bigr) 
+ (2v^2 - 3U_{\rm ext}) \E_{abc}[\n] 
\nonumber \\ & \quad \mbox{}
- \frac{3}{2} v_{\langle a} \E_{bc\rangle d}[\n] v^d
- v_{\langle a} \dot{\E}_{bc\rangle}[\pn] 
- 10 \bigl( g_{\langle a} + g^{\rm tidal}_{\langle a}[\n] \bigr) \E_{bc\rangle}[\n].  
\end{align} 
\end{subequations} 
All external potentials are now evaluated at $\bm{x} = \bm{z}(t)$ after differentiation. The new tensor 
\begin{equation} 
{\cal F}_{ab} := 2 \dot{A}\, \ddot{\E}_{ab}[\pn] + \ddot{A}\, \dot{\E}_{ab}[\pn] 
- 4 \epsilon_{jc(a} \dot{R}^c\, \dot{\E}^j_{\ b)}[\pn] 
- 2 \epsilon_{jc(a} \ddot{R}^c\, \E^j_{\ b)}[\n] 
\end{equation} 
accounts for the time derivatives of the transformation matrix when computing $\ddot{\E}_{ab}[\pn]$ from Eq.~(\ref{M-transf}).  
  
The acceleration vector also is re-expressed in terms of the barycentric tidal moments. We have
\begin{subequations} 
\label{acceleration_bary} 
\begin{align} 
a_a[\n] &= g_a + g_a^{\rm tidal}[\n], \\ 
a_a[\pn] &= \partial_{a} \Psi_{\rm ext} 
+ 4 \partial_{t} U^{\rm ext}_a 
+ 4 v^b \bigl( \partial_b U^{\rm ext}_a - \partial_a U^{\rm ext}_b \bigr) 
+ (v^2 - 4 U_{\rm ext}) g_a 
- \bigl( 3 \partial_t U_{\rm ext} + 4 v^b g_b \bigr) v_a 
\nonumber \\ & \quad \mbox{}
+ g^{\rm tidal}_a[\pn] 
-  (v^2 + 3U_{\rm ext}) g^{\rm tidal}_a[\n]  
- \frac{1}{2} v^b g^{\rm tidal}_b[\n]\, v_a, 
\end{align} 
\end{subequations} 
where 
\begin{subequations} 
\label{gtidal_bary} 
\begin{align} 
g_a &= \partial_{a} U_{\rm ext}, \\ 
g^{\rm tidal}_a[\n] &= \frac{1}{3} k_2 \frac{R^5}{M}\, \EE_a[\pn], \\ 
g^{\rm tidal}_a[\pn] &= \frac{1}{3} k_2 \frac{R^5}{M}\, \EE_a[2\pn]. 
\end{align} 
\end{subequations} 
Again it is understood that the external potentials are evaluated at $\bm{x} = \bm{z}(t)$ after differentiation. 

The physical meaning of the transformation of Eq.~(\ref{M-transf}) is best understood by focusing attention on a fixed direction in the body frame, described by the unit vector $\bar{e}_a$. We may think of this, for example, as the direction of the body's spin, in a context in which the spin is nonvanishing but too small to produce significant departures from the metric of Sec.~\ref{sec:GR}. The corresponding vector in the barycentric frame is 
\begin{equation} 
e_a(t) = {\cal M}_a^{\ j} (\bar{t})\, \bar{e}_j. 
\end{equation} 
Taking $\bar{e}_a$ to be independent of $\bar{t}$, the rate of variation of $e_a$ is given by 
\begin{equation} 
\dot{e}_a = \dot{\cal M}_a^{\ j}\, \bar{e}_j = \epsilon_{abc} \bar{e}^b \dot{R}^c, 
\end{equation} 
or 
\begin{equation} 
\dot{e}_a = \Bigl( 3 g_{[a} v_{b]} - g^{\rm tidal}_{[a}[\n] v_{b]} - 4 \partial_{[a} U^{\rm ext}_{b]} \Bigr) e^b + 2\pn
\label{edot_precession} 
\end{equation}
after inserting Eq.~(\ref{Rdot}) and writing $\bar{e}^a = e^a + \pn$. The first term on the right of Eq.~(\ref{edot_precession}) is recognized as describing geodetic precession; in the context of a spin vector, this effect is also known as ``spin-orbit'' precession (refer, for example to Sec.~9.5.6 of Poisson and Will \cite{poisson-will:14}). The second term provides a tidal contribution to the precession. The third term is recognized as describing Schiff precession, also known as ``spin-spin'' precession in the context of a spin vector.  

\begin{acknowledgments} 
Sam Gralla kindly provided helpful comments that helped me clarify certain aspects of the presentation; he has my gratitude. This work was supported by the Natural Sciences and Engineering Research Council of Canada.  
\end{acknowledgments} 

\appendix 

\section{Regge-Wheeler gauge and its uniqueness} 
\label{app:RW} 

We consider a perturbation of the Schwarzschild spacetime, described by
\begin{equation} 
p_{\alpha\beta} := g_{\alpha\beta} - \bar{g}_{\alpha\beta}, 
\end{equation} 
where $g_{\alpha\beta}$ is the true metric of the spacetime, while $\bar{g}_{\alpha\beta}$ is the reference Schwarzschild metric. We take $p_{\alpha\beta}$ to be small, but we do not think of it as being a quantity ``of the first order'' only. In the context of this paper, for example, the perturbation includes terms of the first order in the tidal interaction, but it includes also terms of the second order. In principle $p_{\alpha\beta}$ could be constructed to any order in perturbation theory. 

We decompose $p_{\alpha\beta}$ in spherical harmonics. Adopting the conventions of Ref.~\cite{martel-poisson:05}, we let $x^a := (t,r)$, $\theta^A := (\theta,\phi)$, $\Omega_{AB} := \mbox{diag}[1,\sin^2\theta]$, and we write 
\begin{subequations}
\label{pert_expansion} 
\begin{align}  
p_{ab} &= \sum_{\ell m} h_{ab}^{\ell m}\, Y^{\ell m}, \\ 
p_{aB} &= \sum_{\ell m} j_a^{\ell m}\, Y_B^{\ell m} + \sum_{\ell m} h_a^{\ell m}\, X_B^{\ell m}, \\ 
p_{AB} &= r^2 \sum_{\ell m} \bigl( K^{\ell m}\, \Omega_{AB} Y^{\ell m} + G^{\ell m}\, Y_{AB}^{\ell m} \bigr) 
+ \sum_{\ell m} h_2^{\ell m}\, X_{AB}^{\ell m}, 
\end{align} 
\end{subequations} 
where $Y^{\ell m}(\theta^A)$ are the usual scalar harmonics, 
\begin{equation} 
Y_A^{\ell m} := D_A Y^{\ell m}, \qquad 
Y_{AB}^{\ell m} := \Bigl[ D_A D_B + \frac{1}{2} \ell(\ell+1) \Omega_{AB} \Bigr] Y^{\ell m} 
\end{equation} 
are even-parity tensorial harmonics, and 
\begin{equation} 
X_A^{\ell m} := -\varepsilon_A^{\ B} D_B Y^{\ell m}, \qquad 
X_{AB}^{\ell m} := -\frac{1}{2} \bigl( \varepsilon_A^{\ C} D_B + \varepsilon_B^{\ C} D_A \bigr) D_C Y^{\ell m} 
\end{equation} 
are odd-parity harmonics. Here $D_A$ is the covariant-derivative operator compatible with $\Omega_{AB}$, and $\varepsilon_{AB}$ is the Levi-Civita tensor on the unit-two sphere ($\varepsilon_{\theta\phi} = \sin\theta$). The fields $h^{\ell m}_{ab}$, $j_a^{\ell m}$, $K^{\ell m}$, and $G^{\ell m}$ make up the even-parity sector of the perturbation; they depend on $x^a$ only. The fields $h_a^{\ell m}$ and $h_2^{\ell m}$ make up the odd-parity sector; they also depend on $x^a$ only. 

A perturbation $p_{\alpha\beta}$ can be altered at will by a (small) coordinate transformation, known in this context as a gauge transformation. At second order, the transformation is generated by a vector field 
\begin{equation} 
\Xi^\alpha = \Xi^\alpha_1 + \Xi^\alpha_2, 
\end{equation} 
in which $\Xi^\alpha_1$ is the first-order contribution, and $\Xi^\alpha_2$ the second-order contribution. The transformation changes the perturbation according to (see, for example, Ref.~\cite{bruni-etal:97}) 
\begin{equation} 
p^{\rm new}_{\alpha\beta} = p^{\rm old}_{\alpha\beta} 
- \Lie_{\Xi_1} \bar{g}_{\alpha\beta}
- \Lie_{\Xi_2} \bar{g}_{\alpha\beta} 
- \Lie_{\Xi_1} p^{\rm old}_{\alpha\beta} 
+ \frac{1}{2} \Lie_{\Xi_1} \Lie_{\Xi_1} \bar{g}_{\alpha\beta}, 
\end{equation} 
where $\Lie_\Xi$ denotes Lie differentiation in the direction of $\Xi^\alpha$. The transformation can be generalized to any order in perturbation theory. 

The Regge-Wheeler gauge is formulated for all $\ell \geq 2$ terms in the multipole expansions of Eq.~(\ref{pert_expansion}); the special cases $\ell = 0$ and $\ell = 1$ are treated separately in the main text. The gauge is defined by setting 
\begin{equation} 
j_a^{\ell m} = 0 = G^{\ell m}, \qquad 
h_2^{\ell m} = 0. 
\label{RWconditions} 
\end{equation} 
These choices are behind the expressions listed in the main text, in Eqs.~(\ref{M0_linear_metric}), (\ref{M0_quad-quad_metric}), (\ref{M0_quad-oct_metric}), (\ref{M0_time_metric}), (\ref{metric_linear}), (\ref{metric_bilinear}), (\ref{metric_first-time}), and (\ref{metric_second-time}). 

We wish to prove that the Regge-Wheeler gauge is unique, by which we mean that the only gauge vector that takes a perturbation in the Regge-Wheeler gauge and returns it to the Regge-Wheeler gauge is $\Xi^\alpha = 0$. In other words, we shall prove that any nontrivial transformation from the Regge-Wheeler gauge necessarily produces a perturbation that violates at least one of the conditions of Eq.~(\ref{RWconditions}). 

We begin at first order, with a change described by 
\begin{equation} 
p^{\rm new}_{\alpha\beta} = p^{\rm old}_{\alpha\beta} - \Lie_{\Xi_1} \bar{g}_{\alpha\beta}. 
\end{equation} 
We decompose the gauge vector as 
\begin{subequations}
\begin{align}  
\Xi^1_a &= \sum_{\ell m} \xi^{\ell m}_a\, Y^{\ell m}, \\ 
\Xi^1_A &= \sum_{\ell m} \xi^{\ell m}_{\rm even}\, Y_A^{\ell m}  
+ \sum_{\ell m} \xi^{\ell m}_{\rm odd}\, X_A^{\ell m}. 
\end{align} 
\end{subequations}   
A straightforward calculation then reveals that the fields featured in Eq.~(\ref{RWconditions}) change according to (we omit the $\ell m$ labels to unclutter the notation)
\begin{subequations} 
\begin{align} 
j_t^{\rm new} &= j_t^{\rm old} - \partial_t \xi_{\rm even} - \xi_t, \\ 
j_r^{\rm new} &= j_r^{\rm old} - \partial_r \xi_{\rm even} - \xi_r + \frac{2}{r} \xi_{\rm even}, \\ 
G^{\rm new} &= G^{\rm old} - \frac{2}{r^2} \xi_{\rm even}, \\ 
h_2^{\rm new} &= h_2^{\rm old} - 2 \xi_{\rm odd}. 
\end{align} 
\end{subequations} 
With the old fields set to zero, it is easy to see that any nonzero $\Xi_1^\alpha$ will produce new fields that violate at least one of the conditions listed in Eq.~(\ref{RWconditions}). Only $\Xi_1^\alpha = 0$ preserves these conditions, and we have established that the Regge-Wheeler gauge is unique at first order. 

We proceed to second order. With $\Xi_1^\alpha = 0$ we have that the second-order transformation is given by 
\begin{equation} 
p^{\rm new}_{\alpha\beta} = p^{\rm old}_{\alpha\beta} - \Lie_{\Xi_2} \bar{g}_{\alpha\beta},  
\end{equation} 
and that it takes exactly the same form as a first-order transformation. The same manipulations allow us to conclude that only $\Xi_2^\alpha = 0$ preserves the conditions of Eq.~(\ref{RWconditions}). The Regge-Wheeler gauge is unique at second order as well. 

The argument can be extended to higher orders, with the same conclusion: the Regge-Wheeler gauge is unique at any order in perturbation theory. 

\section{Hypergeometric and Legendre functions}
\label{app:Legendre} 

We let $z := 2M/r$ and $y := r/M-1$, so that $y = 2/z-1$ and $z = 2/(y+1)$. For $r \geq 2M$ we have that $z \leq 1$ and $y \geq 1$. It is also useful to note that $1-z = (y-1)/(y+1)$. The properties of hypergeometric and Legendre functions required here are summarized in the {\it NIST Handbook of Mathematical Functions} \cite{NIST:10}, to which we refer repeatedly. 

It is easy to verify that the functions
\begin{equation}
h_1 :=z^{-\ell} F(-\ell,-\ell;-2\ell;z), \qquad h_2 := z^{\ell+1} F(\ell+1, \ell+1; 2\ell+2; z)
\end{equation}
and
\begin{equation}
h_3 := P_\ell(y), \qquad h_4 := Q_\ell(y)
\end{equation}
all satisfy the differential equation
\begin{equation}
r^2 f \frac{d^2 h}{dr^2} + 2(r-M) \frac{dh}{dr} - \ell(\ell+1) h = 0.
\end{equation}
Because $h_1$ is a terminating polynomial in $z$, and $h_3$ is a terminating polynomial in $y$, these functions must be proportional to each other. In principle, $h_2$ could be a linear superposition of $h_3$ and $h_4$; an examination of a large number of special cases reveals instead that $h_2$ is simply proportional to $h_4$.

To identify the ratio $h_1/h_3$ we examine the $z \to 0$ behavior of $h_1$, which corresponds to the $y \to \infty$ behavior of $h_3$. The leading-order term in $h_1$ is $z^{-\ell}$, and according to [NIST (14.8.12)], the leading-order term in $h_3$ is $[(2\ell-1)!!/\ell!] y^\ell$. Because $z \sim 2/y$ in this regime, we conclude that
\begin{equation}
z^{-\ell} F(-\ell,-\ell;-2\ell;z) = \frac{(\ell-1)!\, \ell!}{2(2\ell-1)!}\, P_\ell(y).
\label{id1}
\end{equation}
To find the ratio $h_2/h_4$ we examine the $z \to 1$ behavior of $h_2$, which must match the $y \to 1$ behavior of $h_4$. According to [NIST (14.7.7)], $h_4 \sim -1/2 \ln(y-1)$, and from [NIST (15.8.10)] we infer that $h_2 \sim -[(2\ell+1)!/\ell!^2] \ln(1-z)$. With $\ln(1-z) \sim \ln(y-1)$, we have that
\begin{equation}
z^{\ell+1} F(\ell+1,\ell+1;2\ell+2;z) = \frac{2(2\ell+1)!}{\ell!^2}\, Q_\ell(y).
\label{id2}
\end{equation}

The identity [NIST (15.5.3)]
\begin{equation}
z \frac{d}{dz} \bigl[ z^a F(a,b;c;z) \bigr] = a z^a F(a+1, b; c; z)
\end{equation}
allows us to derive other relations between hypergeometric and Legendre functions; we note that $z d/dz = -(y+1) d/dy$. With $a=-\ell$, $b=-\ell$, and $c=-2\ell$ we get
\begin{equation}
z^{-\ell} F(-\ell+1,-\ell;-2\ell;z) = \frac{(\ell-1)!^2}{2(2\ell-1)!} (y+1) P'_\ell(y),
\label{id3}
\end{equation}
in which a prime indicates differentiation with respect to $y$. With $a = \ell+1$, $b=\ell+1$, $c=2\ell+2$ we get instead
\begin{equation}
z^{\ell+1} F(\ell+2,\ell+1;2\ell+2;z) = -\frac {2(2\ell+1)!}{\ell!\, (\ell+1)!} (y+1) Q'_\ell(y). 
\label{id4}
\end{equation}
With $a=-\ell+1$, $b=-\ell$, and $c=-2\ell$ we obtain
\begin{equation}
z^{-\ell} F(-\ell+2,-\ell;-2\ell;z) = \frac{(\ell-2)!\, (\ell-1)!}{2(2\ell-1)!} (y+1)^2 P''_\ell(y). 
\label{id5}
\end{equation}
And with $a=\ell+2$, $b=\ell+1$, $c=2\ell+2$ we arrive at
\begin{equation}
z^{\ell+1} F(\ell+3,\ell+1;2\ell+2;z) = \frac {2(2\ell+1)!}{\ell!\, (\ell+2)!} (y+1)^2 Q''_\ell(y). 
\label{id6}
\end{equation}

\section{Radial functions: $A_\ell$, $B_\ell$, $C_\ell$, and $D_\ell$} 
\label{app:ABCD} 

With $x := M/r$, the explicit form of the radial functions is 
\begin{subequations} 
\label{ABCD2} 
\begin{align} 
A_2 &= 1, \\ 
B_2 &= -\frac{15}{16} \frac{1}{x^5} \ln f - \frac{5}{8} \frac{(1-x)(3-6x-2x^2)}{x^4 f^2}, \\ 
C_2 &= 1 - 2x^2, \\ 
D_2 &= -\frac{15}{16} \frac{1-2x^2}{x^5} \ln f - \frac{5}{8} \frac{3+3x-2x^2}{x^4}, 
\end{align} 
\end{subequations} 
\begin{subequations} 
\label{ABCD3} 
\begin{align} 
A_3 &= 1 - x, \\ 
B_3 &= -\frac{105}{16} \frac{1-x}{x^7} \ln f - \frac{7}{8} \frac{15-60x+65x^2-10x^3-2x^4}{x^6 f^2}, \\ 
C_3 &= 1 - 2x + \frac{4}{5} x^3, \\ 
D_3 &= -\frac{21}{16} \frac{5-10x+4x^3}{x^7}  \ln f - \frac{7}{8} \frac{15-15x-10x^2+2x^3}{x^6}, 
\end{align} 
\end{subequations} 
\begin{subequations} 
\label{ABCD4} 
\begin{align} 
A_4 &= 1 - 2x + \frac{6}{7} x^2, \\ 
B_4 &= -\frac{315}{64} \frac{7-14x+6x^2}{x^9} \ln f 
- \frac{21}{32} \frac{(1-x)(105-420x+440x^2-40x^3-4x^4)}{x^8 f^2}, \\ 
C_4 &= 1 - \frac{10}{3}x + \frac{20}{7}x^2 - \frac{8}{21} x^4, \\ 
D_4 &= -\frac{105}{64} \frac{21-70x+60x^2-8x^4}{x^9}  \ln f 
- \frac{7}{32} \frac{315-735x+270x^2+130x^3-12x^4}{x^8}, 
\end{align} 
\end{subequations} 
and 
\begin{subequations} 
\label{ABCD5} 
\begin{align} 
A_5 &= \frac{1}{3}(1-x)(3 - 6x + 2x^2), \\ 
B_5 &= -\frac{3465}{64} \frac{(1-x)(3-6x+2x^2)}{x^{11}} \ln f 
- \frac{33}{32} \frac{315-1890x+4095x^2-3780x^3+1288x^4-56x^5-4x^6}{x^{10} f^2}, \\ 
C_5 &= 1 - \frac{9}{2}x + \frac{20}{3}x^2 - \frac{10}{3}x^3 + \frac{4}{21} x^5, \\ 
D_5 &= -\frac{495}{128} \frac{42-189x+280x^2-140x^3+8x^5}{x^{11}}  \ln f 
- \frac{33}{64} \frac{630-2205x+2205x^2-420x^3-154x^4+8x^5}{x^{10}}.  
\end{align} 
\end{subequations} 

\section{Radial functions: $\gothm^\ell$, $\gothn^\ell$, and $\gotho^\ell$}  
\label{app:MNO} 

In all equations below, $x := M/r$. 

For $\ell = 0$ we have 
\begin{subequations}
\label{radialf_m0} 
\begin{align} 
\gothm^0_{tt} &= -\frac{15}{64} \frac{(1+2x)f^3}{x^4} (\ln f)^2 
- \frac{5}{16} \frac{3-9x-8x^2+28x^3-6x^4-4x^5}{x^3} \ln f 
\nonumber \\ & \quad \mbox{} 
- \frac{5}{48} \frac{9-36x-3x^2+90x^3-20x^4-20x^5+12x^6}{x^2 f}, \\ 
\gothm^0_{rr} &= -\frac{15}{64} \frac{(1+2x)(3-2x)}{x^4} (\ln f)^2 
- \frac{5}{16} \frac{9-15x-36x^2+58x^3-2x^4-4x^5}{x^3 f^2} \ln f 
\nonumber \\ & \quad \mbox{} 
- \frac{5}{48} \frac{(1-x)(3-6x-2x^2)(9+3x-30x^2-2x^3)}{x^2 f^3} 
\end{align} 
\end{subequations} 
and 
\begin{subequations}
\label{radialf_n0} 
\begin{align} 
\gothn^0_{tt} &= \frac{1}{4} \frac{(1+2x)f^3}{x^4} \ln f 
+ \frac{1}{30} \frac{15-45x-40x^2+190x^3-136x^4-20x^5}{x^3}, \\ 
\gothn^0_{rr} &= \frac{1}{4} \frac{(1+2x)(3-2x)}{x^4} \ln f 
+ \frac{1}{30} \frac{45-75x-180x^2+290x^3-16x^4-20x^5}{x^3 f^2}.  
\end{align} 
\end{subequations} 

For $\ell = 2$ we have 
\begin{subequations}
\label{radialf_m2} 
\begin{align} 
\gothm^2_{tt} &= -\frac{225}{224} \frac{(1+3x-5x^2)f^2}{x^4} (\ln f)^2 
- \frac{75}{448} \frac{24-105x+116x^2+52x^3-112x^4+32x^5}{x^3} \ln f 
\nonumber \\ & \quad \mbox{} 
- \frac{25}{224} \frac{(1-x)(36-315x+780x^2-390x^3-380x^4-8x^5)}{x^2 f}, \\ 
\gothm^2_{rr} &= -\frac{225}{224} \frac{2+x-5x^2}{x^4} (\ln f)^2 
- \frac{75}{448} \frac{48-225x+292x^2+4x^3-144x^4+32x^5}{x^3 f^2} \ln f 
\nonumber \\ & \quad \mbox{} 
- \frac{25}{224} \frac{(1-x)(72-495x+1020x^2-390x^3-460x^4-24x^5)}{x^2 f^3}, \\ 
\gothm^2 &= -\frac{225}{896} \frac{1-4x^2-16x^4}{x^4} (\ln f)^2 
- \frac{75}{448} \frac{6-99x-16x^2+198x^3-64x^4}{x^3} \ln f 
\nonumber \\ & \quad \mbox{} 
- \frac{25}{224} \frac{9-315x+276x^2+834x^3-428x^4+40x^5)}{x^2 f}
\end{align} 
\end{subequations} 
and 
\begin{subequations}
\label{radialf_n2} 
\begin{align} 
\gothn^2_{tt} &= \frac{15}{98} \frac{(7+21x-30x^2)f^2}{x^4} \ln f 
+ \frac{5}{49} \frac{21-251x^2+368x^3-78x^4+38x^5}{x^3}, \\ 
\gothn^2_{rr} &= \frac{15}{98} \frac{14+7x-30x^2}{x^4} \ln f 
+ \frac{5}{49} \frac{42-105x-97x^2+326x^3-106x^4+38x^5}{x^3 f^2}, \\ 
\gothn^2 &= \frac{15}{392} \frac{7-8x^2-152x^4}{x^4} \ln f 
+ \frac{5}{196} \frac{21+21x+4x^2+18x^3-264x^4}{x^3}.   
\end{align} 
\end{subequations} 

For $\ell = 4$ we have 
\begin{subequations}
\label{radialf_m4} 
\begin{align} 
\gothm^4_{tt} &= -\frac{75}{128} \frac{(3-19x+13x^2)f^2}{x^4} (\ln f)^2 
\nonumber \\ & \quad \mbox{} 
+ \frac{25}{1536} \frac{12495-75402x+164682x^2-153456x^3+51816x^4-672x^5-352x^6}{x^4} \ln f 
\nonumber \\ & \quad \mbox{} 
+ \frac{25}{768} \frac{(1-x)(12495-75186x+154336x^2-114664x^3+12476x^4+2056x^5+160x^6)}{x^3 f}, \\ 
\gothm^4_{rr} &= \frac{75}{128} \frac{1+11x-13x^2}{x^4} (\ln f)^2 
\nonumber \\ & \quad \mbox{} 
+ \frac{25}{1536} \frac{12495-74826x+161802x^2-149232x^3+50664x^4-1440x^5-352x^6}{x^4f^2} \ln f 
\nonumber \\ & \quad \mbox{} 
+ \frac{25}{768} \frac{(1-x)(12495-74898x+152896x^2-112744x^3+12476x^4+1416x^5+32x^6)}{x^3 f^3}, \\ 
\gothm^4 &= \frac{25}{384} \frac{9+189x-414x^2+220x^4}{x^4} (\ln f)^2 
\nonumber \\ & \quad \mbox{} 
+ \frac{25}{1536} \frac{12495-41506x+38868x^2-3408x^3-7064x^4+928x^5}{x^4} \ln f 
\nonumber \\ & \quad \mbox{} 
+ \frac{25}{2304} \frac{37485-162219x+211596x^2-58798x^3-34024x^4+7368x^5-384x^6)}{x^3 f}  
\end{align} 
\end{subequations} 
and 
\begin{subequations}
\label{radialf_n4} 
\begin{align} 
\gothn^4_{tt} &= \frac{5}{16} \frac{(6411-12848x+5516x^2)f^2}{x^4} \ln f 
\nonumber \\ & \quad \mbox{} 
+ \frac{5}{72} \frac{57699-288729x+473472x^2-264642x^3+19848x^4+2240x^5}{x^3}, \\ 
\gothn^4_{rr} &= \frac{5}{16} \frac{6403-12832x+5516x^2}{x^4} \ln f 
\nonumber \\ & \quad \mbox{} 
+ \frac{5}{72} \frac{57627-288369x+472944x^2-264498x^3+19944x^4+2240x^5}{x^3 f^2}, \\ 
\gothn^4 &= \frac{5}{144} \frac{57627-192528x+165528x^2-22400x^4}{x^4} \ln f 
\nonumber \\ & \quad \mbox{} 
+ \frac{5}{72} \frac{57627-134901x+49836x^2+24078x^3-2312x^4}{x^3}.   
\end{align} 
\end{subequations} 

For $\ell = 1$ we have 
\begin{subequations}
\label{radialf_m1} 
\begin{align} 
\gothm^1_h &= -\frac{9}{32}  \frac{(5-5x-10x^2+16x^3)f}{x^5} (\ln f)^2 
\nonumber \\ & \quad \mbox{} 
- \frac{1}{8}  \frac{45-180x+105x^2+314x^3-360x^4+72x^5-8x^6}{x^4 f} \ln f
\nonumber \\ & \quad \mbox{} 
- \frac{1}{24}  \frac{135-675x+765x^2+687x^3-1176x^4+112x^5-284x^6-12x^7}{x^3f^2}, \\ 
\gothm^1_j &= \frac{9}{64}  \frac{35-80x-10x^2+88x^3-32x^4}{x^5} (\ln f)^2 
\nonumber \\ & \quad \mbox{} 
+ \frac{1}{16}  \frac{315-1665x+2490x^2+312x^3-3040x^4+1482x^5-72x^6+88x^7}{x^4 f^2} \ln f
\nonumber \\ & \quad \mbox{} 
+ \frac{1}{48}  \frac{945-5940x+11835x^2-3834x^3-12294x^4+10296x^5-848x^6+784x^7-240x^8}{x^3f^3}, 
\end{align} 
\end{subequations} 
\begin{subequations}
\label{radialf_n1} 
\begin{align} 
\gothn^1_h &= \frac{3}{140}  \frac{(5-5x-10x^2+16x^3)f}{x^5} \ln f 
\nonumber \\ & \quad \mbox{} 
+ \frac{1}{1470}  \frac{315-1260x+735x^2+1708x^3-1134x^4+84x^5-1208x^6}{x^4 f}, \\
\gothn^1_j &= -\frac{3}{280}  \frac{35-80x-10x^2+88x^3-32x^4}{x^5} \ln f 
\nonumber \\ & \quad \mbox{} 
- \frac{1}{2940}  \frac{2205-11655x+17430x^2+2184x^3-20888x^4+9240x^5-672x^6+3088x^7}{x^4 f^2},  
\end{align} 
\end{subequations} 
and 
\begin{subequations}
\label{radialf_o1} 
\begin{align} 
\gotho^1_h &= \frac{3}{20}  \frac{(5-5x-10x^2+16x^3)f}{x^5} \ln f 
\nonumber \\ & \quad \mbox{} 
+ \frac{1}{70}  \frac{105-420x+245x^2+896x^3-1302x^4+308x^5+120x^6}{x^4 f}, \\
\gotho^1_j &= -\frac{3}{40}  \frac{35-80x-10x^2+88x^3-32x^4}{x^5} \ln f 
\nonumber \\ & \quad \mbox{} 
- \frac{1}{140}  \frac{735-3885x+5810x^2+728x^3-7224x^4+3836x^5-112x^6-128x^7}{x^4 f^2}.  
\end{align} 
\end{subequations} 

For $\ell=3$ we have 
\begin{subequations}
\label{radialf_m3} 
\begin{align} 
\gothm^3_{tt} &= -\frac{35}{64}  \frac{(10+20x-80x^2+37x^3)f^2}{x^5} (\ln f)^2 
\nonumber \\ & \quad \mbox{} 
- \frac{35}{384}  \frac{240-1395x+2735x^2-1792x^3-220x^4+416x^5-48x^6}{x^4} \ln f
\nonumber \\ & \quad \mbox{} 
- \frac{35}{576}  \frac{360-4185x+16350x^2-27003x^3+16804x^4-490x^5-900x^6-40x^7}{x^3f}, \\ 
\gothm^3_{rr} &= -\frac{35}{64}  \frac{5+35x-90x^2+37x^3}{x^5} (\ln f)^2 
\nonumber \\ & \quad \mbox{} 
- \frac{35}{384}  \frac{120-675x+1255x^2-672x^3-332x^4+320x^5-48x^6}{x^4 f^2} \ln f
\nonumber \\ & \quad \mbox{} 
- \frac{35}{576}  \frac{180-2925x+13170x^2-23703x^3+15908x^4-992x^5-788x^6-24x^7}{x^3f^3}, \\ 
\gothm^3 &= \frac{35}{128}  \frac{5-65x+170x^2-108x^3-24x^5}{x^5} (\ln f)^2 
\nonumber \\ & \quad \mbox{} 
+ \frac{35}{384}  \frac{60+435x-970x^2-176x^3+820x^4-144x^5}{x^4} \ln f
\nonumber \\ & \quad \mbox{} 
+ \frac{35}{576}  \frac{90+2295x-7365x^2+3006x^3+4258x^4-1188x^5+56x^6}{x^3f}, 
\end{align} 
\end{subequations} 
\begin{subequations}
\label{radialf_n3} 
\begin{align} 
\gothn^3_{tt} &= \frac{1}{72}  \frac{(30+60x-1475x^2+1346x^3)f^2}{x^5} \ln f 
\nonumber \\ & \quad \mbox{} 
+ \frac{1}{108}  \frac{90-90x-4845x^2+17613x^3-17894x^4+2674x^5+476x^6}{x^4}, \\
\gothn^3_{rr} &= \frac{1}{72}  \frac{15+105x-1505x^2+1346x^3}{x^5} \ln f 
\nonumber \\ & \quad \mbox{} 
+ \frac{1}{108}  \frac{45+180x-5400x^2+18033x^3-17924x^4+2614x^5+476x^6}{x^4 f^2}, \\
\gothn^3 &= -\frac{1}{144}  \frac{15-195x+2980x^2-5264x^3+1904x^5}{x^5} \ln f 
\nonumber \\ & \quad \mbox{} 
- \frac{1}{216}  \frac{45-540x+8415x^2-7542x^3-5138x^4+856x^5}{x^4},  
\end{align} 
\end{subequations} 
and 
\begin{subequations}
\label{radialf_o3} 
\begin{align} 
\gotho^3_{tt} &= \frac{7}{72}  \frac{(30+60x+205x^2-334x^3)f^2}{x^5} \ln f 
\nonumber \\ & \quad \mbox{} 
+ \frac{7}{108}  \frac{90-90x+195x^2-2547x^3+3994x^4-782x^5-196x^6}{x^4}, \\
\gotho^3_{rr} &= \frac{7}{72}  \frac{15+105x+175x^2-334x^3}{x^5} \ln f 
\nonumber \\ & \quad \mbox{} 
+ \frac{7}{108}  \frac{45+180x-360x^2-2127x^3+3940x^4-794x^5-196x^6}{x^4 f^2}, \\
\gotho^3 &= -\frac{7}{144}  \frac{15-195x-380x^2+1456x^3-784x^5}{x^5} \ln f 
\nonumber \\ & \quad \mbox{} 
- \frac{7}{216}  \frac{45-540x-1665x^2+2538x^3+1822x^4-440x^5}{x^4}.  
\end{align} 
\end{subequations} 

For $\ell=5$ we have 
\begin{subequations}
\label{radialf_m5} 
\begin{align} 
\gothm^5_{tt} &= -\frac{105}{128}  \frac{(10-73x+106x^2-38x^3)f^2}{x^5} (\ln f)^2 
\nonumber \\ & \quad \mbox{} 
+ \frac{7}{256}  \frac{28413-200091x+542736x^2-703570x^3+432440x^4
- 100888x^5+992x^6+288x^7}{x^5} \ln f
\nonumber \\ & \quad \mbox{} 
+ \frac{7}{1920} \frac{}{x^4f} \Bigl( 426195-3418560x+10756575x^2-16586610x^3
\nonumber \\ & \quad \mbox{} 
+12622044x^4 -4000696x^5+205204x^6+25464x^7+1360x^8 \Bigr), \\ 
\gothm^5_{rr} &= \frac{105}{128}  \frac{4+31x-78x^2+38x^3}{x^5} (\ln f)^2 
\nonumber \\ & \quad \mbox{} 
+ \frac{7}{256}  \frac{28413-198411x+532656x^2-682850x^3+416760x^4
- 99320x^5+2336x^6+288x^7}{x^5 f^2} \ln f
\nonumber \\ & \quad \mbox{} 
+ \frac{7}{1920} \frac{}{x^4f^3} \Bigl( 426195-3405960x+10668375x^2-16364010x^3
\nonumber \\ & \quad \mbox{} 
+12391044x^4-3937976x^5+235444x^6+17624x^7+240x^8 \Bigr), \\ 
\gothm^5 &= \frac{105}{128}  \frac{4+53x-212x^2+198x^3-36x^5}{x^5} (\ln f)^2 
\nonumber \\ & \quad \mbox{} 
+ \frac{7}{512}  \frac{56826-254757x+392520x^2-226300x^3+15520x^4
+ 18280x^5-1536x^6}{x^5} \ln f
\nonumber \\ & \quad \mbox{} 
+ \frac{7}{3840}  \frac{852390-4680945x+9045495x^2-6918990x^3+1247958x^4
+471788x^5-68928x^6+2080x^7}{x^4f},  
\end{align} 
\end{subequations} 
\begin{subequations}
\label{radialf_n5} 
\begin{align} 
\gothn^5_{tt} &= \frac{1}{16}  \frac{(62065-186238x+165586x^2-41408x^3)f^2}{x^5} \ln f 
\nonumber \\ & \quad \mbox{} 
+ \frac{1}{120}  \frac{930975-5586495x+12105800x^2-11176600x^3+3809140x^4
-165524x^5-11856x^6}{x^4}, \\
\gothn^5_{rr} &= \frac{1}{16}  \frac{62051-186196x+165558x^2-41408x^3}{x^5} \ln f 
\nonumber \\ & \quad \mbox{} 
+ \frac{1}{120}  \frac{930765-5585235x+12103210x^2-11174640x^3+3809000x^4
-165804x^5+11856x^6}{x^4 f^2}, \\
\gothn^5 &= \frac{1}{32}  \frac{124102-558601x+827824x^2-414096x^3+23712x^5}{x^5} \ln f 
\nonumber \\ & \quad \mbox{} 
+ \frac{1}{240}  \frac{1861530-6517485x+6520385x^2-1243040x^3-455910x^4+23872x^5}{x^4},  
\end{align} 
\end{subequations} 
and
\begin{subequations}
\label{radialf_o5} 
\begin{align} 
\gotho^5_{tt} &= \frac{7}{16}  \frac{(17713-53182x+47314x^2-11840x^3)f^2}{x^5} \ln f 
\nonumber \\ & \quad \mbox{} 
+ \frac{7}{120}  \frac{265695-1594815x+3457160x^2-3193240x^3+1088964x^4
-47412x^5-3408x^6}{x^4}, \\
\gotho^5_{rr} &= \frac{7}{16}  \frac{17699-53140x+47286x^2-11840x^3}{x^5} \ln f 
\nonumber \\ & \quad \mbox{} 
+ \frac{7}{120}  \frac{265485-1593555x+3454570x^2-3191280x^3+1088712x^4
-47468x^5-3408x^6}{x^4 f^2}, \\
\gotho^5 &= \frac{7}{32}  \frac{35398-159433x+236464x^2-118416x^3+6816x^5}{x^5} \ln f 
\nonumber \\ & \quad \mbox{} 
+ \frac{7}{240}  \frac{530970-1860525x+1863425x^2-356000x^3-130846x^4+6896x^5}{x^4}.  
\end{align} 
\end{subequations} 

\section{Radial functions: $\dot{e}^\ell_{tr}$} 
\label{app:edot_tr} 

The radial functions of Eq.~(\ref{edot_tr}) are given explicitly by 
\begin{subequations} 
\label{edot_tr_explicit} 
\begin{align} 
\dot{e}^2_{tr} &= 1 - \frac{3}{2}x - 3x^2 + 3x^3 
+ K_2 \Biggl[ -\biggl( \frac{15}{8} - \frac{45}{16}x - \frac{45}{8}x^2 + \frac{45}{8}x^3 \biggr) \ln f 
- \frac{5}{8} x(6-3x-19x^2) \Biggr], \\ 
\dot{e}^3_{tr} &= 1 - \frac{10}{3}x + 2x^2 + \frac{8}{5}x^3 - \frac{4}{5}x^4 
\nonumber \\ & \quad \mbox{} 
+ K_3 \Biggl[ -\biggl( \frac{105}{8} - \frac{175}{4}x + \frac{105}{4}x^2 + 21x^3 - \frac{21}{2}x^4 \biggr) \ln f 
- \frac{7}{12} x(45-105x+52x^3) \Biggr], \\ 
\dot{e}^4_{tr} &= 1 - \frac{55}{12}x + \frac{45}{7}x^2 - \frac{15}{7}x^3 - \frac{20}{21}x^4 + \frac{2}{7} x^5  
\nonumber \\ & \quad \mbox{} 
+ K_4 \Biggl[ -\biggl( \frac{2205}{32} - \frac{40425}{128}x + \frac{14175}{32}x^2 - \frac{4725}{32}x^3 
- \frac{525}{8}x^4 + \frac{315}{16}x^5 \biggr) \ln f 
\nonumber \\ & \quad \mbox{} 
- \frac{7}{64} x(1260-4515x+4005x^2+220x^3-618x^4) \Biggr], \\ 
\dot{e}^5_{tr} &= 1 - \frac{57}{10}x + \frac{23}{2}x^2 - \frac{28}{3}x^3 + 2x^4 + \frac{4}{7}x^5 - \frac{4}{35}x^6
\nonumber \\ & \quad \mbox{} 
+ K_5 \Biggl[ -\biggl( \frac{10395}{32} - \frac{118503}{64}x  + \frac{239085}{64}x^2 - \frac{24255}{8}x^3
+ \frac{10395}{16}x^4 + \frac{1485}{8}x^5 - \frac{297}{8}x^6 \biggr) \ln f 
\nonumber \\ & \quad \mbox{} 
- \frac{33}{160} x(3150 - 14805x + 22470x^2 - 10815x^3 -630x^4 + 694x^5) \Biggr], 
\end{align} 
\end{subequations} 
where  $x := M/r$. 

\section{Radial functions: $\gotha_\ell$, $\gothb_\ell$, $\gothc_\ell$, $\gothd_\ell$, $\gothf_\ell$, and $\gothg_\ell$}  
\label{app:ABCDFG}

The expressions below involve the polylogarithms 
\begin{subequations} 
\label{polylogs} 
\begin{align} 
\dilog(z) &:= -\int_1^z \frac{\ln t}{t-1}\, dt = \sum_{k=1}^\infty \frac{(1-z)^k}{k^2} = \mbox{Li}_2(1-z), \\ 
\polylog(n,z) &:= \frac{z}{\Gamma(n)} \int_0^\infty \frac{t^{n-1}}{e^t-z}\, dt 
= \sum_{k=1}^\infty \frac{z^k}{k^n} = \mbox{Li}_n(z). 
\end{align} 
\end{subequations} 
An overview of these functions is provided in Sec.~25.12 of Ref.~\cite{NIST:10}. Useful identities are 
\begin{equation} 
\dilog(f) = \dilog (r/2M) + \frac{1}{2} \bigl[ \ln(r/2M) \bigr]^2+ \ln(r/2M) \ln f + \frac{\pi^2}{6} 
\end{equation} 
and 
\begin{equation} 
\polylog(2,f) = -\dilog(r/2M) - \frac{1}{2} \bigl[ \ln(r/2M) \bigr]^2. 
\end{equation} 
We also have that $\polylog(n,1) = \zeta(n)$, with the right-hand side denoting the Riemann zeta function. We make use of the short-hand 
\begin{equation} 
\Theta := 2(\ln f)^2 - 2 \bigl[ \ln(r/2M) \bigr]^2 - 4\,  \dilog(r/2M), 
\label{Theta_def}
\end{equation} 
and the radial functions are expressed in terms of $x := M/r$.  

For $\ell = 2$ we have 
\begin{subequations} 
\label{ABCD_L2}
\begin{align} 
\gotha_2 &= \Theta\, A_2 
- \frac{2}{105} \frac{107-8x-832x^2+560x^3+280x^4}{f^2} \ln(2x) 
+ \frac{8}{3} \frac{x(1-x)(3-6x-2x^2)}{f^2} \ln f 
\nonumber \\ & \quad \mbox{} 
+ \frac{165+410x-4764x^3+4628x^4-1744x^5-1712x^6}{630\, x^2 f^2}, \\ 
\gothb_2 &= \Xi_2\, A_2 
- \frac{107-218x-202x^2+280x^3+140x^4}{28\, f^2} 
\biggl\{ 2\, \dilog\biggl(\frac{1}{2x} \biggr) + \bigl[ \ln(2x) \bigr]^2 \biggr\}
\nonumber \\ & \quad \mbox{} 
- \frac{107}{42} \frac{x(1-x)(3-6x-2x^2)}{f^2} \ln(2x) 
+ \frac{107-8x-832x^2+560x^3+280x^4}{28\, f^2} \ln(2x) \ln f 
\nonumber \\ & \quad \mbox{} 
- \frac{5}{2} \frac{x(1-x)(3-6x-2x^2)}{f^2} ( \ln f )^2 
- \frac{165+410x-4764x^3+4628x^4-1744x^5-1712x^6}{336\, x^2 f^2} \ln f
\nonumber \\ & \quad \mbox{} 
- \frac{495+1086x^2+842x^3-5620x^4-3428x^5}{504\, x f^2}, \\ 
\gothc_2 &= \Theta\, C_2 
- \biggl( \frac{214}{105} + 8x + \frac{412}{105} x^2 - \frac{16}{3}x^3 \biggr) \ln(2x) 
+ \frac{8}{3} x(3+3x-2x^2) \ln f 
\nonumber \\ & \quad \mbox{} 
+ \frac{2}{315} \frac{15+125x+480x^2+192x^3-205x^4+428x^5}{x^2}, \\ 
\gothd_2 &= \Xi_2\, C_2 
- \biggl( \frac{107}{28} + \frac{15}{2}x - \frac{1}{7}x^2 - 5x^3 \biggr) 
\biggl\{ 2\, \dilog\biggl(\frac{1}{2x}\biggr) + \bigl[ \ln(2x) \bigr]^2 \biggr\}
- \frac{107}{42} x(3+3x-2x^2) \ln(2x) 
\nonumber \\ & \quad \mbox{} 
+ \biggl( \frac{107}{28} + 15x + \frac{103}{14}x^2 - 10x^3 \biggr) \ln(2x) \ln f 
- \frac{5}{2} x(3+3x-2x^2) ( \ln f )^2 
\nonumber \\ & \quad \mbox{} 
- \frac{15+125x+480x^2+192x^3-205x^4+428x^5}{84\, x^2} \ln f 
- \frac{180-45x-204x^2+2462x^3+3428x^4}{504\, x}, \\ 
\gothf_2 &= 1 - \frac{3}{2} x - 3x^2 + 3x^3, \\ 
\gothg_2 &= -\biggl( \frac{15}{8} - \frac{45}{16}x - \frac{45}{8} x^2 + \frac{45}{8} x^3 \biggr) \ln f 
- \frac{5}{8} x(6 - 3x-19x^2), 
\end{align} 
\end{subequations} 
where 
\begin{equation} 
\Xi_2 := -15\, \zeta(3) - \frac{107}{84} \pi^2 - \frac{575}{168} 
+ 15\, \polylog(3,f) + \frac{15}{4} \biggl\{ 2\, \dilog\biggl(\frac{1}{2x}\biggr) 
+ \bigl[ \ln(2x) \bigr]^2 \biggr\} \ln f 
- \frac{5}{4} ( \ln f )^3. 
\label{Xi_L2} 
\end{equation}  

For $\ell = 3$ we have
\begin{subequations} 
\label{ABCD_L3}
\begin{align} 
\gotha_3 &= \Theta\, A_3 
- \frac{2}{105} \frac{65+95x-1160x^2+1560x^3-280x^4-56x^5}{f^2} \ln(2x) 
+ \frac{8}{15} \frac{x(15-60x+65x^2-10x^3-2x^4)}{f^2} \ln f 
\nonumber \\ & \quad \mbox{} 
+ \frac{105-766x^3-1664x^4+2404x^5+704x^6+208x^7}{630\, x^2 f^2}, \\ 
\gothb_3 &= \Xi_3\, A_3 
- \frac{65-115x-320x^2+650x^3-140x^4-28x^5}{4\, f^2} 
\biggl\{ 2\, \dilog\biggl(\frac{1}{2x} \biggr) + \bigl[ \ln(2x) \bigr]^2 \biggr\}
\nonumber \\ & \quad \mbox{} 
- \frac{13}{6} \frac{x(15-60x+65x^2-10x^3-2x^4)}{f^2} \ln(2x) 
+ \frac{65+95x-1160x^2+1560x^3-280x^4-56x^5}{4\, f^2} \ln(2x) \ln f 
\nonumber \\ & \quad \mbox{} 
- \frac{7}{2} \frac{x(15-60x+65x^2-10x^3-2x^4)}{f^2} ( \ln f )^2 
- \frac{105-766x^3-1664x^4+2404x^5+704x^6+208x^7}{48\, x^2 f^2} \ln f
\nonumber \\ & \quad \mbox{} 
- \frac{1575-13425x^2+46560x^3-76010x^4+25580x^5+5464x^6}{360\, x f^2}, \\ 
\gothc_3 &= \Theta\, C_3 
- \biggl( \frac{26}{21} + \frac{116}{21}x - 8x^2 - \frac{152}{35}x^3 + \frac{16}{15}x^4\biggr) \ln(2x) 
+ \frac{8}{15} x(15-15x-10x^2+2x^3) \ln f 
\nonumber \\ & \quad \mbox{} 
+ \frac{105+525x+1575x^2+1730x^3-1830x^4-2024x^5-520x^6}{1575\, x^2}, \\ 
\gothd_3 &= \Xi_3\, C_3 
- \biggl( \frac{65}{4} + 20x - \frac{105}{2}x^2 - 22x^3 + 7x^4 \biggr) 
\biggl\{ 2\, \dilog\biggl(\frac{1}{2x}\biggr) + \bigl[ \ln(2x) \bigr]^2 \biggr\}
- \frac{13}{6} x(15-15x-10x^2+2x^3) \ln(2x) 
\nonumber \\ & \quad \mbox{} 
+ \biggl( \frac{65}{4} + \frac{145}{2}x - 105x^2 - 57x^3 + 14x^4 \biggr) \ln(2x) \ln f 
- \frac{7}{2} x(15-15x-10x^2+2x^3) ( \ln f )^2 
\nonumber \\ & \quad \mbox{} 
- \frac{105+525x+1575x^2+1730x^3-1830x^4-2024x^5-520x^6}{120\, x^2} \ln f 
\nonumber \\ & \quad \mbox{} 
- \frac{630+2205x-6810x^2+22590x^3+6856x^4-5464x^5}{360\, x}, \\ 
\gothf_3 &= 1 - \frac{10}{3} x + 2x^2 + \frac{8}{5}x^3 - \frac{4}{5}x^4, \\ 
\gothg_3 &= -\biggl( \frac{105}{8} - \frac{175}{4}x + \frac{105}{4} x^2 + 21x^3 - \frac{21}{2} x^4 \biggr) \ln f 
- \frac{7}{12} x(45 - 105x + 52x^3), 
\end{align} 
\end{subequations} 
where 
\begin{equation} 
\Xi_3 := -105\, \zeta(3) - \frac{65}{12} \pi^2 - \frac{35}{8} 
+ 105\, \polylog(3,f) + \frac{105}{4} \biggl\{ 2\, \dilog\biggl(\frac{1}{2x}\biggr) 
+ \bigl[ \ln(2x) \bigr]^2 \biggr\} \ln f 
- \frac{35}{4} ( \ln f )^3. 
\label{Xi_L3} 
\end{equation}  

For $\ell = 4$ we have
\begin{subequations} 
\label{ABCD_L4}
\begin{align} 
\gotha_4 &= \Theta\, A_4 
- \frac{2}{24255} \frac{10997+31038x-343710x^2+668960x^3-405816x^4+33264x^5+3696x^6}{f^2} \ln(2x) 
\nonumber \\ & \quad \mbox{} 
+ \frac{8}{105} \frac{x(1-x)(105-420x+440x^2-40x^3-4x^4 )}{f^2} \ln f 
\nonumber \\ & \quad \mbox{} 
+ \frac{601965-808794x+9513105x^3-62779005x^4+100987160x^5-39087252x^6
-2390976x^7-351904x^8}{5093550\, x^2 f^2}, \\ 
\gothb_4 &= \Xi_4\, A_4 
\nonumber \\ & \quad \mbox{} 
- \frac{10997-17472x-101160x^2+271640x^3-184056x^4+16632x^5+1848x^6}{176\, f^2} 
\biggl\{ 2\, \dilog\biggl(\frac{1}{2x} \biggr) + \bigl[ \ln(2x) \bigr]^2 \biggr\}
\nonumber \\ & \quad \mbox{} 
- \frac{1571}{1320} \frac{x(1-x)(105-420x+440x^2-40x^3-4x^4)}{f^2} \ln(2x) 
\nonumber \\ & \quad \mbox{} 
+ \frac{10997+31038x-343710x^2+668960x^3-405816x^4+33264x^5+3696x^6}{176\, f^2} \ln(2x) \ln f 
\nonumber \\ & \quad \mbox{} 
- \frac{21}{8} \frac{x(1-x)(105-420x+440x^2-40x^3-4x^4)}{f^2} ( \ln f )^2 
\nonumber \\ & \quad \mbox{} 
- \frac{601965-808794x+9513105x^3-62779005x^4+100987160x^5-39087252x^6
-2390976x^7-351904x^8}{73920\, x^2 f^2} \ln f
\nonumber \\ & \quad \mbox{} 
- \frac{9029475-157269420x^2+640947195x^3-1238515000x^4+954141033x^5
-128241258x^6-14749202x^7}{554400\, x f^2}, \\ 
\gothc_4 &= \Theta\, C_4 
- \biggl( \frac{3142}{3465}+\frac{10348}{2079}x-\frac{77984}{4851}x^2+\frac{48}{7}x^3
+ \frac{215104}{72765}x^4-\frac{32}{105}x^5 \biggr) \ln(2x) 
\nonumber \\ & \quad \mbox{} 
+ \frac{8}{315} x(315-735x+270x^2+130x^3-12x^4) \ln f 
\nonumber \\ & \quad \mbox{} 
+ \frac{787185+2620863x+4623885x^2+37014355x^3-94366530x^4+9892680x^5
+15733044x^6+1055712x^7}{15280650\, x^2}, \\ 
\gothd_4 &= \Xi_4\, C_4 
- \biggl( \frac{10997}{176}+\frac{4445}{66}x-\frac{40885}{88}x^2+\frac{945}{4}x^3+\frac{11873}{132}x^4
- \frac{21}{2}x^5 \biggr) 
\biggl\{ 2\, \dilog\biggl(\frac{1}{2x}\biggr) + \bigl[ \ln(2x) \bigr]^2 \biggr\}
\nonumber \\ & \quad \mbox{} 
- \frac{1571}{3960} x(315-735x+270x^2+130x^3-12x^4) \ln(2x) 
\nonumber \\ & \quad \mbox{} 
+ \biggl( \frac{10997}{176}+\frac{90545}{264}x-\frac{12185}{11}x^2+\frac{945}{2}x^3+\frac{6722}{33}x^4
-21x^5 \biggr) \ln(2x) \ln f 
\nonumber \\ & \quad \mbox{} 
- \frac{7}{8} x(315-735x+270x^2+130x^3-12x^4) ( \ln f )^2 
\nonumber \\ & \quad \mbox{} 
- \frac{787185+2620863x+4623885x^2+37014355x^3-94366530x^4+9892680x^5
+15733044x^6+1055712x^7}{221760\, x^2} \ln f 
\nonumber \\ & \quad \mbox{} 
- \frac{3935925+20142675x-107432010x^2+329379470x^3-239927585x^4
-68936772x^5+14749202x^6}{554400\, x}, \\ 
\gothf_4 &= 1 - \frac{55}{12}x + \frac{45}{7}x^2 - \frac{15}{7}x^3 - \frac{20}{21}x^4 + \frac{2}{7}x^5, \\ 
\gothg_4 &= -\biggl( \frac{2205}{32} - \frac{40425}{128}x + \frac{14175}{32}x^2 - \frac{4725}{32}x^3
- \frac{525}{8}x^4 + \frac{315}{16}x^5\biggr) \ln f 
\nonumber \\ & \quad \mbox{} 
- \frac{7}{64} x(1260 - 4515x + 4005x^3 + 220x^3 - 618x^4), 
\end{align} 
\end{subequations} 
where 
\begin{equation} 
\Xi_4 := -\frac{2205}{4}\, \zeta(3) - \frac{10997}{528} \pi^2 + \frac{9849}{1760} 
+ \frac{2205}{4}\, \polylog(3,f) + \frac{2205}{16} \biggl\{ 2\, \dilog\biggl(\frac{1}{2x}\biggr) 
+ \bigl[ \ln(2x) \bigr]^2 \biggr\} \ln f 
- \frac{735}{16} ( \ln f )^3. 
\label{Xi_L4} 
\end{equation}  

For $\ell = 5$ we have
\begin{subequations} 
\label{ABCD_L5}
\begin{align} 
\gotha_5 &= \Theta\, A_5 
- \frac{2}{45045} \frac{16233+66549x-778064x^2+1963570x^3-1945720x^4+693448x^5
-32032x^6-2288x^7}{f^2} \ln(2x) 
\nonumber \\ & \quad \mbox{} 
+ \frac{8}{315} \frac{x(315-1890x+4095x^2-3780x^3+1288x^4-56x^5-4x^6)}{f^2} \ln f 
\nonumber \\ & \quad \mbox{} 
+ \frac{1}{4054050\, x^2 f^2} \Bigl( 363825-887040x+16705206x^3-93742936x^4 
\nonumber \\ & \quad \mbox{} 
+ 187595165x^5-145588820x^6 +32972548x^7+832992x^8+74208x^9 \Bigr), \\ 
\gothb_5 &= \Xi_5\, A_5 
- \frac{3}{208} \frac{16233-23541x-237524x^2+792400x^3-864640x^4+325080x^5
-16016x^6-1144x^7}{f^2} 
\nonumber \\ & \quad \mbox{} 
\times \biggl\{ 2\, \dilog\biggl(\frac{1}{2x} \biggr) + \bigl[ \ln(2x) \bigr]^2 \biggr\}
- \frac{773}{520} \frac{x(315-1890x+4095x^2-3780x^3+1288x^4-56x^5-4x^6)}{f^2} \ln(2x) 
\nonumber \\ & \quad \mbox{} 
+ \frac{3}{208} \frac{16233+66549x-778064x^2+1963570x^3-1945720x^4+693448x^5
-32032x^6-2288x^7}{f^2} \ln(2x) \ln f 
\nonumber \\ & \quad \mbox{} 
- \frac{33}{8} \frac{x(315-1890x+4095x^2-3780x^3+1288x^4-56x^5-4x^6)}{f^2} ( \ln f )^2 
\nonumber \\ & \quad \mbox{} 
- \frac{1}{12480\, x^2 f^2} \Bigl( 363825-887040x+16705206x^3-93742936x^4
\nonumber \\ & \quad \mbox{} 
+ 187595165x^5-145588820x^6+32972548x^7+832992x^8+74208x^9 \Bigr) \ln f
\nonumber \\ & \quad \mbox{} 
- \frac{1}{655200\, x f^2} \Bigl( 38201625-1040374125x^2+5017512780x^3-11635593900x^4
\nonumber \\ & \quad \mbox{} 
+ 13091688645x^5 -5790089291x^6+409750348x^7+29900942x^8 \Bigr), \\ 
\gothc_5 &= \Theta\, C_5
- \biggl( \frac{1546}{2145}+\frac{3401}{715}x-\frac{29852}{1287}x^2+\frac{32944}{1287}x^3
- \frac{16}{3}x^4-\frac{81904}{45045}x^5+\frac{32}{315}x^6 \biggr) \ln(2x) 
\nonumber \\ & \quad \mbox{} 
+ \frac{4}{315} x(630-2205x+2205x^2-420x^3-154x^4+8x^5) \ln f 
\nonumber \\ & \quad \mbox{} 
+ \frac{1}{56756700\, x^2} \Bigl( 2390850+4826745x-3929310x^2+224296569x^3
\nonumber \\ & \quad \mbox{} 
- 747363680x^4+588063910x^5 - 17957520x^6-36919948x^7-1038912x^8 \Bigr), \\ 
\gothd_5 &= \Xi_5\, C_5 
- \biggl( \frac{48699}{208}+\frac{102249}{416}x-\frac{621285}{208}x^2+\frac{783615}{208}x^3
- \frac{3465}{4}x^4-\frac{28395}{104}x^5+\frac{33}{2}x^6 \biggr) 
\nonumber \\ & \quad \mbox{} 
\times \biggl\{ 2\, \dilog\biggl(\frac{1}{2x}\biggr) + \bigl[ \ln(2x) \bigr]^2 \biggr\}
- \frac{773}{1040} x(630-2205x+2205x^2-420x^3-154x^4+8x^5) \ln(2x) 
\nonumber \\ & \quad \mbox{} 
+ \biggl( \frac{48699}{208} + \frac{642789}{416}x - \frac{783615}{104}x^2 + \frac{216195}{26}x^3
- \frac{3465}{2}x^4 - \frac{15357}{26}x^5 + 33x^6 \biggr) \ln(2x) \ln f 
\nonumber \\ & \quad \mbox{} 
- \frac{33}{16} x(630-2205x+2205x^2-420x^3-154x^4+8x^5) ( \ln f )^2 
\nonumber \\ & \quad \mbox{} 
- \frac{1}{174720\, x^2} \Bigl( 2390850+4826745x-3929310x^2+224296569x^3
\nonumber \\ & \quad \mbox{} 
- 747363680x^4+588063910x^5-17957520x^6-36919948x^7-1038912x^8 \Bigr) \ln f 
\nonumber \\ & \quad \mbox{} 
- \frac{1}{1310400\,x} \Bigl(35862750+218139075x-1660374450x^2+5709727485x^3 
\nonumber \\ & \quad \mbox{} 
- 7753670715x^4+2539628490x^5 + 602388968x^6-59801884x^7 \Bigr), \\ 
\gothf_5 &= 1 - \frac{57}{10}x + \frac{23}{2}x^2 - \frac{28}{3}x^3 + 2x^4 + \frac{4}{7}x^5 - \frac{4}{35}x^6, \\ 
\gothg_5 &= -\biggl( \frac{10395}{32} - \frac{118503}{64}x + \frac{239085}{64}x^2 - \frac{24255}{8}x^3 
+ \frac{10395}{16}x^4 + \frac{1485}{8}x^5 - \frac{297}{8}x^6 \biggr) \ln f 
\nonumber \\ & \quad \mbox{} 
- \frac{33}{160} x(3150-14805x+22470x^2-10815x^3-630x^4+694x^5), 
\end{align} 
\end{subequations} 
where 
\begin{align} 
\Xi_5 &:= -\frac{10395}{4}\, \zeta(3) - \frac{16233}{208} \pi^2 + \frac{34881}{416} 
+ \frac{10395}{4}\, \polylog(3,f) 
\nonumber \\ & \quad \mbox{} 
+ \frac{10395}{16} \biggl\{ 2\, \dilog\biggl(\frac{1}{2x}\biggr)  + \bigl[ \ln(2x) \bigr]^2 \biggr\} \ln f 
- \frac{3465}{16} ( \ln f )^3. 
\label{Xi_L5} 
\end{align}  

\bibliography{/Users/poisson/writing/papers/tex/bib/master}
 \end{document}